\documentclass[footinbib,a4paper,aps,prx,superscriptaddress,reprint,twocolumn,preprintnumbers,amsmath,amssymb,nobalancelastpage,10pt]{revtex4-1}

\usepackage{amsmath}
\usepackage{verbatim}
\usepackage{graphicx}
\usepackage{bbold}

\usepackage[toc,page]{appendix}
\usepackage{color}
\usepackage{tikz}

\usepackage{subfigure}
\usepackage{float}
\usepackage{soul}
\usepackage{mathptmx}
\usepackage{amssymb}
\usepackage{amsmath}
\usepackage{amsfonts}
\usepackage{array}
\usepackage[T1]{fontenc}

\usepackage{bm}
\usepackage{xcolor}
\graphicspath{{figures/}}

\usepackage{braket}

\definecolor{mygrey}{gray}{0.35}
\definecolor{myblue}{rgb}{0.2,0.2,0.8}
\definecolor{myzard}{cmyk}{0,0,0.05,0}
\definecolor{mywhite}{rgb}{1,1,1}
\definecolor{myred}{rgb}{1,0.,0.3}

\usepackage[colorlinks=true,citecolor=myblue,linkcolor=myred]{hyperref}

\def\beq{{\begin{equation}}}
\def\eeq{{\end{equation}}}
\def\k{{\rm k}}

\def\q{{\rm q}}
\def\r{{\rm r}}
\def\s{{\rm s}}
\def\i{{\rm i}}
\def\x{{\rm x}}

 \def\ee{\mathord{\rm e}}
 
 \def\ii{\mathord{\rm i}}

\def\tr{\mathop{\rm Tr}}
\def\half{\textstyle\frac{1}{2}}

\renewcommand{\ii}{{\rm i}}

\def\beq{\begin{equation}}
\def\eeq{\end{equation}}
\def\barray{\begin{eqnarray}}
\def\earray{\end{eqnarray}}

\begin{document}

\title{Interacting Dirac fields in an expanding universe:  dynamical  condensates and particle production}

\author{C. Fulgado-Claudio}
\affiliation{Instituto de F\'isica Te\'orica, UAM-CSIC, Universidad Aut\'onoma de Madrid, Cantoblanco, 28049 Madrid, Spain. }
\author{P. Sala}
\affiliation{Department of Physics and Institute for Quantum Information and Matter,
California Institute of Technology, Pasadena, CA 91125, USA}
\affiliation{Walter Burke Institute for Theoretical Physics, California Institute of Technology, Pasadena, CA 91125, USA }
\author{D. Gonz\'alez-Cuadra}
\affiliation{Institute for Theoretical Physics, University of Innsbruck, 6020 Innsbruck, Austria}
\affiliation{Institute for Quantum Optics and Quantum Information of the Austrian Academy of Sciences, 6020 Innsbruck, Austria}

\author{A. Bermudez}
\affiliation{Instituto de F\'isica Te\'orica, UAM-CSIC, Universidad Aut\'onoma de Madrid, Cantoblanco, 28049 Madrid, Spain.  }

\begin{abstract}

The phenomenon of particle production for quantum field theories  in  curved spacetimes is crucial to understand the large-scale structure of a universe from an inflationary epoch. In contrast to the free and fixed-background case, the  production of particles with   strong interactions and back reaction  is not completely understood, especially in  situations that require going beyond perturbation theory. In this work, we present advances in this direction by focusing on a self-interacting  field theory of Dirac fermions in an expanding Friedmann-Robertson-Walker universe. By using a Hamiltonian lattice regularization with continuous conformal time and rescaled fields,  this model becomes amenable to either a cold-atom analogue-gravity quantum simulation,  or  a dynamical variational approach. Leveraging a family of variational fermionic Gaussian states, we investigate how   dynamical mass generation and the formation of fermion condensates  associated to certain broken symmetries modify some well-known results of the free field theory.  In particular, we study how the non-perturbative  condensates arise and, more importantly, how their real-time evolution has an impact on particle production. Depending on the Hubble expansion rate, we  find  an interesting interplay of interactions and particle production, including a non-trivial back reaction on the condensates  and a parity-breaking spectrum of produced particles. 

\end{abstract}

\maketitle

\setcounter{tocdepth}{2}
\begingroup
\hypersetup{linkcolor=black}
\tableofcontents
\endgroup

  \section{\bf Introduction}

    \begin{figure*}[t!]
    \centering
    \includegraphics[width=1.6\columnwidth]{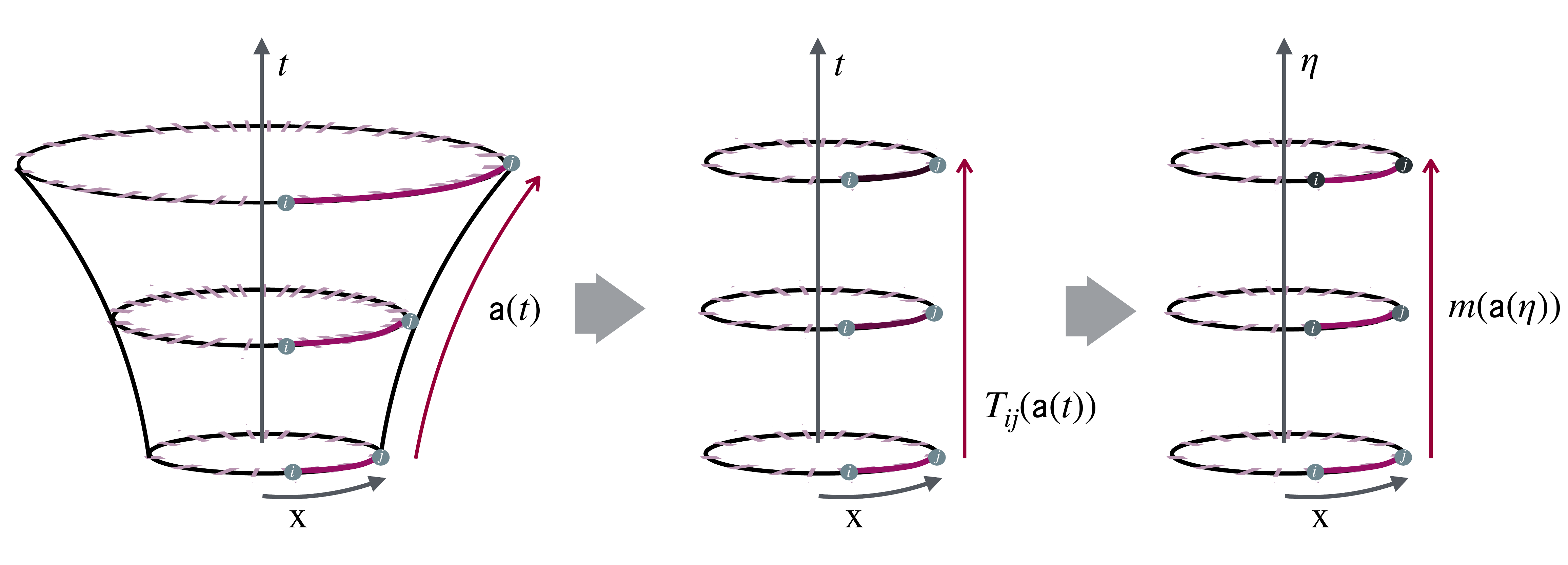}
    \caption{\textbf{Scheme for the simulation of particle production in an expanding universe.} First, we discretize the theory, and the expansion of the universe, here depicted in $(1+1)$-dimensional case with spatial periodic boundary conditions,  is manifested in a growing separation of two lattice points, which  increases in time according to the so-called scale factor $\mathsf{a}(t)$. In the middle panel, we depict how   by rescaling the spinor field with the scale factor, the discretized lattice field theory in the expanding universe would correspond to a  field theory in a static lattice, but endowed with  time-dependent tunnelings $T_{ij}(\mathsf{a}(t))$. Finally, in the rightmost panel,  by performing a coordinate transformation and introducing the so-called conformal time $\eta$, we can encode all the effects of the expanding universe  in a dynamical mass term.}

    \label{fig:scheme}
\end{figure*}

  Quantum field theories (QFTs) in curved spacetimes describe how fundamental particles    evolve under  a classical gravitational field, which can   lead to  various non-trivial effects~\cite{Birrell:1982ix, Parker:2009uva}.  The very notion of  a particle, which is simply the excitation of a quantum field in  a flat spacetime, needs to be carefully revisited in a gravitational scenario, as particles can be created by gravitational fields with either time dependence or  an event horizon~\cite{Birrell:1982ix, Parker:2009uva,Ford_2021}. Some of the related phenomena include the evaporation of black holes due to the emission of thermal Hawking radiation \cite{Hawking:1974rv}; the Unruh effect, describing how inertial and accelerated observers can measure a different particle content~\cite{PhysRevD.14.870}; and the production of particles in cosmological scenarios that involve an expanding universe~\cite{PhysRevLett.21.562, PhysRevD.3.346}. Although these effects already appear at the level of  non-interacting QFTs in a fixed gravitational background, a regime in which  they can be fully  understood~\cite{Ford_2021}, there are important subtleties when departing from this idealised limit. For instance, one  needs to consider   renormalization-group aspects to assess the back reaction of the produced particles  on the gravitational field, as well as   a perturbative loop expansion of the effective action when considering interacting QFTs~\cite{Birrell:1982ix, Parker:2009uva}. Moreover, there can be non-perturbative situations that require  a different approach  to understand the effect of interactions  on the aforementioned gravitational phenomena. In particular,  the vacuum of the QFTs can develop spontaneous symmetry breaking and dynamical  mass generation, which would  have profound consequences in particle production.  These  questions  deserve further attention, as they cannot be addressed using traditional perturbative approaches. 
 
  We note that, in spite of the important foundational character of these questions, a direct measurement of these  gravitational effects  lies far from current experimental capabilities in the  original cosmological scenario~\cite{carroll2003spacetime}.  In the last decades, an alternative possibility to experimentally test the predictions of QFTs in curved spacetimes has seen an increased activity:   \textit{analogue gravity} \cite{Barcelo:2005fc,Jacquet:2020bar}. This approach considers table-top experimental platforms which, although being very different from the fundamental relativistic fields in a gravitational curved spacetime, can effectively mimic the relevant quantum dynamics associated to a particular gravitational problem. The scales in analogue gravity are completely different from the original cosmological ones, and novel   protocols to measure observables of interest can be developed.  This can provide an interesting experimental test of the foundational effects predicted for QFTs in curved spacetimes, and actually  provide useful information to guide our understanding of   
  interacting regimes that are beyond the reach of traditional theoretical tools. Among other results, these analog simulators have successfully observed the Hawking radiation of bosonic fields using light propagation in non-linear media 
  \cite{Philbin2008, Belgiorno2010, Drori2019}, and  sound waves in Bose-Einstein condensates \cite{Garay2000, Lahav2010, Steinhauer2016, Kolobov2021}. The   Unruh effect and  particle production  in expanding universes  have also been experimentally proposed and tested  with ultracold atoms, trapped ions, and microcavity polaritons \cite{Hung:2012nc,Eckel2018,PhysRevA.109.013316, Wittemer2019, Hu2019, Steinhauer:2021fhb, Viermann2022, Tolosa-Simeon2022,PhysRevLett.130.111501,doi:10.1073/pnas.2301287120}.

  Among the various possible experimental platforms for quantum simulations, ultra-cold atoms in optical lattices are particularly promising to  test the effect of interactions in a non-trivial analogue-gravity scenario. First of all,  they offer the possibility of implementing  tunable interactions via the external control of the  atomic scattering in the dilute gas~\cite{Jaksch_2005, Lewenstein_2007,RevModPhys.80.885, Gross_2017}. Additionally, the optical lattice  provides a natural connection to the Hamiltonian lattice regularization of QFTs~\cite{PhysRevD.11.395}, which was originally introduced in Euclidean space to account for non-perturbative effects such as the confinement of quarks~\cite{PhysRevD.10.2445,gattringer_lang_2010}. Inspired by recent experimental progress with ultracold fermionic atoms in Raman optical lattices~\cite{Zhang_2018,PhysRevLett.110.076401,PhysRevLett.112.086401,PhysRevLett.113.059901,doi:10.1126/science.aaf6689,PhysRevLett.121.150401,doi:10.1126/sciadv.aao4748,https://doi.org/10.48550/arxiv.2109.08885}, we proposed in a previous work~\cite{FulgadoClaudio2023fermionproduction} that this setup  could be used for the analogue-gravity  quantum simulation of particle production of fermionic Dirac fields in an expanding universe. As emphasized in that work, using conformal time coordinates and rescaled fields allows to express  the expanding universe QFT by an effectively-flat one of massive Dirac fermions with a  dynamical multiplicative renormalization of the bare fermion mass. This feature, which is also common to scalar QFTs~\cite{Mukhanov}, simplifies considerably the implementation of the target QFT (see Fig.~\ref{fig:scheme}). As noted  in the outlook of our work~\cite{FulgadoClaudio2023fermionproduction}, the cold-atom platform \cite{RevModPhys.80.885} also allows to naturally implement  four-Fermi contact interactions~\cite{Fermi1934,doi:10.1119/1.1974382,PhysRev.122.345,PhysRev.124.246,PhysRevD.10.3235,hep-lat/9706018,Tong_2024} in the quantum simulator, which can be  tuned by Feshbach resonances \cite{Cazalilla:2014wfa}. Such four-Fermi QFTs are  nowadays considered as effective theories that capture some key properties of gauge theories, such as  dynamical mass generation by the spontaneous breakdown of chiral symmetry~\cite{RevModPhys.64.649}. The impact of these non-perturbative effects on the real-time evolution of fermionic fields in an expanding universe presents, however, a more challenging task.

\begin{figure}[t!]
    \centering
    \includegraphics[width=0.8\columnwidth]{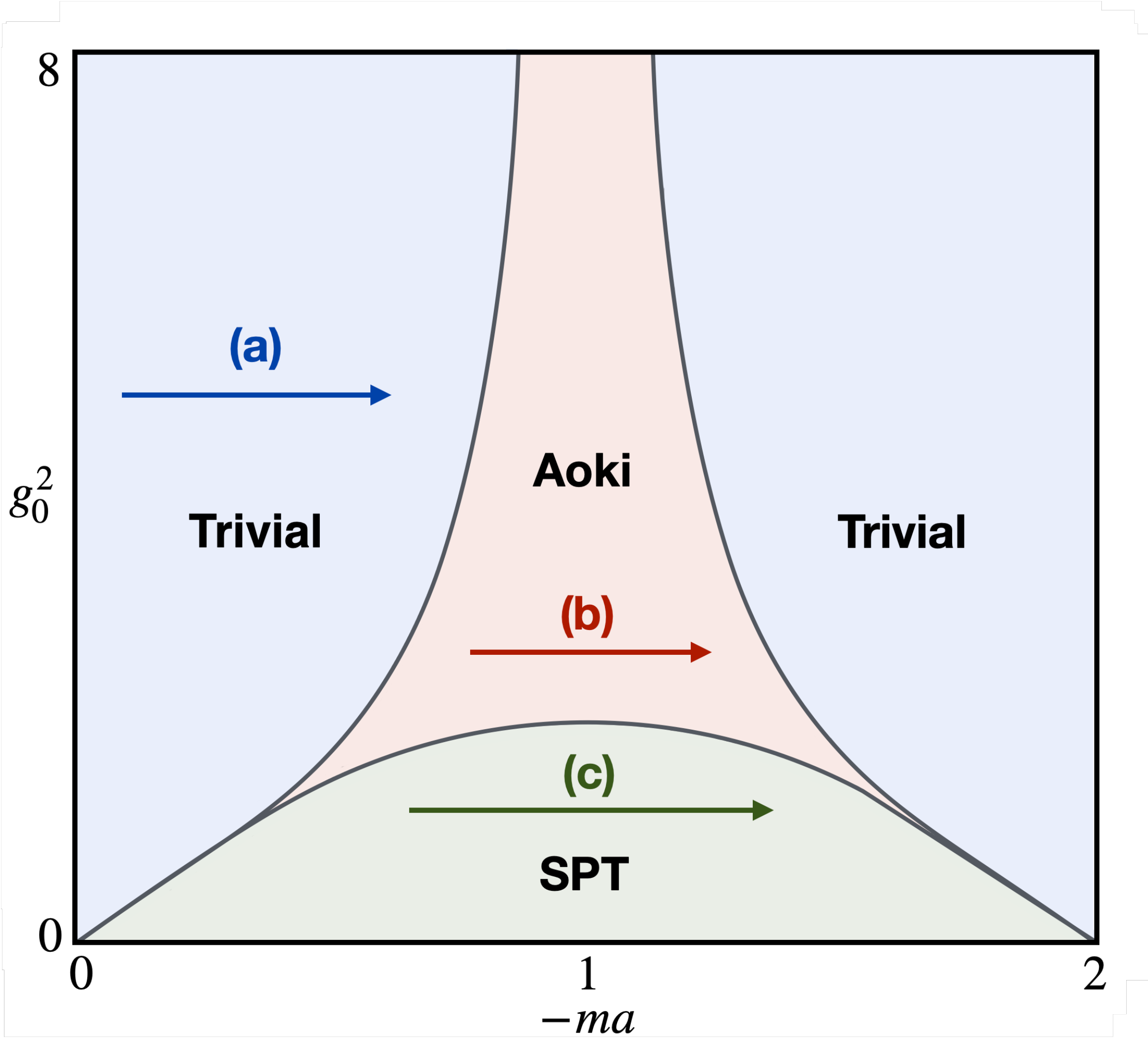}
    \caption{{\bf Phase diagram for the Gross-Neveu-Wilson QFT. }We depict the phase diagram of the lattice QFT as a function of the bare mass $ma$ ($a$ being the lattice spacing) and four-Fermi coupling strength $g_0^2$.The expanding universe amounts to an effective dynamical renormalization of the mass with the scale factor $\mathsf{a}(t)$, whereas the dimensionless quartic coupling $g_0^2$ remains static.  Therefore,  the system dynamically goes through different trajectories  in parameter space as the universe expands, with a velocity determined by $\mathsf{H}$. The arrows \textbf{(a)}, \textbf{(b)} and \textbf{(c)} depict the evolution of the parameters for an expansion within the trivial, Aoki, and SPT phases, respectively.}
    \label{fig:trajectories_explanation}
\end{figure}

In this work, we take a first step in this direction and study non-perturbative phenomena in a Gross-Neveu model in an expanding universe~\cite{PhysRevD.10.3235}, assuming that there is a vanishing back-reaction on the gravitational field from the creation of particles in the Dirac field. This corresponds to a $(1+1)$-dimensional QFT of Dirac fermions in an expanding universe described by a Friedmann-Robertson-Walker spacetime. Resorting to a Hamiltonian Wilson-type lattice formulation of the Dirac fermions~\cite{Wilson1977,GOLTERMAN1993219,PhysRevLett.83.2636,Kaplan:2009yg,PhysRevLett.105.190404,PhysRevLett.108.181807,PhysRevX.7.031057,alejandroGrossNeveu2018,PhysRevB.106.045147}, we develop a variational approach based on fermionic Gaussian states~\cite{RevModPhys.77.513,WANG20071,Kraus_2009,Kraus_2010,doi:10.1142/S1230161214400010,SHI2018245,Surace22}, showing that it can account for non-perturbative effects such as dynamical mass generation and spontaneous broken symmetries characterised by the appearance of fermion condensates, and their effects on the real-time dynamics of the QFT. We find that, by tuning the Hubble parameter $\mathsf{H}$ that controls the rate of a de Sitter expansion in the  Gross-Neveu-Wilson QFT, different regimes of adiabaticity can be explored. This, along with the rich phase diagram of the interacting lattice field theory~\cite{PhysRevX.7.031057,alejandroGrossNeveu2018}, which is depicted in Fig.~\ref{fig:trajectories_explanation} 
 and includes both  topologically-trivial and  symmetry-protected topological (SPT) phases~\cite{originaltopophases,doi:10.1063/1.3149495,doi:10.1063/1.3149495,classification_spt} with broken chiral symmetry,  as well as a parity-breaking Aoki phase~\cite{PhysRevD.30.2653}, leads to an interesting interplay between interactions and particle production. First, we show that in the quench limit, namely for sudden expansions, non-zero interactions lead to a non-perturbative shift in the parameters of the lattice field theory 
via a static value  given by the fermion condensates. This results in a change of the particle production in the Dirac QFT when compared to its non-interacting counterpart, which can actually depend  on the specific phase in which the expansion occurs (see Fig.~\ref{fig:trajectories_explanation}). As detailed below, we find interesting effects such as a  modification of the mass of particles at which maximal particle production is attained; a decrease in the density of produced particles in the trivial phase as the interactions increase; or an increase of particle production in both the SPT and Aoki phases. 
On the other hand, when moving away from the quench limit, the situation becomes more interesting, as the condensates display a non-trivial real-time evolution that affects the density of produced particles. In turn, this leads to substantial modifications of the observables, such as an inversion in the dependence of the density of produced particles with the interaction strength and a back-reaction in the form of periodic synchronised oscillations of the condensates and the produced particles. Finally, we show that for expansions within the Aoki phase, parity-breaking is manifested at the level of the spectra of production, which become asymmetric with respect to the zero-momentum axis. We note that all of these interesting effects are predicted within the fermionic Gaussian state formalism, which has limitations in terms of the amount of correlations that can be accounted for. Although, as we show, this formalism agrees with a large-$N$ expansion that captures correctly the shape of the static phase diagram (see Fig.~\ref{fig:trajectories_explanation}), it would be very interesting to test these  predictions for the real-time dynamics with other methods such as tensor networks~\cite{RevModPhys.77.259,SCHOLLWOCK201196,RevModPhys.93.045003} and, ultimately, with the potential experimental results of the Raman-lattice cold-atom quantum simulator.  

The article is organized as follows. In Sec.~\ref{sec:2}, we first present the QFT of  Dirac fermions with  four-Fermi interactions on an expanding universe. We show how this QFT can be connected  
to the one-flavor Gross-Neveu model with a bare mass that depends on the expansion scale factor $\mathsf{a}(\eta)$. After performing a Wilson-type discretization, we introduce a 
variational family of fermionic Gaussian states (FGSs) to analyze the full interacting Hamiltonian field theory. Specifically, we use both imaginary-time evolution (ITE) and real-time evolution (RTE) to compute the corresponding correlation matrix, and use it to obtain the spectra and the density of produced particles. 
In Sec.~\ref{sec:3}, we relate the phenomenon of particle production to the adiabatic approximation by changing the Hubble constant $\mathsf{H}$ that sets the de Sitter expansion  $\mathsf{a}(\eta)=-1/\mathsf{H}\eta$,  allowing us to explore different regimes of adiabaticity. Subsequently, and using this variational approach, we explore various non-perturbative phenomena in the presence of an expanding universe, exploring distinct regions of the phase diagram as well as various regimes of adiabaticity. Finally, in Sec.~\ref{sec:4}, we present the conclusions and the outlook of this work. 

  \section{\bf Variational Gaussian states for self-interacting Dirac fermions in curved spacetimes}
\label{sec:2}
  \subsection{ Gross-Neveu model in an expanding universe}
  \label{subsec:part_prod}
  
  Let us introduce the Hamiltonian field theory under consideration. For a general spacetime with metric $g^{\mu\nu}$, the action for a QFT of self-interacting Dirac fermions is given by 
  \beq
  \label{eq:action}
S=\!\!\int\!\! {\rm d}^Dx\;\sqrt{-g}\left(\overline{\psi}(x)\!\left(\!\!-\tilde\gamma^\mu\nabla_\mu-m\right)\!\psi(x)+\frac{g_0^2}{2}\left(\overline{\psi}\psi\right)^{\!2}\!\right)\!,
\eeq
where $x=(t,\bf{x})$ are the spacetime coordinates,  and we use Einstein's convention of summation over repeated indexes and natural units $\hbar=c=1$. 
In this action, $\tilde\gamma^\mu$ are generalized Dirac matrices that satisfy $\{\tilde\gamma^\mu,\;\tilde\gamma^\nu\}=2g^{\mu\nu}$, where $g^{\mu\nu}$ is the metric and $g=\det(g^{\mu\nu})$. Here,  $\nabla_\mu$ is the covariant derivative for fermions, which acts over spinor fields and their adjoint $\psi(x),\overline{\psi}(x)=\ii\psi^\dagger(x)\gamma^0$ (where the $\ii$ factor in the definition of the adjoint field stems from the mostly-plus convention) 
as $\nabla_\mu\psi=\left(\partial_\mu+\omega_\mu\right)\psi$ and $\nabla_\mu\bar\psi=\partial_\mu\bar\psi-\bar\psi\omega_\mu$, and $\omega_\mu$ is the so-called spin connection \cite{Shapiro:2016pfm}.
In the present work, we focus on self-interacting Dirac fermions with mass $m$ and interaction strength $g_0^2$ evolving in the background of a $(1+1)$-dimensional Friedmann-Robertson-Walker (FRW) spacetime. Neglecting  back-reaction effects on the gravitational field, the FRW expansion is controlled by a single parameter, the scale factor $\mathsf{a}(t)$. It will prove crucial to work in conformal time, which is expressed as a function of the cosmological time via ${\rm d}\eta={\rm d}t/\mathsf{a}(t)$
  (see  Appendix \ref{app:dirac_in_CS}). 

As discussed in Appendix \ref{app:dirac_in_CS}, 
  the corresponding Hamiltonian field theory can be effectively expressed as a 
  flat-spacetime QFT with a time-dependent fermion mass, which is actually a generic result for any FRW spacetime in $D=d+1$ spacetime dimensions. Focusing on the case $d=1$, and working with a mostly-plus convention for the metric, one finds
  \beq
  \label{eq:GN_ham_QFT}
  H=\int\!{\rm d}{\rm x}\bigg(\overline{\chi}(x)\left(\gamma^1\partial_{\rm x}+m\mathsf{a}(\eta)\right)\chi(x)-\frac{g_0^2}{2}\big(\overline{\chi}\chi\big)^2\bigg).
  \eeq

Here, $\chi(x)$ and $\overline{\chi}(x)=\i\chi^\dagger(x)\gamma^0$ are the fermionic spinor  and adjoint fields, respectively, satisfying canonical anti-commutation relations, and   we have introduced the  Dirac matrices  $\gamma^0=-\ii\sigma^z$, $\gamma^1=\sigma^y$. We note that in the mostly-plus convention, the matrix $\gamma^0$ is anti-Hermitian, while $\gamma^i$, with $i=1,\;...,\;d$, are Hermitian. 
We also note that  ${\chi}(x)$ is connected to the original field operator in curved spacetime $\psi(x)$ via the rescaling ${\chi}(x)=\sqrt{\mathsf{a}(\eta)}{\psi}(x)$. Being $g_0^2$ dimensionless, this rescaling leaves the interaction term unchanged, as discussed in more detail in the Appendix.

  In our previous work~\cite{FulgadoClaudio2023fermionproduction}, we considered 
  the following time dependence for the scale factor 
\begin{equation}
\label{eq:scale_factor}
    \mathsf{a}(\eta) = 
     \begin{cases}
       -\frac{1}{{\mathsf{H}}\eta_{\rm 0} },&\quad\text{if }\eta<\eta_{\rm 0},\\-\frac{1}{{\mathsf{H}}\eta},&\quad\text{if }\eta_{\rm 0}\leq\eta\leq\eta_{\rm f}, \\ -\frac{1}{{\mathsf{H}}\eta_{\rm f}},&\quad\text{if }\eta_{\rm f}<\eta, 
     \end{cases}
\end{equation}
which corresponds to a de Sitter exponential expansion $\mathsf{a}(t)=\ee^{\mathsf{H}t}$~\cite{Baumann:2022mni} between two asymptotically flat regions in the remote past and distant future. This fixes the initial and final sizes of the universe, $\mathsf{a}_0$ and $\mathsf{a}_{\rm f}$, and the initial and final times depend on the so-called Hubble parameter $\mathsf{H}$ via $\eta_{0}=-{1}/{\mathsf{Ha}_{0}}$ and $\eta_{\rm f}=-{1}/{\mathsf{Ha}_{\rm f}}$. Hence, $\mathsf{H}$ controls the relative rate of expansion between the two in- and out- asymptotic Minkowski spacetimes, both of which have a well-defined static vacuum state and Fock spaces~\cite{Peskin} and, thus, a well-defined notion of particle \cite{preskill_notes_QFT_curved}. Note that, by tuning $\mathsf{H}$, one can control the duration of an expansion from $\mathsf{a}_0$ to $\mathsf{a}_{\rm f}$, 
\beq
\Delta\eta=\eta_{\rm f}-\eta_0=-\frac{1}{{\mathsf{H}}\mathsf{a}_{\rm f}}+\frac{1}{{\mathsf{H}}\mathsf{a}_0}=\frac{1}{\mathsf{H}}\frac{\mathsf{a}_{\rm f}-\mathsf{a}_0}{\mathsf{a}_0\mathsf{a}_{\rm f}},
\label{eq:duration_expansion}
\eeq
which is inversely proportional to $\mathsf{H}$.
The focus of~\cite{FulgadoClaudio2023fermionproduction} was to explore how, in the absence of interactions $g_0^2=0$, an inertial observer in the asymptotic out region would detect a production of particles as a consequence of the expansion. In order to study this problem, we solved the equations of motion for the so-called mode functions $u_\k(x)$, $v_\k(x)$ of the field

\begin{equation}
    {\psi}(x)=\!\!\int\!\frac{{\rm d}\mathrm{k}}{2\pi}\frac{1}{\sqrt{\mathsf{a}(\eta)}}\left({a}^{\phantom{\dagger}}_\mathrm{k} u^{\phantom{\dagger}}_{\k}(x)+{b}^\dag_{-\mathrm{k}} v^{\phantom{\dagger}}_{-\k}(x)\right),
    \label{eqn:decompositionmodefunctions}
\end{equation}
where the creation and annihilation operators satisfy the usual anti-commutation algebra $\{{a}^{\phantom{\dagger}}_\mathrm{k},{a}^\dagger_{\mathrm{k}'}\}=\{{b}^{\phantom{\dagger}}_{\mathrm{k}},{b}^\dagger_{\mathrm{k}'}\}=2\pi \delta(\mathrm{k}-\mathrm{k}')$. The time dependence  of the mode functions $u_k(\eta,{\rm x}), v_k(\eta,{\rm x})$  can be used to obtain the Bogoliubov coefficients connecting the evolved state and the  groundstate (vacuum) of the instantaneous final Hamiltonian (see Appendix~\ref{app:part_prod_FGS}), which determine a corresponding   canonical transformation of the creation and annihilation operators
\beq 
\label{eq:evolutionop}
\begin{split}
&{a}_{\mathrm{k}}(\eta_{\rm f})=\alpha_{\mathrm{k}}(\eta_{\rm f}){a}_{\mathrm{k}}(\eta_{0})-\beta_{\mathrm{k}}(\eta_{\rm f}){b}_{\mathrm{-k}}^\dagger(\eta_{0}),\\&{b}_{\mathrm{k}}^\dagger(\eta_{\rm f})=\beta_{-\mathrm{k}}^{*}(\eta_{\rm f}){a}_{-\mathrm{k}}(\eta_{0})+\alpha_{-\mathrm{k}}^*(\eta_{\rm f}){b}_{\mathrm{k}}^\dagger(\eta_{0}).
\end{split}
\eeq
The  so-called Bogoliubov coefficients $\alpha_{k},\beta_{k}$, which define  a fermionic two-mode squeezing \cite{PhysRevLett.65.3341} and  satisfy $
|\alpha_{\mathrm{k}}(\eta_{\rm f})|^2+|\beta_{\mathrm{k}}(\eta_{\rm f})|^2=1$~\cite{valatin1958,bogolyubov1958},   fully determine the
mean  density of produced particles and antiparticles  
\begin{equation}
\label{eq:density_produced}
\begin{split}
    &n_a=\frac{1}{ \mathsf{a}(\eta_{\rm f})}\int\frac{{\rm d}\mathrm{k}}{2\pi}\langle a_\k^\dagger a^{\phantom{\dagger}}_\k\rangle=\frac{1}{ \mathsf{a}(\eta_{\rm f})}\int{\rm d}\mathrm{k}\,|\beta_{\mathrm{k}}(\eta_{\rm f})|^2,\\
    &n_b=\frac{1}{ \mathsf{a}(\eta_{\rm f})}\int\frac{{\rm d}\mathrm{k}}{2\pi}\langle b_{-\k}^\dagger b^{\phantom{\dagger}}_{-\k}\rangle=\frac{1}{ \mathsf{a}(\eta_{\rm f})}\int{\rm d}\mathrm{k}\,|\beta_{\mathrm{k}}(\eta_{\rm f})|^2,
    \end{split}
\end{equation}
which are actually equal as a consequence of the expansion conserving the total particle charge. Therefore, in this limit, particle production amounts to the solution of a set of differential equations for the mode functions~\footnote{As discussed in~\cite{FulgadoClaudio2023fermionproduction}, the time-dependence of the mode functions is completely determined by a pair of exactly-solvable  Bessel differential equations, which admit a solution in closed form.}.

Moving to the interacting case $g_0^2\neq 0$, the Hamiltonian field theory~\eqref{eq:GN_ham_QFT} can be identified with that of a  Gross-Neveu model \cite{Schnetz:2005ih}, in which the mass  has acquired an explicit time dependence that encapsulates the FRW expansion of the spacetime. The static Gross-Neveu model in a flat spacetime was introduced in \cite{Gross-Neveu} as a low-dimensional QFT that shares characteristic features with higher-dimensional non-Abelian gauge theories~\cite{Peskin:1995ev}, namely asymptotic freedom and chiral symmetry breaking. The latter occurs via the so-called dynamical mass generation and the formation of a scalar condensate $\Sigma\propto\langle\overline{\chi}{\chi}\rangle$ \cite{Gross-Neveu}. 
Even for the massless Gross-Neveu model, fermions acquire a finite mass even for infinitesimally-small interactions, which is described by $\mu=\Lambda\ee^{-2\pi/g_0^2}$ and cannot be accounted for perturbatively. This mass   also yields an example of dimensional transmutation~\cite{Coleman1982}, as a dimensionless coupling $g_0^2$ induces a dimensionful parameter $\mu$ via the ultraviolet cutoff $\Lambda$. Note that for the effective (1+1) QFT in the expanding universe~\eqref{eq:GN_ham_QFT}, when restricting to the massless limit, the non-perturbative dynamically-generated mass $\mu$ would not depend on the scale factor $\mathsf{a}(\eta)$, making the problem fully static such that all the effects of the expanding spacetime are reduced to  the rescaling of the fields. 
On the other hand, when allowing for a finite bare mass $m\neq 0$, the model becomes time-dependent, and an interplay between the scale factor $\mathsf{a}(\eta)$ and the interactions arises due to the dependence of the condensates with $m$. As a consequence, the scalar condensate can display interesting dynamics $\Sigma\mapsto\Sigma(\eta)$, which can change the phenomenon of particle production beyond the $g_0^2=0$ free limit studied in~\cite{FulgadoClaudio2023fermionproduction}. The goal of this article is to explore quantitatively how interactions modify non-perturbatively the phenomenon of particle production. 
 
In the presence of interactions, even if the above exact solution in terms of a Bogoliubov transformation~\eqref{eq:evolutionop} no longer holds, the formalism of fermionic Gaussian states allows us to retain a similar expression, albeit using parameters that now depend on the non-perturbative fermion condensates and can thus display a very different time dependence. It is very natural to adopt this formalism, as Eq.~\eqref{eq:evolutionop} in the free case   transforms the vacuum state within the family of fermionic Gaussian states \cite{SHI2018245, Surace22}. By generalizing to a fully variational approach using fermionic Gaussian states (FGSs) \cite{RevModPhys.77.513,WANG20071,Kraus_2009,Kraus_2010,doi:10.1142/S1230161214400010,SHI2018245,Surace22}, we can thus  analyse the  interacting QFT while always checking that the previous non-interacting results are recovered as $g_0^2\to 0$. Hence, we continue 
referring to the Bogoliubov transformations \eqref{eq:evolutionop}, 
baring in mind that  the Bogoliubov parameters $\alpha_\k$ and $\beta_\k$ will now be obtained variationally using the FGS formalism. These can be fully characterized by their two-point correlation functions, which allow to  calculate any $n$-point correlation functions~\cite{GAUDIN196089} as well as e.g., the entanglement entropy~\cite{Peschel_2003}. Hence, the number of parameters necessary to describe these states grows only quadratically, and not exponentially, with the number of modes. Note that other potential alternative approaches such as Matrix Product states (MPS) can face some limitations when addressing the particular time evolution hereby discussed. Although MPS have already been used for accurate studies of the static properties of the Gross-Neveu-Wilson model \cite{gross_neveu_wilson},  their use for real-time phenomena  with long evolution times, as studied in our work, may be limited  due to the linear growth of entanglement \cite{Calabrese_2005} and the so-called entanglement barrier. 

Before delving into these details, let us discuss in more detail the Wilson-type  lattice regularization mentioned in the introduction, and consider a momentum space formulation that will be useful to frame the FGS approach.

\subsection{Wilson discretization and fermion condensates}
\label{subsec:static_GN}

We follow  Wilson's prescription to deal with  fermion doublers~\cite{Wilson1977} when putting the Dirac fields on a spatial lattice, namely the additional low-energy excitations that appear around  Dirac cones located at the boundaries of the Brillouin zone. We thus discretize the fermion fields as $\chi(x)\mapsto\chi_i(\eta)$, where $i\in\{1,\cdots N_{\rm S}\}$ labels the lattice sites~\cite{PhysRevD.16.3031}. 
The details of this discretization are presented in Appendix \ref{app:Gross-Neveu-Wilson}, and amount to substituting the partial derivatives in Eq.~\eqref{eq:GN_ham_QFT} by finite differences, and augment the bare mass by a second derivative that leads to a shift of the fermion-doubler mass with the inverse lattice spacing $a$. The quasi-momentum takes then values within the first Brillouin zone ${\rm BZ}=\{\k=-{\pi}/{a}+{2\pi n}/{ N_{\rm S} a}:\,\, n\in\mathbb{Z}_{N_{\rm S}}\}$. The resulting Hamiltonian is given by
\beq
H=\sum_{\k}\chi_\k^\dagger h_\k(\eta)\chi^{\phantom{\dagger}}_\k+\frac{g_0^2}{2aN_{\rm S}}\sum_{\k\q \rm{Q}}\!\!\!\!\big({\chi}^{{\dagger}}_{\k}\gamma^0{\chi}^{\phantom{\dagger}}_{\q}\big)\!\big({\chi}^\dagger_{{\rm Q}-\k}\gamma^0{\chi}^{\phantom{\dagger}}_{{\rm Q}-\q}\big),
\label{eqn:ham_BZ_main}
\eeq
where we use the Fourier-transformed lattice fields ${\chi}_{\k}=\sqrt{\frac{a}{N_{\rm S}}}\sum_{j}{\chi}_{j}\,{\rm e}^{\ii\k ja}$, and the  single-particle Hamiltonian
\beq
h_\k\big(\eta\big)=\frac{\sin(\k a)}{a}\gamma^0\gamma^1+\ii m_\k(\eta)
\gamma^0.
\label{eqn:hamSP_BZ_main}
\eeq
Here, we have introduced  a momentum-dependent Wilson mass
\beq
 m_{\k}(\eta)=m\mathsf{a}(\eta)+\frac{1-\cos(\k a)}{a},
\label{eq:ham_SP_fourier}
\eeq
which depends on the conformal time via the scale factor.

The effect of this specific discretization on the phase structure of the static model, i.e. setting $\mathsf{a}(\eta)=1$ has been depicted in Fig.~\ref{fig:trajectories_explanation}. This structure is discussed in detail in \cite{gross_neveu_wilson}, and we now review it. As a consequence of the quartic interactions, the fermions can develop vacuum expectation values corresponding to a pair of condensates, the so-called
scalar  $\Sigma_0$ and pseudo-scalar $\Pi_0$ condensates, which are customarily assumed to be homogeneous, and defined as
\beq
\label{eqn:sigma_pi_main}
\begin{split}
\Sigma_0=\frac{\ii g_0^2}{2aN_{\rm S}}\sum_\k\left<\chi_\k^\dagger\gamma^0\chi_\k\right>,\hspace{1.5ex}
\Pi_0=\frac{g_0^2}{2aN_{\rm S}}\sum_\k\left<\chi^\dagger_\k\gamma^1\chi_\k\right>.
\end{split}
\eeq

Note that for finite chemical potential, one could also consider inhomogeneous condensates, connecting with some previous works that address these four-Fermi QFTs at finite particle densities \cite{PhysRevD.79.105012, BUBALLA201539, Koenigstein_2022}. 
However, this would break the translational invariance of the Hamiltonian, spoiling its diagonal form in reciprocal space, and  leading to a larger number of equations to be solved, increasing the computational cost of our FGS approach. As we will restrict to the half-filled case (i.e., one particle per site), corresponding to the vacuum QFT, such finite-density inhomogeneities are expected to be absent, and we will explore the production of particle and anti-particle pairs as a consequence of the universe expansion, and its interplay with the homogeneous fermion condensates.

The homogeneous scalar condensate acquires non-zero values for any pair of microscopic couplings $(m,g_0^2)$, except along the lines $ma=-1$ and $g_0^2=0$. 
This is connected to the dynamical mass generation of the continuum Gross-Neveu QFT \cite{Gross-Neveu}, and leads in this case to a pair of critical lines that  separate a trivial band insulator from a topological one (see Fig.~\ref{fig:trajectories_explanation}). The latter  displays a non-vanishing topological invariant, the Zak's phase \cite{PhysRevLett.62.2747}, and topological edge states when considering open boundary conditions. This topological phase is protected by time-reversal and charge conjugation symmetries, respectively denoted as $\mathsf{T}$ and $\mathsf{C}$. Time-reversal can be represented by a anti-unitary operator $D(\mathsf{T})=T\mathcal{K}$, where $T=\ii\gamma^0$ and $\mathcal{K}$ represents complex conjugation. Similarly, charge conjugation can be represented by an anti-unitary operator $D(\mathsf{C})=C\mathcal{K}$, where $C=\ii\gamma^0\gamma^1=\gamma^5$. Time-reversal symmetry yields $T^{-1} h_{-\k}^{*}T=h_{k}$, and charge conjugation yields $\mathsf{C}^{-1} h_{-\k}^{*}\mathsf{C}=-h_{\k}$. The combination of both yields the so-called sub-lattice symmetry $\mathsf{S}=\mathsf{TC}$, which can be represented by a unitary operator $D(\mathsf{S})=-\gamma^1$, and which yields  $\mathsf{S}^\dagger h_{\k}\mathsf{S}=-h_{\k}$. This  phase is called the symmetry-protected topological (SPT) phase henceforth. Since $\mathsf{T}^2=\mathsf{C}^2=\mathsf{S}^2=1$, this SPT phase is in the $\mathsf{BDI}$ class~\cite{RevModPhys.88.035005,Schnyder_2009}.

The role of the pseudo-scalar condensate is very different, as its non-zero value is associated to the spontaneous breakdown of parity, $\chi(\eta,\;{\rm x})\rightarrow{\rm e}^{\ii\theta}\gamma^0\chi(\eta,\;-{\rm x})$, where ${\rm e}^{\ii\theta}$ is a phase factor, which requires sufficiently strong interactions  $g_0^2$. The pseudo-scalar condensate is no longer consistent with the protecting symmetries of the SPT phase, such that its onset signals the appearance of a new phase, the so-called Aoki phase \cite{
PhysRevD.30.2653, PhysRevD.58.114507}. Altogether, one finds  the following three possible static phases, depicted in Fig. \ref{fig:trajectories_explanation},
\beq
\label{eq:phases}
\begin{split}
&ma\in(-2+\Sigma_0a,-\Sigma_0a), \Pi_0=0,\hspace{2ex} {\rm SPT\,\ phase},\\
&ma\notin[-2+\Sigma_0a,-\Sigma_0a], \,\,\Pi_0=0,\,\hspace{2ex} {\rm trivial\,\ insulator},\\
&ma\in(-2+\Sigma_0a,-\Sigma_0a), \Pi_0\neq0,\,\hspace{2ex} {\rm Aoki\,\ phase}.\\
\end{split}
\eeq

We remark that the values of these condensates, which fully determine  the phase diagram in Fig.~\ref{fig:trajectories_explanation},  can be found in Appendix \ref{app:Gross-Neveu-Wilson}. In that Appendix,  we compare the predictions of the FGSs with a large-$N$ expansion in the fully static case, finding a perfect agreement. One expects that, when considering the time-dependent scale factor $\mathsf{a}(\eta)$ for the background FRW spacetime, the fermion condensates will acquire some explicit dependence on the conformal time $\Sigma(\eta)$, $\Pi(\eta)$. Its nature will crucially depend on the Hubble expansion rate $\mathsf{H}$ and how it compares to the other microscopic parameters $m,g_0^2$. This condensate dynamics will play a key role in determining how the phenomenon of particle production gets modified by the quartic interactions, leading to novel effects when compared to our previous study for a free Dirac QFT~\cite{FulgadoClaudio2023fermionproduction}.

  \subsection{Fermionic  Gaussian states and the  production of particles and antiparticles}

   In this section, we present more details of how the formalism of FGSs connects to the phenomenon of particle production described in Sec.~\ref{subsec:part_prod}, and how the evolution of the condensates introduced in Sec.~\ref{subsec:static_GN} can affect it. We start by introducing the variational family of FGSs, first discussing how to use them for real-time evolution. Finally, we address how to connect to the previous Bogoliubov transformations and the rate of particle production~\eqref{eq:density_produced} due to the expanding spacetime.

   As briefly mentioned above, FGSs  are fully characterized by their two-point correlation functions, which  are collected in the so-called correlation matrix $\Gamma$. The dimension of this matrix scales quadratically with the total number of modes  $N_{\rm tot}$, which can be understood as variational parameters  $\{\ket{\psi(\Gamma)}\}$ for the family of FGSs.
  Using the dimensionless operators of the 2-component spinors, 
  $\chi_i(\eta)=\left(c_{\k,\,\uparrow}(\eta),\;c_{\k,\,\downarrow}(\eta)\right)^{\rm t}$, the correlation matrix  $\Gamma\in{\rm GL}(N_{\rm tot})$ has the following components
\beq
  \Gamma_{\k,\,\q}=\begin{pmatrix}
      \bra{\psi(\Gamma)} {c}^{\phantom{\dagger}}_{\k,\uparrow}{c}_{\q,\uparrow}^\dagger\ket{\psi(\Gamma)}&&\bra{\psi(\Gamma)} {c}^{\phantom{\dagger}}_{\k,\uparrow}{c}_{\q,\downarrow}^\dagger\ket{\psi(\Gamma)}\\\bra{\psi(\Gamma)} {c}^{\phantom{\dagger}}_{\k,\downarrow}{c}_{\q,\uparrow}^\dagger\ket{\psi(\Gamma)}&&\bra{\psi(\Gamma)} {c}^{\phantom{\dagger}}_{\k,\downarrow}{c}_{\q,\downarrow}^\dagger\ket{\psi(\Gamma)}
  \end{pmatrix},
  \eeq
where $N_{\rm tot}=2N_{\rm S}$  in our case, 
since $\bra{\psi(\Gamma)} {c}^{\phantom{\dagger}}_{\k,\sigma}{c}_{\q,\upsilon}\ket{\psi(\Gamma)}=\bra{\psi(\Gamma)} {c}^{{\dagger}}_{\k,\sigma}{c}_{\q,\upsilon}^\dagger\ket{\psi(\Gamma)}=0$ for all $\omega, \upsilon=\uparrow,\downarrow$. Considering homogeneous condensates, Eqs.~\eqref{eqn:sigma_pi_main}, leads to a translationally-invariant Hamiltonian, resulting in vanishing $\k\neq\q$ elements of the correlation matrix.  We note that it is also frequent to work with the so-called covariance matrix  in the literature~\cite{SHI2018245}. 
  Although this is a completely equivalent description, we stick to  the correlation matrix as it allows for a more transparent analysis of particle production. 

As discussed in Appendix~\ref{app:FGS}, FGSs are typically defined as Gibbs states of a so-called parent Hamiltonian, which fully determines the correlation matrix and, in turn, the particular FGS. In the case of variational states for the groundstate and the dynamics of an interacting model, this formalism allows for an alternative approach that does not require any prior knowledge about the specific form of the mode functions in \eqref{eqn:decompositionmodefunctions}. The variational family of FGSs can be defined by a two-mode squeezing unitary, responsible for the  Bogoliubov transformation on the  particle $a_\k$ and antiparticle $b_{-\k}$ operators,  acting on the vacuum. For our particular model, taking advantage of the translational invariance of the Hamiltonian~\eqref{eqn:ham_BZ_main} and of the symmetries of the correlation matrix, there are $4N_{\rm S}$ non-vanishing real variational parameters encoded in the elements of $\Gamma$. The variational problem can be parameterized as a two-mode squeezed state using the particle and antiparticle operators $a_\k$ and $b_{-\k}$, connecting with the formalism of particle production \cite{FulgadoClaudio2023fermionproduction}, namely 
\beq
\label{eq:FGS_def}
\ket{\psi(\Gamma)}=\bigotimes_{\k\in{\rm BZ}}\left(\alpha_\k(\eta_{\rm f},\;\Gamma)-\beta_\k(\eta_{\rm f},\;\Gamma)a_\k^\dagger b_{-\k}^\dagger\right)|0\rangle.
\eeq
Here, the complex $\alpha,\beta$ parameters  are the generalization to the interacting regime of the Bogoliubov coefficients in \eqref{eq:evolutionop}, and  thus depend both on the conformal time and the correlation matrix, which is obtained as we now describe.

Note that the formalism of FGSs allows one to use imaginary-time evolution (ITE) in Eq.~\eqref{eqn:ITE_int} as an alternative to 
the direct variational minimization. As detailed in Appendix~\ref{app:FGS},   ITE also 
makes  use of Wick's theorem~\cite{GAUDIN196089} to project the evolved states onto the family of FGSs. 
 We take the asymptotic state of the ITE as the initial condition in the asymptotic in-region for the initial conformal time $\eta_0$~\eqref{eq:scale_factor}. We do so by evolving the correlation matrix in imaginary time $\tau$, while fixing the mass at $m_0=m\mathsf{a}_0$ and the coupling $g_0^2$, until we reach a stable stationary value of the correlation matrix $\Gamma_0$. We check that the derivative of the components of the correlation matrix with respect to $\tau$ is below some numerical threshold. As shown in Appendix~\ref{app:Gross-Neveu-Wilson}, the imaginary-time FGS provides a solution that agrees exactly with previous static results for the Gross-Neveu-Wilson model that use a large-$N_{\rm f}$ expansion (see Fig.~\ref{fig:pi_FGS}).

\label{sec:FGS_part_prod}
  \begin{figure}[t!]
      \centering
      \includegraphics[width=\columnwidth]{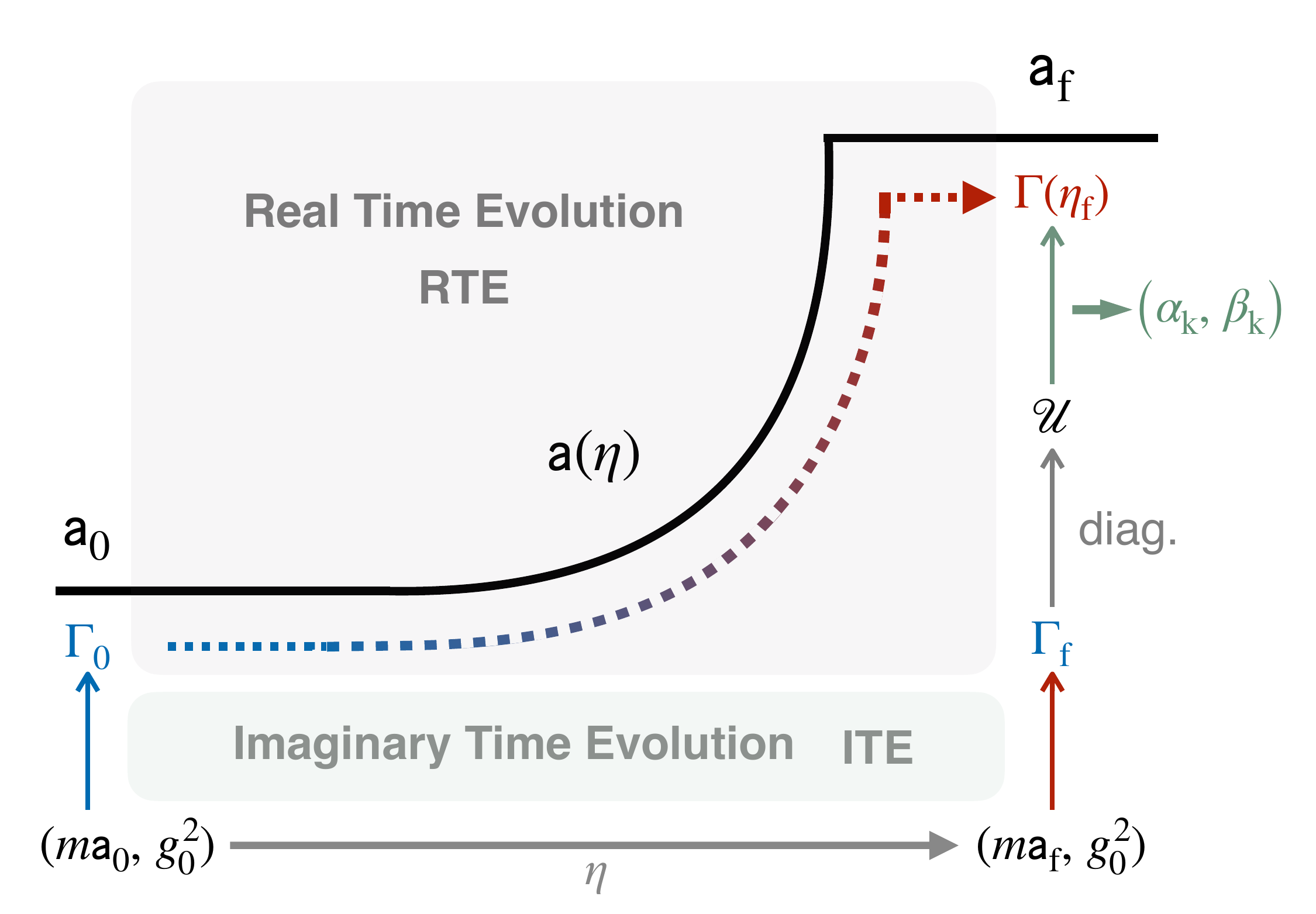}
      \caption{\textbf{Protocol for computing the production of particles after the expansion of the universe using FGSs}. We compute the groundstate correlation matrix for both the initial and final set of parameters, $(m\mathsf{a}_0,\;g_0^2)$ and $(m\mathsf{a}_{\rm f},\;g_0^2)$, via the ITE equations, obtaining respectively the correlation matrices $\Gamma_0$ and $\Gamma_{\rm f}$. Then, we solve the RTE equations using $\Gamma_0$ as the initial condition. Finally, we diagonalize $\Gamma_{\rm f}$ via the matrix $\mathcal{U}$, that connects the fermionic operators $c_{i\alpha}$ with the correct particle/antiparticle operators defined at $\eta=\eta_{\rm f}$, $a_\k$ and $b_{-\k}^\dagger$ respectively. Acting with $\mathcal{U}$ on the real-time evolved correlation matrix $\Gamma(\eta_{\rm f})$ one can obtain the Bogoliubov coefficients $\alpha_\k$ and $\beta_\k$ that contain the information about particle production.}
      
      \label{fig:protocol_explanation}
  \end{figure}

As depicted in Fig.~\ref{fig:protocol_explanation}, this serves as the initial condition for the real time evolution (RTE)  with the formalism of FGSs~\eqref{eqn:RTE_int}. Generalizing the discussion of  Appendixes \ref{app:FGS} and~\ref{app:Gross-Neveu-Wilson}  to the current situation, where there is an explicit dependence on conformal time, the RTE  leads to the following  set of differential equations 
\beq
\label{eq:RTE}
\ii\frac{\rm d}{{\rm d}\eta}\Gamma(\eta)=\left[\,\tilde{h}\big(\eta,\;\Gamma(\eta)\big),\;\Gamma(\eta)\right].
\eeq
Here, we have introduced the following single-particle self-consistent 
Hamiltonian, which is a $2N_{\rm S}\times 2N_{\rm S}$ matrix with  a block structure 
due to translational invariance
\beq
\tilde{h}\big(\eta,\Gamma(\eta)\big)=\bigoplus_{\k}\left(h_{\k}(\eta)-\ii\Sigma(\eta)\gamma^0-\Pi(\eta)\gamma^1\right).\label{eqn:int_ham_SP}
\eeq
The fermion condensates $\Sigma(\eta)$ and $\Pi(\eta)$ are given by specific elements of the correlation matrix 
\beq
\label{eq:condensates}
\begin{split}
\Sigma(\eta)=\frac{g_0^2}{2aN_{\rm S}}\sum_\k\biggl( \Gamma_{\k,\uparrow,\k,\uparrow}(\eta)-\Gamma_{\k,\downarrow,\k,\downarrow}(\eta)\biggr),\\
\Pi(\eta)=\frac{\ii g_0^2}{2aN_{\rm S}}\sum_\k\biggl( \Gamma_{\k,\downarrow,\k,\uparrow}(\eta)-\Gamma_{\k,\uparrow,\k,\downarrow}(\eta)\biggr),
\end{split}
\eeq
which lead to a  self-consistency,
and result in a non-linearity of the set of ordinary differential equations~\eqref{eq:RTE}. Additionally, although the Hamiltonian is block-diagonal in momentum space, each block depends on all other blocks  due to the definition of the condensates \eqref{eq:condensates}. In essence, 
when the scalar and/or pseudo-scalar condensates $\Sigma(\eta)$ and $\Pi(\eta)$ acquire a non-zero value,  the equations for each mode become coupled to  the equations for the rest of the modes, such that the full system needs to be solved simultaneously.

Let us now explain how to connect this FGS framework to the phenomenon of particle production, which will allow us to extend our previous  results \cite{FulgadoClaudio2023fermionproduction} to the interacting case. The numerical solution $\Gamma(\eta)$ is evaluated at a time $\eta=\eta_{\rm f}$, connecting to the out region in which the spacetime becomes  Minkowski flat, and the bare mass acquires the final value $m\mathsf{a}_{\rm f}$ (see Fig.~\ref{fig:protocol_explanation}). If the adiabatic approximation is fulfilled due to a slow evolution of the mass term, the state would remain in the instantaneous groundstate of the system for any latter time, and the correlation matrix obtained from the RTE differential equations would reduce to that for the instantaneous groundstate. For a faster evolution, however, non-adiabatic deviations will lead to a time-evolved correlation matrix $\Gamma(\eta_{\rm f})$ that  no longer coincides with the instantaneous  correlation matrix $\Gamma_{\rm f}$ for $m\mathsf{a}(\eta_{\rm f})$. This will  result in the generation of excitations, i.e. particle production. To calculate those, we need to obtain  $\Gamma_{\rm f}$ again by solving the ITE within the FGS manifold, but this time for the final  mass $m_{\rm f}=m\mathsf{a}_{\rm f}$, and the interactions $g_0^2$. To obtain the spectrum of produced particles, we use the fact that the instantaneous  correlation matrix $\Gamma_{\rm f}$ can be brought into a diagonal form by a unitary matrix 
\beq
\mathcal{U}\Gamma_{\rm f}\,\mathcal{U}^{-1}=\bigoplus_{\k}\begin{pmatrix}
    \langle b_{-\k}^\dagger b^{\phantom{\dagger}}_{-\k}\rangle&\langle b_{-\k}^\dagger a_{\k}^\dagger\rangle\\\langle a^{\phantom{\dagger}}_{\k}b^{\phantom{\dagger}}_{-\k}\rangle&\langle a^{\phantom{\dagger}}_{\k}a^\dagger_\k\rangle\end{pmatrix}=\bigoplus_{\k}\begin{pmatrix}
        0&0\\0&1\end{pmatrix}.
    \eeq
In the last equality, we have considered a half-filled lattice system corresponding to a filled Dirac sea in the lowest energy band.
This matrix  connects the  fermion fields ${c}^{\phantom{\dagger}}_{\k\alpha}$ with the  particle and antiparticle operators ${a}_\k, {b}_{\k}$ used to define the free-field evolution~\eqref{eqn:decompositionmodefunctions}. Acting with this  matrix $\mathcal{U}$ on the correlation matrix obtained from the RTE equations 
\begin{equation}
    \mathcal{U}\Gamma(\eta_{\rm f})\mathcal{U}^{-1}\!=\bigoplus_{\k}\!\begin{pmatrix}
        |\alpha_{\k}(\eta_{\rm f},\Gamma)|^2&\alpha_{\k}^{*}(\eta_{\rm f},\Gamma)\beta_{\k}^{*}(\eta_{\rm f},\Gamma)\\\alpha_{\k}(\eta_{\rm f},\Gamma)\beta_{\k}(\eta_{\rm f},\Gamma)&|\beta_{\k}(\eta_{\rm f},\Gamma)|^2
    \end{pmatrix}\!,
    \label{eq:diag_obtain_bogol}
\end{equation}
one can readily obtain 
the Bogoliubov coefficients  introduced in Eq. \eqref{eq:evolutionop} for the non-interacting case. In the current derivation, however, they correspond to the self-consistent values for the interacting problem, and fully define the FGs~\eqref{eq:FGS_def}. This is depicted in the right part of Fig.~\ref{fig:protocol_explanation}. At this point, the density of produced particles and antiparticles for the interacting problem can be obtained via Eq.~\eqref{eq:density_produced}.

   In Appendix~\ref{app:part_prod_FGS}, we benchmark  this approach with the exact solutions found for the non-interacting regime, where one can use the standard formalism of mode functions presented in publication~\cite{FulgadoClaudio2023fermionproduction}, finding a perfect agreement. Together with the benchmark of the FGSs method with the large-$N_{\rm f}$ method for the static values of the fermion condensates shown in Appendix~\ref{app:Gross-Neveu-Wilson}, this gives us confidence that both imaginary- and real-time FGS algorithms are correctly formulated, such that we can embark on the study of the real-time dynamics and particle production of  self-interacting Dirac fermions in the expanding FRW spacetime.

  \section{\bf Particle production for correlated Dirac fermions}
\label{sec:3}

\subsection{Adiabaticity and dynamical condensates}
\label{sec:cond_dyn}
The adiabatic approximation states that a quantum system with time-dependent parameters that is initially prepared  in the groundstate of a gapped Hamiltonian will evolve by populating the subsequent instantaneous groundstates corresponding to  specific values of the dynamical parameter, provided that its rate of change  is sufficiently small. The formulation of this approximation  finds different versions, 
see e.g. \cite{RevModPhys.90.015002}. 
For a time-dependent Hamiltonian $H(t)$ with instantaneous eigenstates $|n,\;t\rangle$ and instantaneous eigenenergies $E_n(t)$, we follow the adiabatic approximation that requires  
$
\langle m,\;t|\frac{\rm d}{{\rm d}t}|n,\;t\rangle\ll |E_m(t)-E_n(t)|\;\;\forall n,\;m,\;t
$. 
Considering our use of the conformal time and the notation of this work, 
we can define an adiabatic parameter  as 
\beq
\xi_\k=\frac{1}{2\epsilon_{\k}(\eta)}\left({\rm v}_{\k}^{-}(\eta)\right)^\dagger\frac{{\rm d}}{{\rm d}\eta}{\rm v}_{\k}^{+}(\eta),
\eeq 
where $\pm\epsilon_\k(\eta)$ and ${\rm v}_{\k}^{\pm}(\eta)$ are respectively the instantaneous eigenvalues and eigenvectors of the free single-particle Hamiltonian \eqref{eqn:hamSP_BZ_main}, such that $h_\k(\eta){\rm v}_\k^{\pm}(\eta)=\pm\epsilon_\k(\eta){\rm v}_\k^{\pm}(\eta)$. The adiabaticity condition is fulfilled as long as $\xi_\k\ll1$. From this perspective, particle production can be understood as a measure of  deviations from the adiabatic approximation. 
We can gain some insight by inspecting the form of this adiabatic parameter in the non-interacting limit.

According to Eq.~\eqref{eq:scale_factor}, we consider  fixed initial and final sizes of the universe, given respectively by the scale factors $\mathsf{a}_0$ and $\mathsf{a}_{\rm f}$. 
These values set the boundary conditions of a  path in parameter space connecting   $(m\mathsf{a}_0,\;g_0^2)$ and $(m\mathsf{a}_{\rm f},\;g_0^2)$, 
which keep fixed the value of the interaction strength as depicted in Fig.~\ref{fig:trajectories_explanation}. 
According to Eq.~\eqref{eq:duration_expansion}, the rate at which the system moves in parameter space is 
quantified by the Hubble parameter $\mathsf{H}$, such that the limit ${\mathsf{H}}\rightarrow\infty$ yields a vanishingly small  duration $\Delta\eta\rightarrow0$. In this limit, the  universe expansion occurs suddenly, which  amounts to a so-called quench  in a more general context of non-equilibrium many-body systems~\cite{Calabrese_2005, PhysRevLett.96.136801}. 
On the other hand, in the opposite limit ${\mathsf{H}}\rightarrow0$ the expansion lasts for an infinite conformal time $\Delta\eta\rightarrow\infty$, implying that changes in the mass are infinitely slow and  the adiabaticity condition should hold. We can  make this argument more rigorous for $g_0^2=0$ by computing the adiabaticity coefficient 
\beq
\xi_\k=\mathsf{H}a\left|\frac{\mathsf{a}(\eta)\,{\sin(\k a)}\,m_{\k}(\eta)a}{4\left({\sin^2(\k a)}+m^2_{\k}(\eta)a^2\right)^{3/2}}\right|\ll1.
\label{eqn:adiab_cond_wilson_free}
\eeq

For a particular case, i.e. a particular set of parameters  $m$, $\mathsf{a}_0$, $\mathsf{a}_{\rm f}$ and $\mathsf{H}$, it is sufficient to check that the adiabaticity condition is satisfied for 
\beq
{\rm max}_{\begin{subarray}{l}\k\in{\rm BZ}\\
    \mathsf{a}\in[\mathsf{a}_0,\;\mathsf{a}_{\rm f}]\end{subarray}}\xi_\k({\mathsf{H}a},\;ma)\ll1.
\eeq

Since the parameter $\xi_\k$ depends linearly on $\mathsf{H}a$, we confirm our previous intuition: the smaller the Hubble rate is, the further one gets into the adiabatic regime, which should therefore result in no particles being produced. On the opposite limit, sufficiently large Hubble coefficient results in no mode $\k$ satisfying the adiabatic condition for any combination of parameters, therefore reaching a maximal production of particles. Hence, the value of $\mathsf{H}a$ will allow to 
interpolate between a vanishing production for $\mathsf{H}a\to 0$ and its maximum value in the quench limit $\mathsf{H}a\to \infty$. 

Let us  start the FGS analysis of the interacting case by studying  how the scalar $\Sigma (\eta)$ and pseudo-scalar  $\Pi (\eta )$ condensates adapt dynamically to the spacetime expansion. In the adiabatic regime $\mathsf{H}a\to 0$, one expects the condensates to adapt to the instantaneous groundstate of the interacting Hamiltonian, such that their evolution with the conformal time attains different values of $\Sigma(\eta)\in[\Sigma_0,\Sigma_{\rm f}]$, $\Pi(\eta) \in[\Pi_0,\Pi_{\rm f}]$. For each point in parameter space considering $M$  intermediate times $\{(m_i,g_0^2): i\in\{0,\cdots,M\}, m_i=m\mathsf{a}(\eta_i)\}$, we solve the ITE in the large-$\tau$ limit. 
In this regime of adiabaticity, we expect that the production of particles should become negligible. On the contrary, for $\mathsf{H}a\to \infty$, we expect a quench regime in which the evolution is so fast that the condensates do not have time to change, and simply retain their initial value for the whole evolution $\Sigma(\eta)=\Sigma_0, \;\Pi(\eta)=\Pi_0$. Since the initial state is not an equilibrium state of the final many-body Hamiltonian, we expect a maximal density of  particles/antiparticles produced during the diabatic expansion.

\begin{figure}[t!]
    \centering
    \includegraphics[width=0.85\columnwidth]{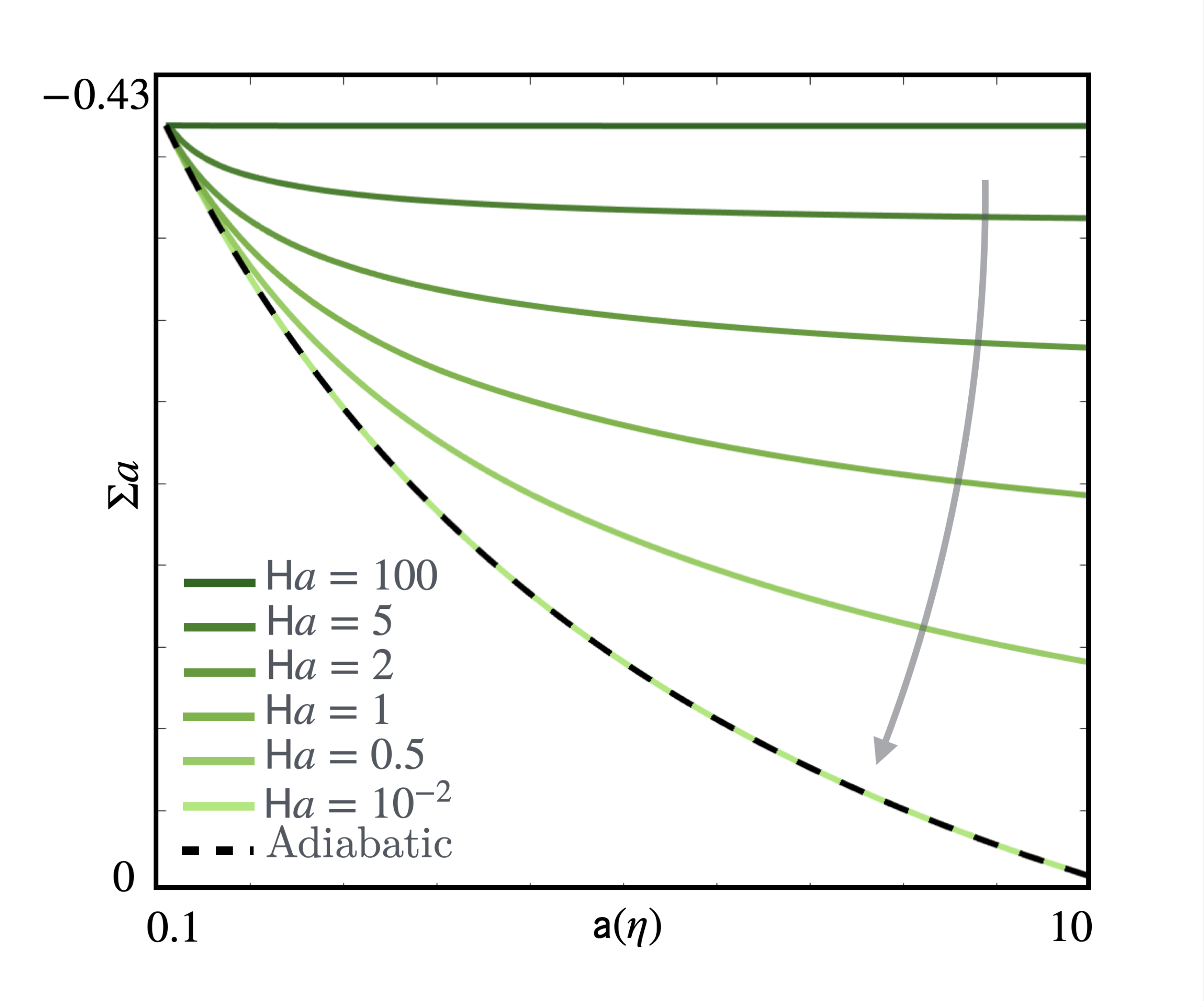}
    \caption{\textbf{Dynamical evolution of the $\Sigma$ condensate as a function of the scale factor for different values of ${\mathsf{H}}a$. Trivial insulator phase}.  The conformal time range $[\eta_0,\;\eta_{\rm f}]$ depends on the value of $\mathsf{H}$. For this reason, and in order to compare different regimes of adiabaticity, we plot the value of $\Sigma a$ as a function of the scale factor $\mathsf{a}(\eta)$ and not of the conformal time $\eta$. The grey arrow points towards increasing adiabaticity. $ma=0.1$, $\mathsf{a}_0=0.1$, $\mathsf{a}_{\rm f}=10$, $g_0^2=1$ and $N_{\rm S}=65$.}
    \label{fig:sigmadyn}
\end{figure}

We can now use the  formalism of FGSs to corroborate these limiting expectations and, in addition,  study an intermediate regime where the evolution of the condensates 
lies between the adiabatic and quench limits.  First, we analyze the case for which $\Pi_0=0$, where the paths in parameter space lie within the SPT and trivial-band insulator phases (see Fig.~\ref{fig:trajectories_explanation}). Note that we only consider  expansions that do not cross any critical line, as these would lead to a Kibble-Zurek phenomenon where excitations can be produced even for infinitely-slow expansions in the thermodynamic limit~\cite{Kibble:1976sj,Zurek:1985qw}. We  consider a FRW  expansion of the universe  from $\mathsf{a}_0=0.1$ to $\mathsf{a}_{\rm f}=10$ for a discretization with $N_{\rm S}=65$ sites. The bare parameters  are $ma=0.1$, $g_0^2=1$, such that the interactions are sufficiently strong so that large deviations from the free case are to be expected. This expansion lies within the trivial band insulator phase (see blue line Fig.~\ref{fig:trajectories_explanation}\textbf{(a)}), in which only the scalar condensate $\Sigma$ becomes non-vanishing in the interacting regime. We solve the real-time evolution of the correlation matrix in Eq.~\eqref{eq:RTE}, leading to the  results for the scalar condensate $\Sigma(\eta)$ shown in Fig.~\ref{fig:sigmadyn}. 
One can see that, for small Hubble parameters, the evolution is adiabatic, and the scalar condensate shows a  perfect agreement with the  self-consistent instantaneous values obtained by solving the imaginary-time equations (represented by the black dashed line). On the other hand, for a much larger Hubble parameter, $\Sigma(\eta)$ remains frozen, which agrees with the previous  description of the quench limit. The results for intermediate values of the Hubble parameter show that one interpolates smoothly between these two limiting cases.

\begin{figure}[t!]
    \centering
    \includegraphics[width=0.85\columnwidth]{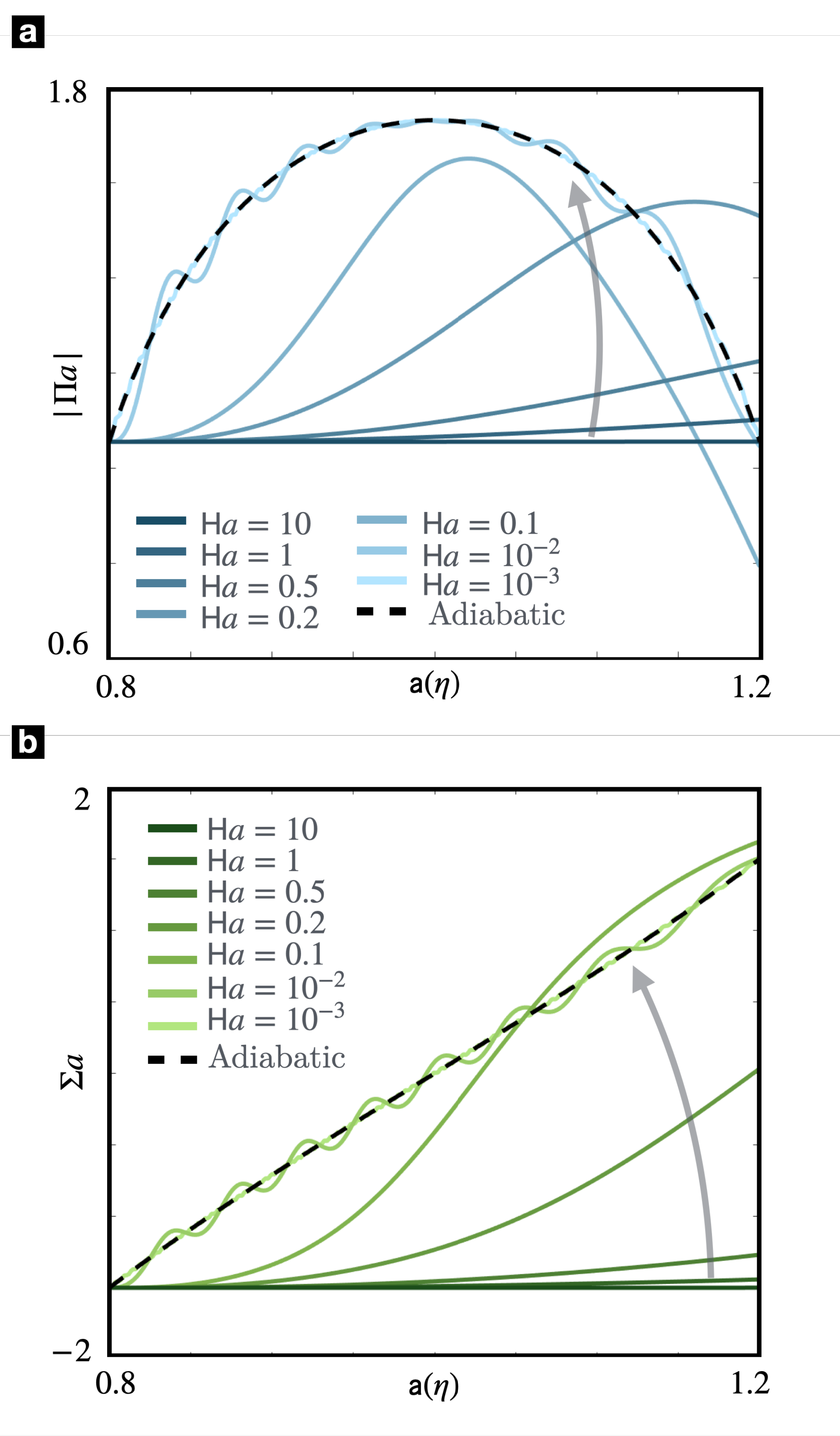}
    \caption{\textbf{Evolution of condensates within the Aoki phase.} We show here the dynamical evolution of the condensates as a function of the scale factor for different values of ${\mathsf{H}}a$. The rest of the parameters are given by $ma=-1$, $\mathsf{a}_0=0.8$, $\mathsf{a}_{\rm f}=1.2$, $g_0^2=4$ and $N_{\rm S}=65$. The grey arrows points towards increasing adiabaticity. \textbf{(a)} Evolution of the pseudo-scalar condensate $\Pi$. We represent the absolute value of the condensate due to the ambiguity in its sign due to the $\mathbb{Z}_2$ symmetry breaking. \textbf{(b)} Evolution of the scalar condensate $\Sigma$.}
    \label{fig:Pidyn}
\end{figure}

We can also explore expansions in which the pseudo-scalar  condensate displays some dynamics $\Pi(\eta)$. To avoid critical crossings, 
we consider the case where both the initial and final sets of parameters 
lie in the Aoki phase (see red line in Fig~{\ref{fig:trajectories_explanation}\textbf{(b)}}). In this situation, both the scalar and pseudo-scalar condensates are non vanishing, and their dynamics can depend on the values of each other. 
We choose $ma=-1$,and $g_0^2=4$, setting the FRW expansion between $\mathsf{a}_0=0.8$ and $\mathsf{a}_{\rm f}=1.2$ for a discretization with $N_{\rm S}=65$ sites. Note that to remain within the Aoki phase, we are restricted to study shorter periods of expansion given its bounded size. 
By solving the real-time FGS equations, we obtain the results  shown in Fig. \ref{fig:Pidyn}, both for the pseudo-scalar condensate $\Pi$ in {\bf{(a)}} and for the scalar condensate $\Sigma$ in {\bf{(b)}}. Here the grey arrows point in the direction of increasing adiabaticity.
We observe similar features as before. For the smallest value of ${\mathsf{H}a}$, the evolution is essentially  adiabatic, and the condensates evolves through   their instantaneous groundstate values, the latter corresponding to the black dashed lines. In the other limit, that of large values of  ${\mathsf{H}a}$, the condensates remain constant. 

\begin{figure}[t!]
    \centering
    \includegraphics[width=0.85\columnwidth]{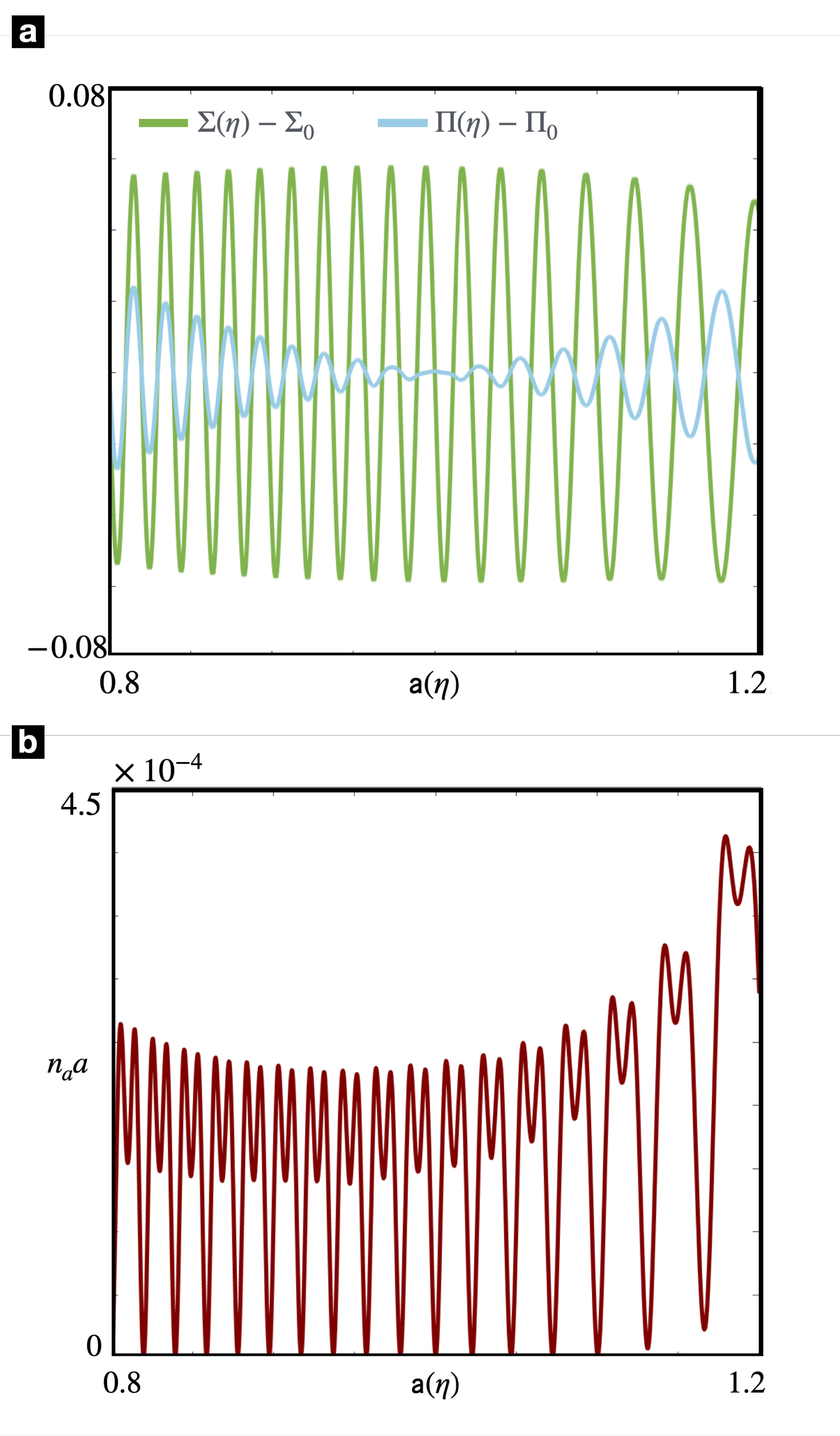}
    \caption{\textbf{(a) Difference between the dynamical evolution of the condensates and their instantaneous groundstate value in the Aoki phase.} We observe that the oscillations of $\Pi(\eta)$ vanish at $\mathsf{a}(\eta)=1$, which corresponds for this particular plot to the mass $ma=-1$. This is a symmetry line of the phase diagram (see Fig.~\ref{fig:trajectories_explanation}). For $\mathsf{a}(\eta)<1$, the oscillations of both condensates with respect to their groundstate values is on phase. Conversely, for $\mathsf{a}(\eta)>1$ the oscillations become out of phase, and $\Pi(\eta)-\Pi_0$ reaches its local maxima at the same time that $\Sigma(\eta)-\Sigma_0$ reaches its minima. This imply that low-energy excitations of the model respect the underlying symmetry of the groundstate condensates that can be seen in Fig.~\ref{fig:Pidyn}, with $\Sigma_0$ and $\Pi_0$ being respectively anti-symmetric and symmetric with respect to the line $ma=-1$. $ma=-1$, $g_0^2=4$, $\mathsf{a}_0=0.8$, $\mathsf{a}_{\rm f}=1.2$, $\mathsf{H}a=5\cdot10^{-3}$, $N_{\rm S}=65$. \textbf{(b) Dynamical evolution of the density of produced particles as a function of the scale factor $\bf \mathsf{a}(\eta)$}. For this regime of adiabaticity, particles are produced displaying an oscillatory evolution synchronous with that  of the condensates. This is a manifestation of the back-reaction of the dynamical evolution of the condensates on particle production and vice versa. $ma=-1$, $g_0^2=4$, $\mathsf{a}_0=0.8$, $\mathsf{a}_{\rm f}=1.2$, $\mathsf{H}a=5\cdot10^{-3}$, $N_{\rm S}=65$.}
    \label{fig:12}
\end{figure}

It is interesting to note that in both cases the interpolation between these two regimes when modifying $\mathsf{H}a$ is not monotonic, which contrasts the previous expansions within the trivial phase. Indeed,  there are some non-trivial dynamical effects taking place. The most evident one are the oscillations 
about the groundstate instantaneous values, which are synchronous for the two condensates, and also their amplitudes appear to be correlated.  To compare the oscillations of both condensates more precisely, we show in Fig.~\ref{fig:12} the deviations of the dynamical evolution of the condensates 
from their instantaneous groundstate values, namely $\Pi(\eta)-\Pi_0$ and $\Sigma(\eta)-\Sigma_0$. For the dynamical evolution of the condensates, we choose a small value of the Hubble parameter, $\mathsf{H}a=5\cdot10^{-3}$, for which we expect  the deviations from the adiabatic approximation to be small. 
We observe that for values of 
$\mathsf{a}(\eta)<1$ both condensates oscillate on phase. The amplitude of the oscillation of $\Pi(\eta)-\Pi_0$ is damped, and vanishing for $\mathsf{a}(\eta)=1$. Then, for values $\mathsf{a}(\eta)>1$ its amplitude starts to grow again, but its oscillations become out of phase with respect to those of $\Sigma(\eta)-\Sigma_0$. As we can see, this feature results from the fact that $\Sigma(\eta)-\Sigma_0$ and $\Pi(\eta)-\Pi_0$ are, respectively, odd and even with respect to $\mathsf{a}(\eta)=1$. 
To understand this, note that the expansion associated to our choice of the mass, $ma=-1$, and of the initial and final sizes of the universe, $\mathsf{a}_0=0.8$ and $\mathsf{a}_{\rm f}=1.2$ respectively, is symmetric with respect to the line $ma=-1$ and correspond to a horizontal red line in the phase diagram depicted in Fig.~\ref{fig:trajectories_explanation}\textbf{(a)} that occurs within the Aoki phase. The line $ma=-1$ is a symmetry axis of the phase diagram. Similarly, the groundstate values of the condensates $\Sigma_0$ and $\Pi_0$ are anti-symmetric and symmetric, respectively, with respect to this line, i.e. $\Sigma_0(ma=-1+\delta,\;g_0^2)=-\Sigma_0(ma=-1-\delta,\;g_0^2)$ and $\Pi_0(ma=-1+\delta,\;g_0^2)=\Pi_0(ma=-1-\delta,\;g_0^2)$, as can be seen in the dashed lines in Fig.~\ref{fig:Pidyn}. 
Comparing these features to those of the dynamical evolution of the condensates in Fig.~\ref{fig:12}, we conclude that the low-energy excitations maintain the odd/even nature of the groundstate values of the condensates. 
On the other hand, the oscillating behaviour and the coupling between the oscillations of both condensates arises due to the aforementioned self-consistent nature of the problem, and must be taken into account when analyzing the subsequent numerical results. 

Additionally, we show in Fig.~\ref{fig:12} that this oscillating evolution in the condensates also result in a synchronous oscillation in the evolution of density of produced particles. This reflects the fact that the interacting problem becomes self-consistent, with the self-consistent evolution of the condensates modifying the production of particles and vice versa, as will be analyzed in more detail in the next subsection.  

  \subsection{Effect of interactions on particle production}

We now study the effect of the interactions and the dynamical fermion condensates  on the density of produced particles. 
When interactions are switched on, there are two main differences with respect to the free case: \textit{(i)} the appearance of non-vanishing values for the scalar $\Sigma_0$ and pseudo-scalar $\Pi_0$ condensates in the groundstate,  which lead to a self-consistent shift of the Hamiltonian parameters and modify the particle production; and \textit{(ii)} the specific time-evolution of these condensates $\Sigma(\eta),\Pi(\eta)$, which will have a back-reaction contribution from  the density of produced particle and antiparticles. As we will see,  the specific form of the solutions of the dynamical equations is modified in the interacting regime. 
The question we now address is to what extent 
these effects 
are relevant for the 
production of particles. Following the analysis of 
Sec.~\ref{sec:cond_dyn}, we expect that the  effect {\it (i)}, associated to the static value of the condensates, will be the only relevant one in the quench regime. Away  from this limit, the dynamics of the condensates  becomes non-trivial, and we expect the effect {\it (ii)} 
to impact 
the particle production.

\begin{figure}[t!]
    \centering
    
    \includegraphics[width=0.85\columnwidth]{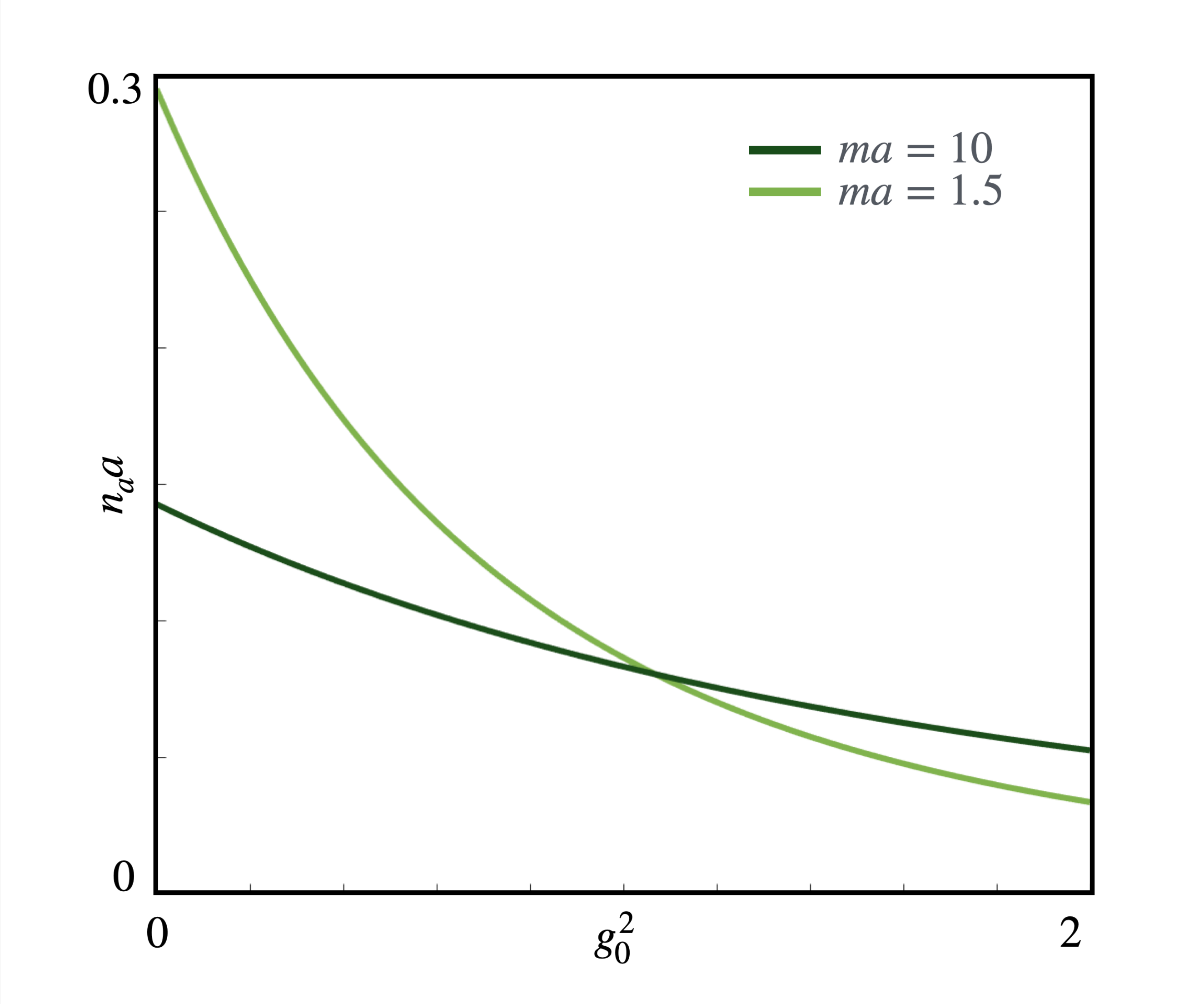}
    \caption{\textbf{Density of production as  a function of the coupling strength $g_0^2$. Topologically trivial phase.} We calculate this for the limit of a quench and for the regime of $m>0$, namely within the trivial phase. The effect of interactions in this case is to generate a static scalar condensate, which for this regime increases the gap of the model thus contributing to the adiabaticity of the problem and to a decrease in particle production. $\mathsf{a}_0=0.1$, $\mathsf{a}_{\rm f}=1$, $N_{\rm S}=65$. }
    \label{fig:5}
\end{figure}

\begin{figure}[t!]
    \centering
    \includegraphics[width=0.85\linewidth]{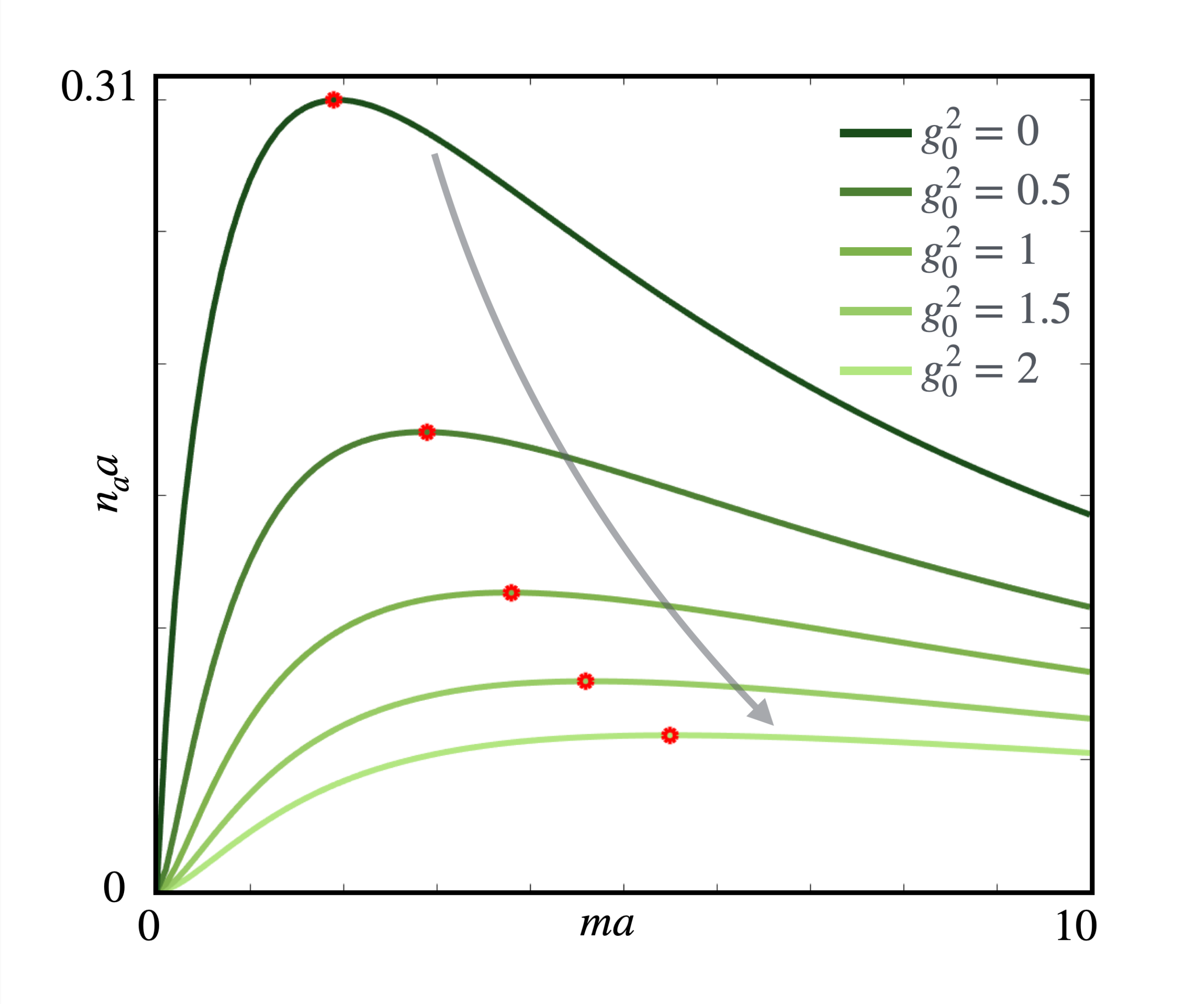}
    \caption{\textbf{Density of produced particles in the quench limit as a function of the mass, for parametrically varying interaction strengths $\mathbf{g_0^2}$. Topologically trivial phase.} The grey arrow points towards increasing interaction strengths. We observe two main effects: \textit{(i)} the density of produced particles for all masses decreases as the interaction strength increases; \textit{(ii)} the peak of the distributions, marked with red circles, happens for bigger masses as interactions increases. This effect is also accompanied by the fact that the distributions gets increasingly flatter. $\mathsf{a}_0=0.1$, $\mathsf{a}_{\rm f}=1$, $N_{\rm S}=65$.}
    \label{fig:14}
\end{figure}

First, let us consider an expansion in which $\Pi(\eta)=0$, particularly focusing on paths within the trivial phase (see Fig.~\ref{fig:trajectories_explanation}\textbf{(c)}),
and start by focusing on the quench limit. According to the non-perturbative phenomenon of dynamical mass generation, 
 interactions will induce  a mass  shift $\delta m\propto\Sigma_0$ which, in contrast to the bare mass 
$m\mathsf{a}(\eta)$, does not have an explicit time dependence. 
In the quench limit ${\mathsf{H}}\rightarrow\infty$, the scalar condensate is effectively frozen,  and we thus expect from Eq.~\eqref{eqn:int_ham_SP} that the shifted mass will simply increase the energy gap by a static constant $m\mathsf{a}(\eta)\mapsto m\mathsf{a}(\eta)=m\mathsf{a}(\eta)+|\Sigma_0|$, leading to an  overall decrease in the density of produced particles. Eventually, when $|\Sigma_0|\gg m\mathsf{a}(\eta)$ $\forall \mathsf{a}(\eta)\in\left[\mathsf{a}_0,\;\mathsf{a}_{\rm f}\right]$, the interaction-induced gap will be so large that the groundstate is no longer sensitive to finite spacetime expansions. 
These 
arguments are indeed validated by 
Fig. \ref{fig:5}, where we can see that 
increasing interactions in the quench regime leads to a decrease in the density of produced particles, eventually tending towards zero in the strongly-interacting limit. 
It is interesting to note that 
the interactions can modify the hierarchy between the production of particles with different masses. 
In Fig.~\ref{fig:5}, this is particularly evident by inspecting the curves corresponding to $ma=10$ and $ma=1.5$. For $g_0^2\lesssim 1$, lighter particles 
are produced more abundantly, 
whereas for $g_0^2\gtrsim1$, the trend reverses, and the FRW expansion produces massive particles more abundantly.
This can be  explored in more depth by plotting the dependence of the density of produced particles with the mass, while varying parametrically the interaction strength (see Fig.~\ref{fig:14}). The  maximum of each curve in this figure is marked with a red circle, and denotes the  optimal  mass for maximal particle production, which 
grows as the interactions increase. This effect is also accompanied by the fact that the curves 
flatten as the interactions get very large, so that there is an increasingly wider range of masses that get similarly produced.

Let us now move to a FRW expansion 
that takes place within either the SPT or the Aoki phases (see Fig.~\ref{fig:trajectories_explanation}).  {We note that, for these expansions, 
one can obtain a decrease }of the gap for increasing interactions, in contrast with 
the previous expansions within the trivial phase. As already explained and shown in Fig.~\ref{fig:trajectories_explanation}, both the SPT and Aoki phase lie within a bounded region along the mass domain. Hence we must restrict  to  shorter expansions, since for a given $g_0^2$ the initial and final masses $m\mathsf{a}_0$ and $m\mathsf{a}_{\rm f}$ cannot be arbitrarily distant.  We thus consider $\mathsf{a}_0=0.8$ and $\mathsf{a}_{\rm f}=1.2$ 
and a  bare mass $ma=-1$,  exploring a range of values of $g_0^2$ to study the dependence of particle production with  interactions. 

\begin{figure}[t!]
    \centering
    \includegraphics[width=0.85\columnwidth]{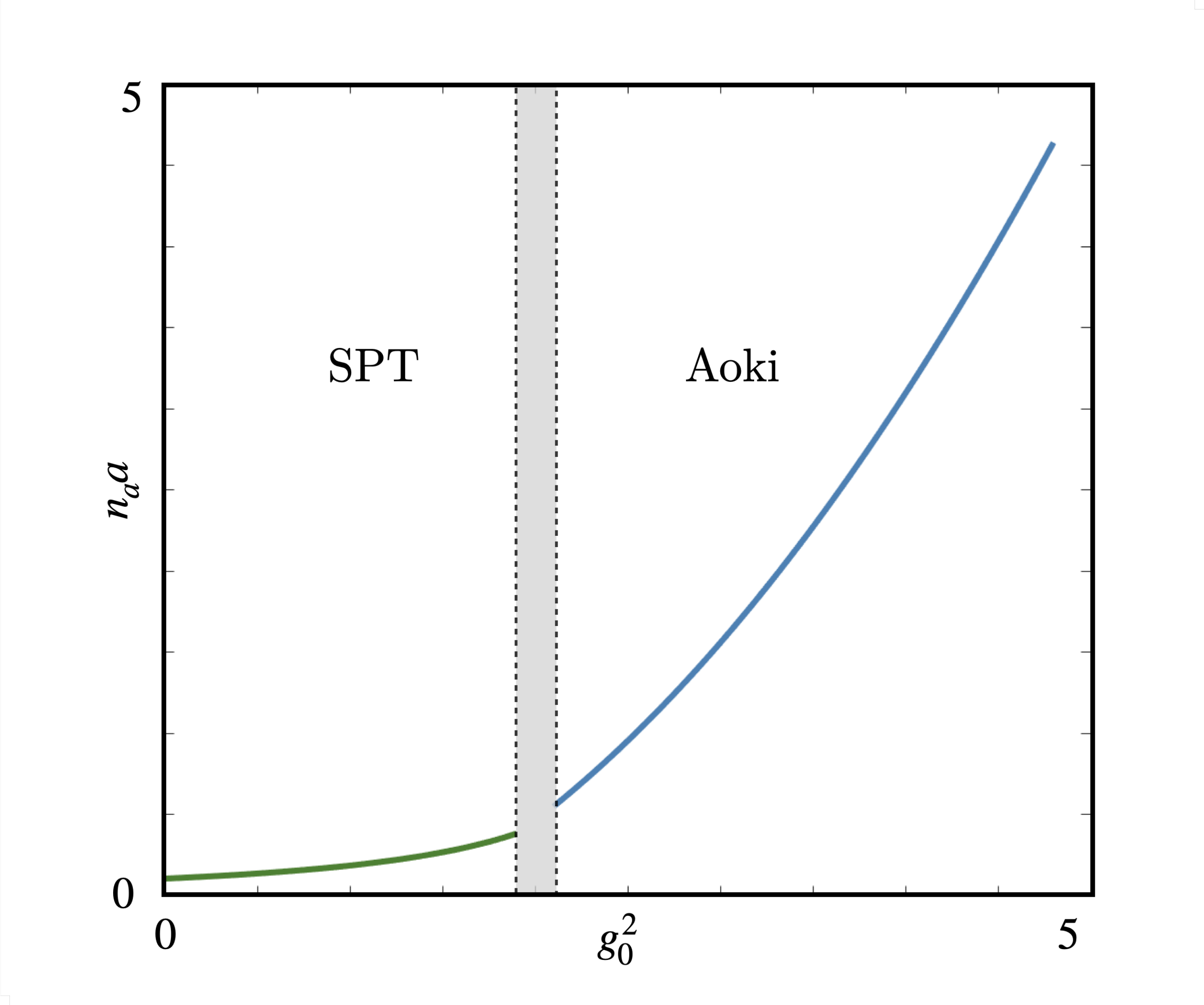}
    \caption{\textbf{Particle production as a function of $\mathbf{g_0^2}$}. We calculate this for the limit of a quench and for expansions within the SPT phase (first segment) and the Aoki phase (second segment). The effect of interactions in this case is to generate condensates that lead to an increase in the density of produced particles. $ma=-1$, $\mathsf{a}_0=0.8$, $\mathsf{a}_{\rm f}=1.2$, $N_{\rm S}=65$. 
    }
    \label{fig:13}
\end{figure}

First, we address  the quench limit and consider the effects of the static values of the condensates. The numerical results using FGSs are presented in Fig. \ref{fig:13}. The green and blue segments correspond, respectively, to expansions within the SPT phase and the Aoki phase. The shaded region around $g_0^2=2$  corresponds to the regime  close to the critical line that separates the SPT and Aoki phases, and that we have avoided. We observe that, in contrast to expansions within the trivial insulator (see Fig.~\ref{fig:5}), 
there is an increasing production of particles with an increasing interaction strength $g_0^2$. The rate of production within the Aoki phase reaches very high rates of production, surpassing those appearing for the trivial insulator and for the SPT phase. This is a manifestation of the fact that, for fixed $m\mathsf{a}_0$ and $m\mathsf{a}_{\rm f}$ within the Aoki phase, an increasing interaction strength imply that trajectories begin and finish gradually closer to the critical lines, yielding a larger production due to the decrease in the gap (see Fig.~\ref{fig:trajectories_explanation}\textbf{(b)}). This situation contrasts with that of expansions occurring within the trivial band insulator phase, where, after fixing the initial and final masses, increasing interaction strengths lead to trajectories gradually further from the critical line (see Fig.~\ref{fig:trajectories_explanation}\textbf{(a)}).  

Let us now consider expansions which are neither in the quench, nor in the fully-adiabatic limit, both within the SPT and Aoki phases. In this case, the dynamics of the condensates can become relevant and can lead to a modification of the overall behaviour of interactions in particle production. Furthermore, in the Aoki phase the pseudo-scalar condensate $\Pi_0$ can also be non-vanishing, getting its dynamics $\Pi (\eta)$  coupled to that of the scalar condensate $\Sigma(\eta)$. As shown  in Figs.~\ref{fig:Pidyn} and \ref{fig:12}\textbf{(a)}, the condensates can display a synchronous  oscillatory evolution for certain values of $\mathsf{H}a$. These oscillations modify the time-dependence of the problem, and hence can have non-trivial effects on the production of particles. This can be further understood by analysing the dynamical evolution of particle production. The numerical results are presented in Fig.~\ref{fig:12}\textbf{(b)}. As the universe expands, the density of produced particles oscillates with a frequency correlated to that present in the evolution of the condensates. This feature manifests the back-reaction of the dynamics of the condensates on particle production, and vice versa, constituting a fundamental difference with respect to the free regime. For $g_0^2>0$, the problem of particle production becomes dependent upon the specific time evolution of the condensates, which is in turn affected by the underlying regime of adiabaticity. This two-way dependency can lead to substantial modifications that cannot be accounted for by a static shift of the parameters, and therefore introduces a qualitative difference compared to its free counterpart and to the quench regime.  

\begin{figure}[t!]
    \centering
       \includegraphics[width=0.85\columnwidth]{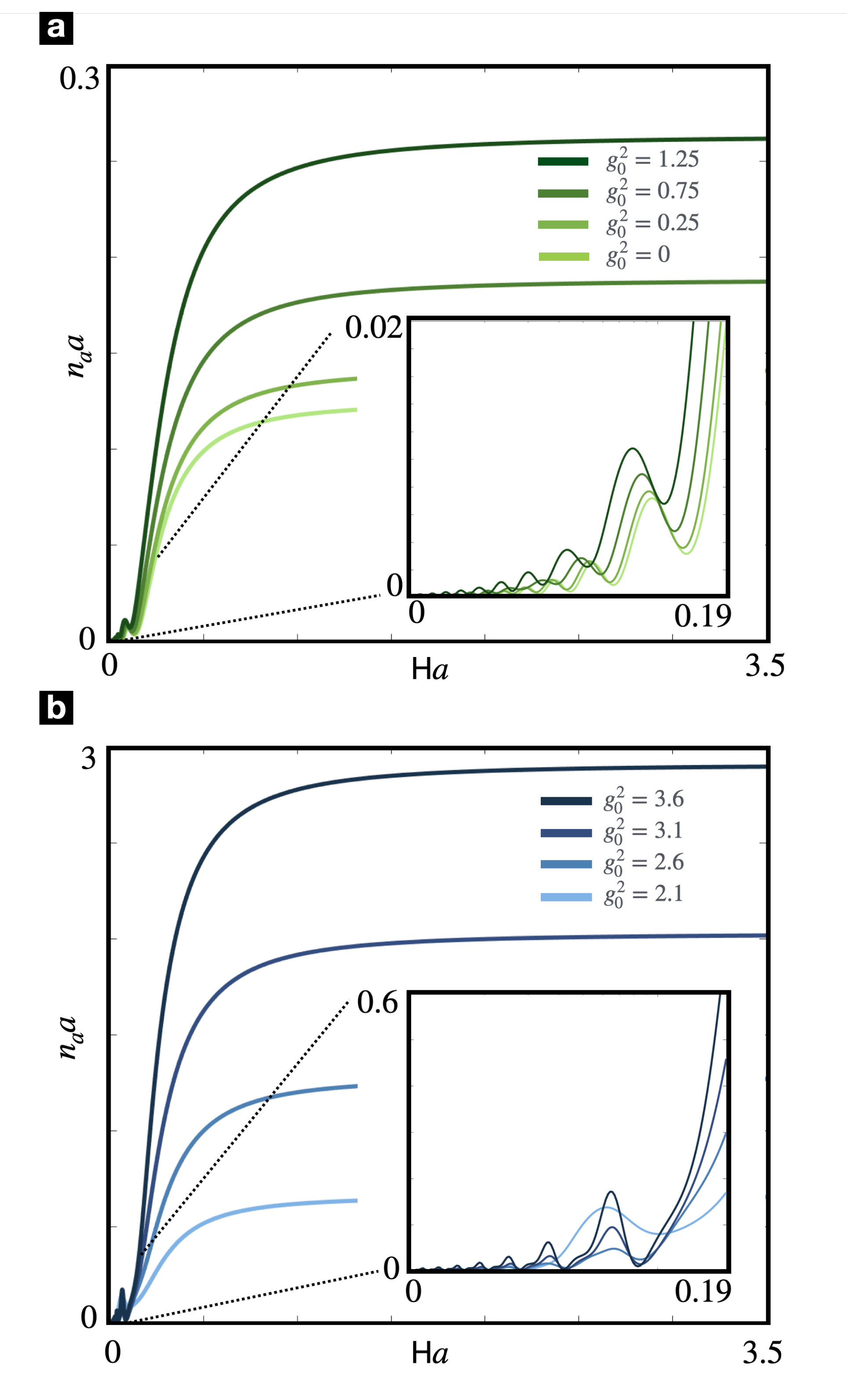}
    \caption{\textbf{Density of produced particles as a function of the Hubble parameter, ${\mathsf{H}}a$, for parametrically varying coupling strength $\mathbf{g_0^2}$}. \textbf{(a)} SPT phase, corresponding to green trajectories in Fig.~\ref{fig:trajectories_explanation}\textbf{(c)}. We see that each curve interpolates between the adiabatic regime of no production and the saturation value of production occurring for a quantum quench, where increasing values of the interaction strength favours the density of produced particles. The qualitative shape of the curves is preserved for increasing values of $g_0^2$, which only contributes to a decreasing scaling of the overall density of produced particles, and also to an earlier reaching of its saturation value. \textbf{(b)} Aoki phase, corresponding to red trajectories in Fig.~\ref{fig:trajectories_explanation}\textbf{(b)}. The curves do not share their qualitative shape, resulting in non-trivial effects such as the inversion of the dependence of the production of particles with respect to $g_0^2$ for small values of $\mathsf{H}a$. This is related to the role played by the dynamics of the condensates in  particle production. $m=-1$, $\mathsf{a}_0=0.8$, $a_{\rm f}=1.2$, $N_{\rm S}=64$.}
    \label{fig:15}
\end{figure}

Now, to understand how this behaviour can modify the overall results, 
    we use the FGS formalism to calculate the density of produced particles as a function of ${\mathsf{H}}a$, which allows us to understand how the density interpolates between the adiabatic and the quench regimes in the presence of interactions. The results are presented in Fig. \ref{fig:15}, both for the SPT, in \textbf{(a)} and for the Aoki, in \textbf{(b)}, phases. Before delving in the detailed analysis of each figure, let us note that a general feature is present in both figures, namely the oscillations as a function of $\mathsf{H}a$ for small values of the Hubble rate. The fact that this oscillations are present even in the free case (see the curve corresponding to $g_0^2=0$ in Fig.~\ref{fig:15}\textbf{(a)}) prevents us from attributing them to the dynamics of the condensates or to any other effect of the interactions, and it suggests that these oscillations can be interpreted as due to the finite expansion duration, connecting with a similar behaviour~\cite{vitanov_landau-zener_1996} that appears in the so-called finite Landau-Zener model \footnote{In the case of the Landau-Zener model, the Schrödinger equation can be analytically solved by means of parabolic cylinder functions \cite{abramowitz+stegun}. If the duration of the coupling lasts from $t_0\rightarrow-\infty$ to $t_{\rm f}\rightarrow\infty$, one recovers the well known Landau-Zener formula \cite{Zener:1932ws}. However, when the duration of the coupling becomes finite, one finds the appearance of oscillations as a function of the detuning due to the behaviour of the parabolic cylinder functions.}.

In our model, due to the block diagonal form of the Hamiltonian \eqref{eqn:int_ham_SP}, each mode can be treated as an effective two-level system,  where the role of the time-dependent  detuning in the two level system is given by $ma(\eta)=-m/(H\eta)$. In the free case, every mode is independent of the others, allowing for particle production to be calculated for each one independently. This perspective allows to connect our system to the Landau-Zener model, differing only in the specific time-dependence of the parameters. 
However, some general features such as the oscillatory behaviour of a finite-time duration of the time-dependence, are expected to be shared by both systems. 
This analogy can be further extended to the interacting regime. When $g_0^2>0$, the single-particle Hamiltonian \eqref{eqn:int_ham_SP} depends on the condensates $\Sigma(\eta)$ and $\Pi(\eta)$ which, according to Eqs.~\eqref{eq:condensates}, have contributions from every mode. Therefore, when interactions are switched on, one can no longer solve the RTE for each mode independently. 
Nonetheless, one can  still solve the RTE equations to find the specific evolution of the condensates in conformal time, and then plug them back into  the self-consistent Hamiltonian~ \eqref{eqn:int_ham_SP}. This allows to treat the condensates as effective parameters that lead to  a specific change of the time dependence. At this level, 
the system can still be understood as a collection of $N_{\rm S}$ independent two-level systems with a more complicated time-dependence, connecting again with a Landau-Zener-like model with a more convoluted time-dependence, which now stems both from the dynamical mass $m\mathsf{a}(\eta)$ and from the self-consistent time-dependence of the condensates $\Sigma(\eta)$ and $\Pi(\eta)$.
 
Now, let us consider expansions within the SPT phase in Fig.~\ref{fig:15}\textbf{(a)}, where 
there is only a scalar condensate $\Sigma(\eta)$. 
Consistent with our previous results, we observe that increasing interactions leads to larger densities of produced particles in the quench limit. The same result holds for intermediate values of $\mathsf{H}a$. This only changes for small values of $\mathsf{H}a$, where the oscillations due to the finite size of the expansion appear. When interactions are switched on, these get rescaled in a similar fashion as the quench values for the density, increasing their amplitude for larger interaction strengths. Additionally, the oscillations appear for progressively smaller values of $\mathsf{H}a$, which can be seen as a manifestation of the fact that interactions contribute to an earlier departure of the adiabatic approximation (see an extension of the adiabatic approximation to the interacting regime in Appendix~\ref{app:adiab_int}). 
These combined effects can yield an effective inversion of the dependence of density with respect to $g_0^2$ due to the overlap of the oscillations, as can be seen in the inset of Fig.~\ref{fig:15}\textbf{(a)}, for some specific values of $\mathsf{H}a$. 
However, these results for small and intermediate values of the Hubble rate can be understood as a generalization of the quench behaviour, with the neat effect of interactions being an overall rescaling of the curves. This manifests that the dominant effect of interactions for the considered parameters and expansions is to introduce static shift given by $\Sigma_0$. Let us stress that this does not imply that the specific dynamical evolution of the scalar condensate $\Sigma(\eta)$ is absent, but its role has less impact on particle production than the static shift induced by the interactions.

Next, we represent the particle production for expansions within the Aoki phase in Fig.~\ref{fig:15}\textbf{(b)}. 
In this case, we also observe the oscillations for small values $\mathsf{H}a$, but we note that this expansion cannot be reproduced in the free case   $g_0^2=0$ as there is no Aoki phase there. 
We can see that as interactions increase, the shape of these oscillations changes with respect to the finite Landau-Zener ones that were found for the SPT expansion. Eventually, these modifications lead to an inversion in the dependence of the density of produced particles with respect to $g_0^2$ and we observe that, for the selected curves, lower coupling strengths can lead to higher productions. We note that this inversion cannot be accounted for by an overall rescaling of the curves due to the static value of the condensates. 
 This behavior is not observed for trajectories within the SPT phase and it significantly modifies the quench results, so this mechanism 
 is driven by the intertwined dynamics of both the scalar and the pseudo-scalar condensates, as depicted in Fig.~\ref{fig:12}, where their dynamics  generate an oscillatory behaviour in the evolution of the production of particles. Both the oscillations in the evolution of the condensates and of the produced particles become coupled, manifesting the self-consistent nature of the interacting regime.  Hence, these oscillations of the condensates have non-trivial effects on the adiabaticity of the evolution, and are therefore responsible for the interesting effects depicted in this picture.

The overall interpretation is as follows. When the interactions increase, both the scalar $\Sigma$ and pseudo-scalar $\Pi$ condensates acquire a finite value. In the quench limit, these introduce a static shift in the parameters of the free model according to \eqref{eqn:int_ham_SP}, with the scalar condensate $\Sigma_0$ shifting the dynamically renormalized mass and the pseudo-scalar condensate $\Pi_0$ introducing a new term in the Hamiltonian. This results in an quantitative modification of the density of produced particles. If this static nature of the condensates is extended for all values of $\mathsf{H}a$, and their dynamical modulations are negligible, the outcome will be an overall rescaling of the density of produced particles for all values of $\mathsf{H}a$ that can be understood as a generalization of the quench results. On the other hand, if away from the quench limit the dynamical evolution of the condensates becomes strong, not only are the parameters of the model shifted, but also its time dependence gets modified self-consistently, $\Sigma\rightarrow\Sigma(\eta)$ and $\Pi\rightarrow\Pi(\eta)$. This will alter the free results in a non-trivial way. As a consequence, for values of $\mathsf{H}a$ in which the dynamics of the condensates becomes strong one expects substantial changes in the dependence of the density of produced particles with respect to $g_0^2$. This explains why the most notable change in the behaviour of particle production as interactions increase happens within the Aoki phase, as it is the phase in which the condensates display an intertwined, non-trivial dynamical evolution, resulting in a modification of the evolution of produced particles, as depicted in Figs.~\ref{fig:Pidyn} and \ref{fig:12}. Note also that the extent to which the dynamical evolution of the condensates is relevant relies on the structure of the groundstate values of the condensates along the trajectories considered. As can be seen in Fig.~\ref{fig:pi_FGS} of Appendix~\ref{app:FGS}, the Aoki phase is the region for which the condensates display a more convoluted groundstate structure, this having a fundamental impact on the relevance acquired by their dynamical evolution for small values of $\mathsf{H}a$.

\subsection{Parity-breaking particle production}
\begin{figure}[t!]
    \centering
    \includegraphics[width=0.85\columnwidth]{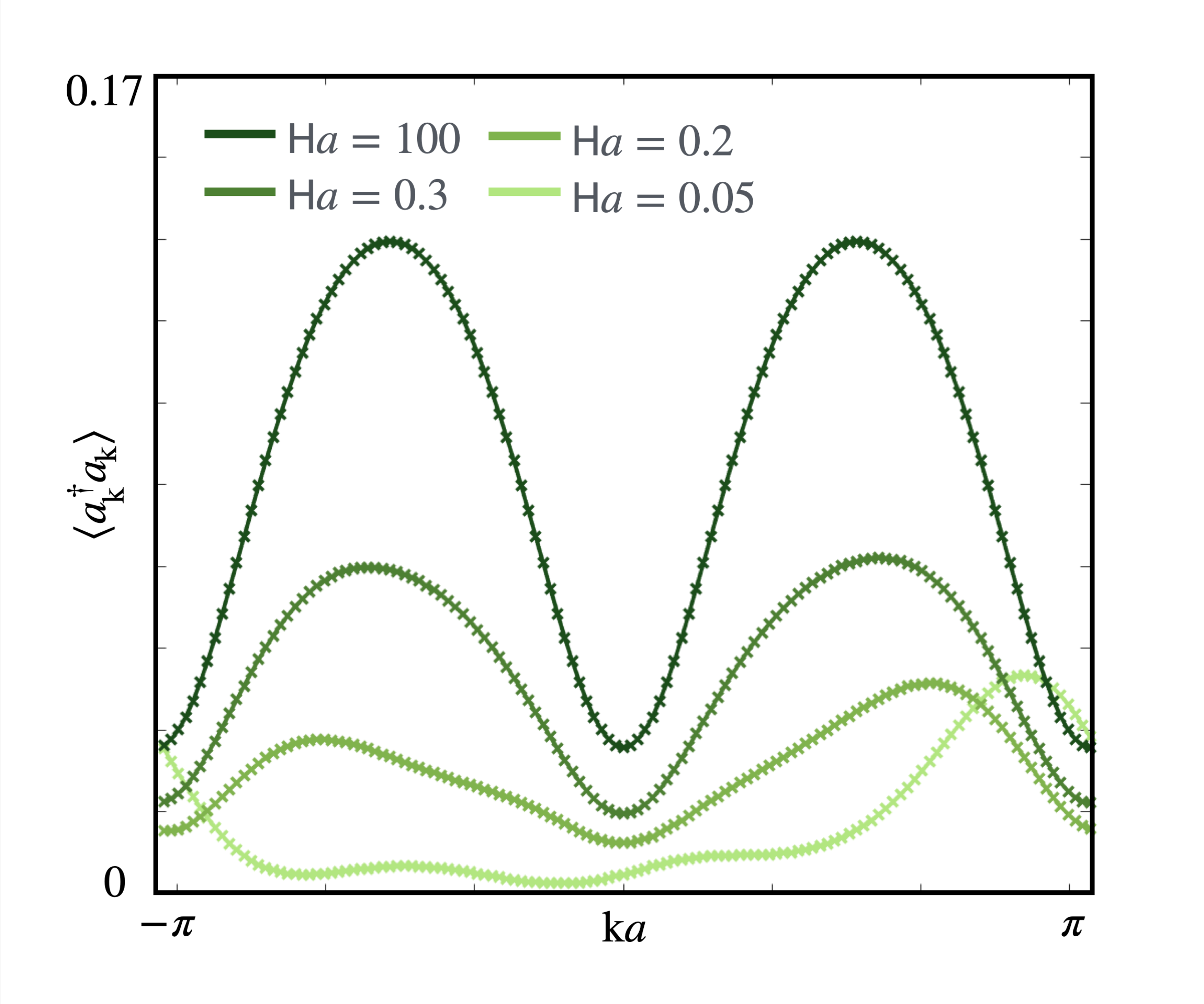}
    \caption{\textbf{Spectra of production with $ \Pi{\bf \neq0}$}. We observe a breakdown of the parity symmetry relating the production for modes $\k$ and $-\k$, which is however recovered for large enough values of ${\mathsf{H}}a$. $ma=-1$, $g_0^2=2.1$, $a_0=0.8$, $a_{\rm f}=1.2$ and $\rm N_{\rm S}=128$.}
    \label{fig:10}
\end{figure}

\begin{figure}[t!]
    \centering
    \includegraphics[width=0.85\columnwidth]{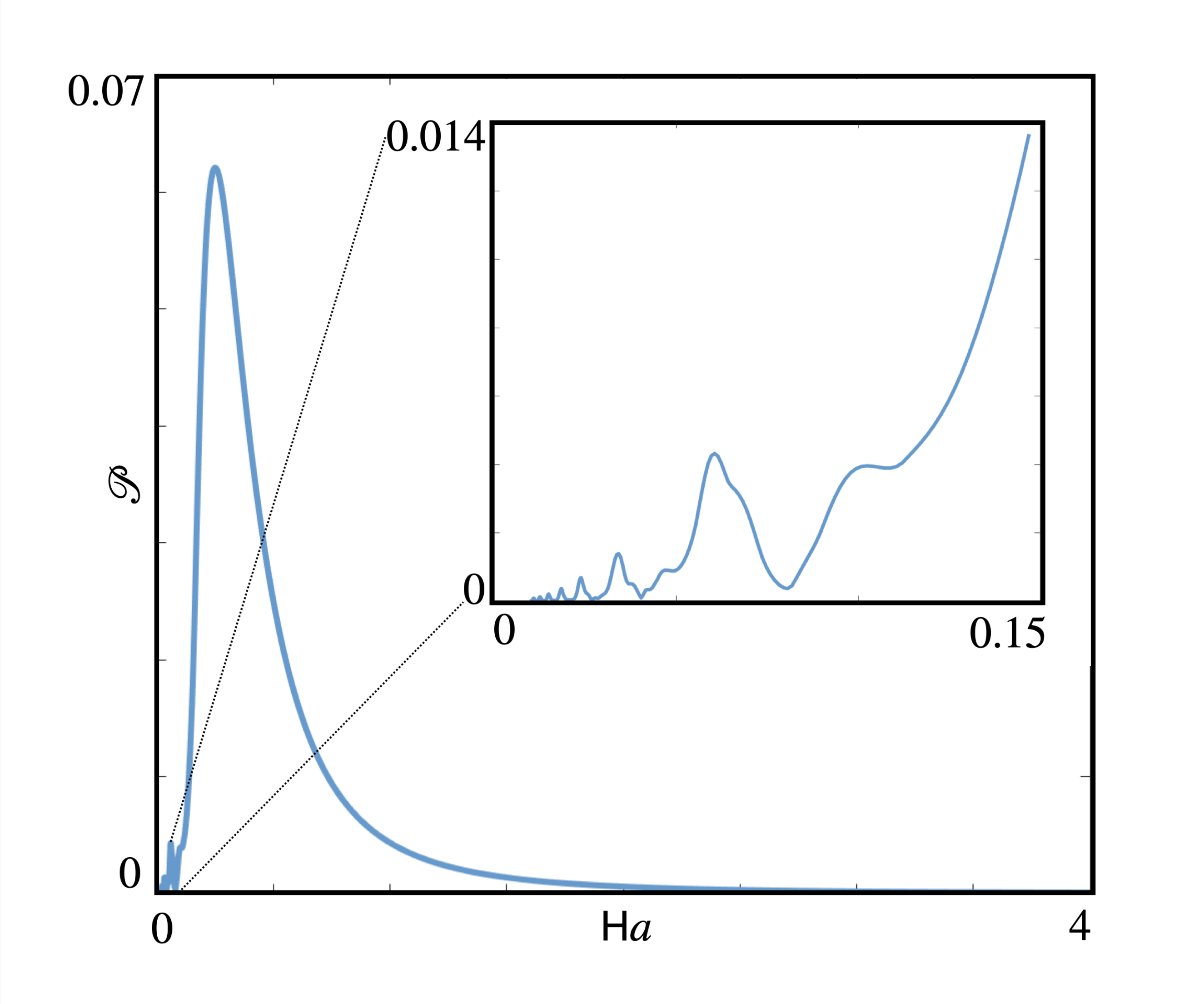}
    \caption{\textbf{Parity-breaking coefficient of the spectra for different values of $\mathbf{\mathsf{H}a}$}. We calculate this for an expansion taking place from $\mathsf{a}_0=0.8$ to $\mathsf{a}_{\rm f}=1.2$ and $ma=-1$, $g_0^2=4$, which ensures that the expansion is always within the parity-breaking Aoki phase. We observe that the parity-breaking is manifested for intermediate values of adiabaticity. In the adiabatic regime, particle production is suppressed and parity-breaking is not manifested due to a vanishing spectrum. On the other hand, in the quench limit parity is restored. This suggests that the parity-breaking mechanism is carried out by the dynamical evolution of the pseudo-scalar condensate. In the inset we also show some oscillations taking place, related to those appearing for a similar range of values of $\mathsf{H}a$ in Fig.~\ref{fig:15} and due to the self-consistency of the problem. $N_{\rm S}=129$.}
    \label{fig:pari_brea}
\end{figure}

A final relevant aspect for expansions within the parity-breaking Aoki phase becomes manifest at the level of the spectra of produced particles. Expansions in which  $\Pi=0$ display a spectrum that is always parity even,
\begin{equation}
n_a(\k)=\langle{a}_\k^\dagger{a}^{\phantom{\dagger}}_\k\rangle=\langle{a}_{-\k}^\dagger{a}^{\phantom{\dagger}}_{-\k}\rangle=n_a(-\k).
\label{eq:parity_condition}
\end{equation}
On the other hand, as noted in Sec.~\ref{subsec:static_GN}, parity is broken in the Aoki phase, which may have a manifestation 
in particle production, in particular in the particle production  spectrum. 
In particular, we check whether the condition \eqref{eq:parity_condition} is respected for expansions within the Aoki phase. 
The effects of this breakdown on particle production are explored in Fig.~\ref{fig:10},  where we explore various values of the Hubble parameter to connect from the adiabatic to the quench limits. It is  interesting to note that, when the Hubble parameter is not very large, the spectra are no longer symmetric with respect to $\k=0$. On the other hand,  parity symmetry of the spectra  is recovered for larger values of $\mathsf{H}a$ towards the quench limit.
We can study this 
phenomenon by introducing a coefficient of parity-breaking imbalance in particle production  $\mathcal{P}$, which we define by 
\begin{equation}
    \mathcal{P}=\frac{1}{2N_{\rm S}}\sum_{\k}\left|\langle a_\k^\dagger a_\k\rangle-\langle a_{-\k}^\dagger a_{-\k}\rangle\right|.
\end{equation}
This vanishes for spectra that are symmetric with respect to $\k=0$, but becomes non-zero when there is a breakdown of parity. We depict this coefficient as a function of $\mathsf{H}a$ in Fig.~\ref{fig:pari_brea}, where we observe quantitatively the behaviour that was already manifested in Fig.~\ref{fig:10}. The spectra become asymmetric with respect to $\k=0$ for intermediate values of $\mathsf{H}a$, away from both the quench and the adiabatic limits, signaling a breakdown of parity. In the limit $\mathsf{H}a\rightarrow0$ the coefficient $\mathcal{P}$ tends to $0$ because as we approach the adiabatic limit, all the elements $\langle a_\k^\dagger a_\k\rangle\rightarrow 0$. Therefore, a vanishing $\mathcal{P}$ in the adiabatic limit is only a manifestation of a vanishing production, and does not give any information about the breakdown of parity. On the other hand, when $\mathsf{H}a\rightarrow\infty$, the coefficient $\mathcal{P}\rightarrow0$. In this opposite limit, the production of particles is maximal, and therefore a vanishing $\mathcal{P}$ implies that there is a   restoration of   symmetry in the spectra. It is also interesting to note that $\mathcal{P}$ displays some oscillations for small $\mathsf{H}a$, which are reminiscent of those appearing also for the Aoki phase in Fig.~\ref{fig:15}\textbf{(b)}. 

Following the analysis done in Sec. \ref{sec:cond_dyn}, these results suggests that the effect of the parity-breaking of the spectra of production is related to the non-trivial dynamics of the condensate. In the quench limit, the value of the condensate is effectively frozen, and  the resulting spectra  are parity even. 
However, when the dynamics of the condensates becomes manifest, the spectra  show how particles with opposite momentum have different rates of production. From the point of view of the total momentum of the system, it should be stressed that this parity breaking does not imply that the state develops a non-vanishing overall momentum. This is due to the 
particle-hole symmetry 
which leads to $\langle{a}_\k^\dagger{a}_\k\rangle=\langle{b}_{-\k}^\dagger{b}_{-\k}\rangle=|\beta_\k|^2$, see Eq.~\eqref{eq:density_produced}. Therefore, the imbalance between positive- and negative-momenta in the production of particles is compensated by the inverse imbalance between negative- and positive-momenta production of anti-particles, resulting in the conservation of the overall momentum. 

\section{\bf Conclusions and Outlook}
\label{sec:4}

In this work, we have studied the effects of interactions on the production of Dirac fermions in an expanding FRW universe. We have focused on a Wilson-type lattice regularization that leads to various symmetry-broken fermion condensates, and developed a variational approach based on fermionic Gaussian states to  account for non-perturbative phenomena such as dynamical mass generation. By solving the corresponding set of imaginary- and real-time  self-consistent differential equations for the  correlation matrix, we have extracted the dynamics of both scalar and pseudo-scalar fermion condensates, and evaluated the spectra and the density of produced fermions due to the expanding universe.

Specifically, we have investigated the impact of a quartic Gross-Neveu-type interaction in the phenomenon of particle production, a cornerstone in the study of QFTs in curved spacetimes which had been primarily focused on non-interacting or perturbative regimes~\cite{Ford_2021}. We have concluded that, tuning the Hubble parameter to a quench limit, the fermion condensates become purely static, such that increasing the interaction strength $g_0^2$ simply modifies the density of produced particles through a static shift of the effective energy gap of the system. In the topologically-trivial phase, increasing interactions lead to a decrease of the density of produced particles; whereas particle production  is enhanced by the interactions in both the SPT and Aoki phases. Moreover, we have found that another consequence of this static shift of the parameters is the modification of the optimal mass at which the Dirac fermions are produced by the  expansion. 

Moving away from the quench limit by decreasing the Hubble parameter, we have also explored the effect of the non-trivial dynamical evolution of the condensates. In contrast to  the topologically trivial and SPT phases, we have found that the dynamics of the condensates becomes more relevant in the Aoki phase, in connection with the richer real-time dynamics displayed by the   condensates, which can show synchronous oscillations. This non-trivial time evolution can lead to an inversion of the dependence of the density of produced particles with the interaction strength for certain values of the Hubble parameter. Furthermore, we have explored the consequences of a non-vanishing pseudo-scalar condensate, finding that it leads to a parity breaking of the spectra of the produced particles for certain regimes of adiabaticity. Since parity is recovered in the quench limit, these results suggest that the parity-breaking mechanism is carried out by the dynamical evolution of the pseudo-scalar condensate. 

In summary, this work constitutes our first step towards the understanding of the interplay between non-perturbative phenomena and particle production of QFTs in curved spacetimes. An important open question for future work is the accuracy of the variational Gaussian approach developed in this work. Even if this variational scheme agrees with previous large-$N$ predictions of the static phase diagram~\cite{gross_neveu_wilson}, and also recovers the predictions of the more standard  mode-function approach in non-interacting QFTs~\cite{FulgadoClaudio2023fermionproduction}, it would be interesting to 
test and refine our predictions 
using other numerical techniques such as MPS, although we note that they would most likely be limited to shorter time scales. Alternatively, a promising approach would be the exploration of these phenomena using cold-atom quantum simulators following our previous proposal~\cite{FulgadoClaudio2023fermionproduction}. Although  that work focused on the non-interacting regime, the Gross-Neveu type contact interactions can be readily implemented in the Raman-lattice quantum simulator, and even increased by  a Feshbach resonance. 

It would also be interesting to explore higher-dimensional field theories, which can connect to other topological phases with a self-interacting Dirac fermion description, such as correlated Chern insulators~\cite{PhysRevB.99.125106,PhysRevResearch.4.L042012,ZIEGLER2022168763} and higher-order topological phases~\cite{10.21468/SciPostPhys.17.1.003}.  Another interesting aspect for future work  is that of inhomogeneous condensates, which are expected to appear for non-vanishing temperatures and chemical potentials \cite{BUBALLA201539}. Recent work has presented a finite-temperature topological invariant~\cite{huang2024interactioninducedtopologicalphasetransition},  which can be used in conjunction with the possibly inhomogeneous fermion condensates to chart the full phase diagram of the model. The present  variational formalism could  be applied to explore real-time dynamics in this regime of finite temperature and chemical potential. Finally, we  also note that FGSs are also suitable to study entanglement propagation during the expansion, which could shed light on different aspects of particle production and mass generation. These include, for example, its interplay with criticality and the extent to which the quasi-particle picture is suitable.

\section{Acknowledgments}
We acknowledge support from  PID2021- 127726NB-I00 (MCIU/AEI/FEDER, UE), from the Grant IFT Centro de Excelencia Severo Ochoa CEX2020-001007-S, funded by MCIN/AEI/10.13039/501100011033, from the grant QUITEMAD+ S2013/ICE-2801, and from the CSIC Research Platform on Quantum Technologies PTI- 001. P.S. acknowledges support from the Caltech Institute for Quantum Information and Matter, an NSF Physics
Frontiers Center (NSF Grant PHY-1733907), and the Walter Burke Institute for Theoretical Physics at Caltech. D.G.-C. is supported by the European Union's Horizon Europe research and innovation program under Grant Agreement No. 101113690 (PASQuanS2.1). This research was funded in whole, or in part, by the Open-Access-Fund of the Austrian Academy of Sciences. For the purpose of open access, the author has applied a CC-BY public copyright licence to any Author Accepted Manuscript version arising from this submission.

\appendix

\renewcommand{\thesection}{\Alph{section}}
\renewcommand{\thesubsection}{\thesection\arabic{subsection}}

\section {\bf Interacting Dirac fermions in curved spacetimes}
\label{app:dirac_in_CS}

For the sake of completeness and to set our notation, we review the Hamiltonian field  theory description of interacting Dirac fermions in curved spacetimes,  clarifying certain aspects related to the Hermiticity of the Hamiltonian, and the scalar product in curved spacetimes \cite{Birrell:1982ix, Parker:2009uva}.  The action describing Dirac fermions with a four-Fermi interaction  invariant under discrete chiral rotations  in a general $D=(d+1)$-dimensional spacetime is given by Eq.~\eqref{eq:action} in the main text, where the mostly-plus convention for the metric tensor has been chosen. One of the representative effects of this model is that, as a consequence of the expansion, fermion pairs can be excited from the vacuum leading to non-trivial dynamics even for a free quantum field theory (QFT). In this article, we explore how the fermion-fermion interactions of strength $g_0^2$ can modify this particle-production non-perturbatively.   

The study of interacting four-Fermi QFTs, such as the Nambu-Jona-Lasinio or the Gross-Neveu model, in curved spacetimes has been already performed in a number of works \cite{inagaki_dynamical_1997,ebert_pion_2008, flachi_chiral_2011,forkel_dynamical_1992,buchbinder_gross-neveu_2012,saharian_fermionic_2021,flachi_dual_2013,inagaki_nambu-jona-lasinio_1993,elizalde_chiral_1994}. We will focus on a $(1+1)$-dimensional expanding universe, described by a Friedman-Robertson-Walker (FRW) line element \cite{Friedmann1924berDM, Friedman1922berDK} given by 
\beq
{\rm d}s^2=g^{\mu\nu}(x){\rm d}x_\mu {\rm d}x_\nu=-{\rm d}t^2+\mathsf{a}^2(t){\rm d}\x^2,
\eeq
where $\mathsf{a}(t)$ is the  scale factor introduced below Eq.~\eqref{eq:action}, and is a monotonically increasing function that depends on the source that causes the expansion of the universe \cite{PhysRevD.23.347}. Making use of the conformal time introduced also below Eq.~\eqref{eq:action}, we can recast the line element as conformally equivalent to a Minkowski spacetime 
\beq
{\rm d}s^2=\mathsf{a}^2(\eta)\left(-{\rm d}\eta^2+{\rm d}\x^2\right).
\label{eqn:line_element}
\eeq
As discussed in \cite{FulgadoClaudio2023fermionproduction}, the spin connection for  the line element \eqref{eqn:line_element} reads $\omega_0=0$, $\omega_1=-\frac{1}{2}\partial_\eta\log\mathsf{a}(\eta)$. Together with the curved Dirac matrices $\tilde\gamma^\mu=\mathsf{a}^{-1}(\eta)\gamma^a\delta^{a\mu}$ 
and $\sqrt{-g}=\mathsf{a}^2(\eta)$, this yields the following action for our self-interacting Dirac fermions
\begin{equation}\nonumber
\begin{split}
S_0&=\!\!\int \!\!{\rm d}^2x\;\mathsf{a}(\eta)\overline{\psi}\bigg(-\gamma^a\partial_a-\half\partial_\eta\log\mathsf{a}(\eta)\gamma^0-m\mathsf{a}(\eta)\bigg)\psi\!,\\
S_{\rm int}&=\int \!{\rm d}^2x\;\frac{g_0^2\mathsf{a}^2(\eta)}{2}\big(\overline{\psi}\psi\big)^2,
\end{split}
\end{equation}
where $x=(\eta,{\rm x})$ is defined in terms of the conformal time, and we integrate over ${\rm d}^2x={\rm d}\eta{\rm d}{\rm x}$. The Hamiltonian field theory associated to this action is  $H=\int {\rm dx}\,(\mathcal{H}_0+\mathcal{H}_{\rm int})$ with a Hamiltonian density given by 
\begin{equation}
\begin{split}
\nonumber
    \mathcal{H}_0=&\,\overline{\psi}(x)\bigg(\mathsf{a}(\eta)\gamma^1\partial_{\rm x}+\half\partial_\eta\mathsf{a}(\eta)\gamma^0+m\mathsf{a}^2(\eta)\bigg)\psi(x),\\
    \mathcal{H}_{\rm int}=&-\frac{g_0^2\mathsf{a}^2(\eta)}{2}\big(\overline{\psi}(x)\psi(x)\big)^2.
\end{split}
\end{equation}
Let us, for the  moment, consider the  equations of motion  from this Hamiltonian in the free case, $g_0^2=0$, which can be expressed in a Schrödinger-like form $\i\partial_\eta\psi(x)=H^{\rm SP}\psi(x)$ with the following  single-particle Dirac Hamiltonian
\beq
H^{\rm SP}=\i\gamma^0\gamma^1\partial_x-\frac{\i}{2}\partial_\eta\log\mathsf{a}(\eta)+\i m\mathsf{a}(\eta)\gamma^0.
\eeq
In the mostly-plus metric, the flat Dirac matrices fulfill $(\gamma^a)^\dagger=(1-2{\delta_{a,0}})\gamma^a$, e.g. $\gamma^0=-\ii\sigma^z$, $\gamma^1=\sigma^y$, with $\{\sigma^a\}_{a=x,y,z}$ being the Pauli matrices. One can thus see that  $H^{\rm SP}$ becomes non-Hermitian 
due to the second spin-connection term, a fact that has been previously discussed when dealing with Dirac fermions in general curved spacetimes~\cite{Huang:2008kh}. The non-Hermiticity can lead to unwanted effects, such as non-real energies, which is a consequence of the definition of the fermionic field as $\psi(x)=\langle x|\psi\rangle$, and the time dependence of the spatio-temporal basis $\{|x\rangle\}$.

Several approaches to overcome these problems in a Hamiltonian formulation for Dirac fermions in curved spacetimes have been proposed, such as introducing  extra terms~\cite{Huang:2008kh} or rescaling the fermionic fields \cite{Minar:2013hva}, which also tackles the anti-commutation relations between the spinor field and its Hermitian conjugate. We follow the approach in~\cite{Gorbatenko:2010bh, Gorbatenko:2011rd}, which  employs the formalism of pseudo-Hermitian quantum mechanics to arrive to a unique final  Hamiltonian field theory that is manifestly Hermitian.

For a general time-dependent metric, 
one can get rid of the non-Hermiticity by defining a new fermionic field $\chi(x)=\zeta\psi(x)$, and by modifying the single-particle Hamiltonian as
\beq
H^{\rm SP}_\zeta=\zeta H^{\rm SP}\zeta^{-1}+\i\frac{\partial\zeta}{\partial\eta}\zeta^{-1}.
\label{eqn:ham_eta}
\eeq
The factor $\zeta$ appearing in these expressions can be uniquely determined by identifying $\rho=\zeta^\dagger\zeta$, where $\rho$ is the  weight operator that appears in Parker's definition of the scalar product between spinor fields \cite{Parker:1980kw} 
\beq
\begin{split}
\left(\psi_1,\;\psi_2\right)&=-\!\int\!\! {\rm d}^D x\sqrt{-g}\psi_1^\dagger(x)\gamma^0\tilde\gamma^0\psi(x) =\!\int\!\! {\rm d}^Dx \psi_1^\dagger(x)\rho\psi_2(x).
\end{split}
\eeq
From this expression, one identifies $\rho=-\sqrt{-g}\gamma^0\tilde{\gamma}^0$, which becomes  $\rho=\mathsf{a}(\eta)$ in our case, such that $\zeta=\sqrt{\mathsf{a}(\eta)}$. 

The transformed Hamiltonian $H^{\rm SP}_\zeta$ can be proved to be Hermitian \cite{Parker:2009uva, Gorbatenko:2010bh, Gorbatenko:2011rd}, 
and the Dirac equation for the new field $\chi(x)$ reads
\beq
i\partial_\eta\chi(x)=H^{\rm SP}_\zeta\chi(x).
\eeq

Consequently, the transformations that  must be performed are: \textit{(i)}  rescale the field as $\psi(x)\rightarrow\chi(x)=\sqrt{\mathsf{a}(\eta)}\psi(x)$, and \textit{(ii)}  modify the Hamiltonian as in \eqref{eqn:ham_eta}, which turns out to be
\beq
H^{\rm SP}_\zeta=\ii\gamma^0\gamma^1\partial_x+\i m\mathsf{a}(\eta)\gamma^0.
\eeq

$H^{\rm SP}_\eta$ is now manifestly Hermitian. As a consequence of the above transformations, the rescaled fields  have a    scalar product  that coincides with that in flat spacetime, 
\beq
\left(\chi_1,\;\chi_2\right)=\int {\rm d}^2x\;\chi_1^\dagger(x)\chi_2(x).
\eeq
In the main text, we work with these rescaled fields and with the modified Hamiltonian, omitting  the subscript $\zeta$ to simplify the notation. Switching on the interactions, and working with the rescaled field $\chi(x)$, we find the Hamiltonian field in Eq.~\eqref{eq:GN_ham_QFT} of the main text. Let us note that, at this level,  we are equipped with the flat-spacetime scalar product, and can  conclude that all the effects of the expanding background are codified in a dynamical renormalization of the mass term $m\mapsto m\mathsf{a}(\eta)$. In contrast, and due to its dimensionless nature, the coupling strength $g_0^2$ is not dynamically renormalized with the scale factor.

\section {\bf Variational fermionic Gaussian states}\label{app:FGS}

In this Appendix, we review the variational formalism  of fermionic Gaussian states (FGSs) \cite{RevModPhys.77.513,WANG20071,Kraus_2009,doi:10.1142/S1230161214400010,SHI2018245}, and fix the conventions used in the main text. Fermionic Gaussian states are those whose density matrix can be written as the exponential of a quadratic fermionic operator, $\rho=\ee^{-{{H}_{\rm P}}}/{\rm Tr}\{\ee^{-{{H}_{\rm P}}}\}$~\cite{Surace22}, usually referred to as the parent Hamiltonian. For models in which a U(1) symmetry remains unbroken, this  reads 
\beq 
\label{eq:parent}
{H}_{\rm P}=\sum_{\alpha,\beta}\sum_{\k,{\rm q}}{c}^\dagger_{\k\alpha}[h_{\rm P}]^{\phantom{\dagger}}_{\k\alpha,\,{\rm q}\beta}{c}^{\phantom{\dagger}}_{{\rm q}\beta},\hspace{2ex}h_{\rm P}\in\mathsf{Herm}(N_{\rm tot}).
\eeq
We consider  $N_{\rm tot}$  modes in which fermions can be created and annihilated by  ${c}^\dagger_{\k\alpha},{c}^{\phantom{\dagger}}_{\k\alpha}$, which fulfill $\{{c}^\dagger_{\k\alpha},{c}^{\phantom{\dagger}}_{\q\beta}\}=\delta_{\k\q}\delta_{\alpha\beta}$. In our case, these fermionic operators are labeled by a first sub-index that will connect to $N_{
\rm S}$ discretized spatial momenta, and a second one that will connect to the $N_{\rm int}= 2$ internal spinor components of the Dirac fermions. 
In principle, since each of the modes can be empty or occupied, the number of components of the density matrix is exponential in the number $N_{\rm tot}=N_{\rm S}N_{\rm int}$ of modes, $\#=2^{N_{\rm tot}}$. However, due to their Gaussian nature, FGSs  fulfill  Wick's theorem, and  can be thus fully characterized  via their correlation matrix 
\beq
\Gamma_{\k\alpha,\,{\rm q}\beta}=\tr\left(\rho{c}^{\phantom{\dagger}}_{\k\alpha}{c}_{{\rm q}\beta}^\dagger\right).
\label{eqn:def_correlation_matrix}
\eeq
In contrast with the full density matrix, $\Gamma$ scales only quadratically in the number of modes $N_{\rm tot}$, and since it  determines the  state completely, it leads to an efficient parametrization of the family of FGSs that will become very useful in connection to variational calculations \cite{Kraus_2010,Shi_2023}. In fact, the correlation matrix is related to the parent Hamiltonian as follows
\beq
\Gamma=\big(\mathbb{1}+\ee^{2h_{\rm P}}\big)^{-1},
\eeq
showing that it fully characterizes the FGS $\rho\mapsto\rho_{\Gamma}$. For pure states $\rho_{\Gamma}^2=\rho_{\Gamma}$, this amounts to a restriction of the correlation matrix $\Gamma^2=\Gamma$, and one typically uses $\ket{\psi(\Gamma)}$ to represent the fermionic Gaussian state $\rho_{\Gamma}=\ket{\psi(\Gamma)}\bra{\psi(\Gamma)}$. For our particular model, these states have been parameterized in Eq.~\eqref{eq:FGS_def}. 

A particular property of this family of states  is  that evolution under a quadratic physical  Hamiltonian preserves the Gaussian nature of the initial FGS \cite{Bravyi05}. It should be noted that this physical  Hamiltonian  does not coincide with the parent Hamiltonian~\eqref{eq:parent}, and will be described by a different, possibly time-dependent Hermitian matrix $h(t)\in\mathsf{Herm}(N_{\rm tot})$. 
Since the transformed state is still Gaussian, the time evolution can be seen as a mapping between two correlation matrices related by a set of ordinary differential equations (ODEs). 
Considering that  
 the fermionic operators under the quadratic physical Hamiltonian fulfill
\beq
\ii\frac{\rm d}{{\rm d}t}{{c}}_{\k\alpha}=\sum_{\rm q\beta}h_{k\alpha,\,\rm q \beta}(t)\,{c}_{\rm q \beta},
\eeq
one arrives at the set of ODEs for real time evolution (RTE) 
\beq    
\ii\frac{\rm d}{{\rm d}t}\Gamma=\left[\,h(t),\;\Gamma\,\right].
\label{eqn:RTE}
\eeq

 Note that, contrary to a generic quantum many-body system, the system of ODEs for the correlations does not involve a hierarchy of higher-order  correlations and  is, at most, described by $(N_{\rm tot})^2$ coupled equations with $N_{\rm tot}$ being the total number of modes.  A useful property of FGSs under a quadratic Hamiltonian is that one can use imaginary-time evolution (ITE)  to obtain directly the instantaneous groundstate   at any fixed instant of time $\tilde{t}$, at which the physical Hamiltonian depends on $\tilde{h}=h(\tilde{t})$~\cite{Motta:2019yya}. 
 Again, this procedure 
 can be 
 expressed at the level of the correlation matrix, yielding a smaller number of equations \cite{SHI2018245}. The resulting  ODEs are 
\beq
\frac{\rm d}{{\rm d}\tau}\Gamma=\{\tilde{h},\;\Gamma\}-2\Gamma \tilde{h}\Gamma,
\label{eqn:ITE}
\eeq
which yields  $(N_{\rm tot})^2$ coupled non-linear ODEs. 
Letting  $\tau\rightarrow\infty$, the solution converges to the groundstate correlation matrix  for non-degenerate Hamiltonians. From a numerical perspective, one can  solve these  ODEs up to  $\tau=\tau_{\rm max}$ for which the derivative with respect to imaginary time of the components of $\Gamma$ is below some numerical threshold.

So far, the formalism of FGSs only provides an alternative approach for non-interacting quadratic problems. The real interest,  however, lies on the use of FGSs from a variational perspective to address   interacting problems. The results will then be only approximate, as this procedure is equivalent to a self-consistent Hartree-Fock \cite{1930ZPhy...61..126F,1935RSPSA.150....9H,PhysRev.81.385} or large-$N_{\rm f}$ \cite{Coleman1982} methods. 
 When interested in  the groundstate, one can follow different approaches \cite{Kraus_2010}: \textit{(i)} solve a direct minimization of the energy $\Gamma_0={\rm argmin}\{\bra{\psi(\Gamma)}H\ket{\psi(\Gamma)}/\langle{\psi(\Gamma)}\ket{\psi(\Gamma)}:\Gamma\in{\rm GL}(N_{\rm tot})\}$, \textit{(ii)} discretize the imaginary time in the ITE equations, and project at each finite time step the resulting state onto the set of FGSs, and \textit{(iii)} obtain, via Wick's theorem, a quadratic but state-dependent Hamiltonian and use it to get the ITE equations.  All of these approaches can be proven to be equivalent \cite{Kraus_2010}. In this article, we follow the third approach. An additional advantage is that the same state-dependent Hamiltonian will be used to construct the variational real-time evolution (RTE), which will be our concern to discuss the phenomenon of particle production in the expanding spacetime. We thus present these details for a quartic physical Hamiltonian
\begin{align}\nonumber
H=&\sum_{\k\alpha,\,\q\beta}h_{\k\alpha,\,\q\beta}{c}^\dagger_{\k\alpha}{c}^{\phantom{\dagger}}_{\q\beta}\\+&\half\sum_{\substack{\k\alpha,\,\q\beta,\\\r\lambda,\,\s\delta}}V^{\phantom{\dagger}}_{\k\alpha,\,\q\beta,\,\r\lambda,\,\s\delta}{c}^\dagger_{\k\alpha}{c}^\dagger_{\q\beta}{c}^{\phantom{\dagger}}_{\r\lambda}{c}^{\phantom{\dagger}}_{\s\delta},
\label{eqn:interacting_Hamiltonian}
\end{align}
where $V_{\k\alpha,\,\q\beta,\,\r\lambda,\,\s\delta}$ denotes  the generic interaction strength. 

The real-time evolution of  the correlation matrix is 
\begin{align}
\ii\frac{{\rm d}}{{\rm d}t}\Gamma_{\k\alpha,\,\q\beta}=\bra{\Psi(\Gamma)}[H,\;{c}^{\phantom{\dagger}}_{\k\alpha}{c}^\dagger_{\q\beta}]\ket{\Psi(\Gamma)}.
\end{align}
In the right-hand-side of this equation the free and interacting terms of the Hamiltonian \eqref{eqn:interacting_Hamiltonian} yield expectation values of four and six fermionic operators. However, since we are considering the variational family of FGSs, these terms can be expressed in terms of the correlation matrix using Wick's theorem. 
As a  result, we obtain an evolution for the variational parameters encoded in the correlation matrix that is dictated by an effective quadratic Hamiltonian that depends  on the correlation matrix itself, i.e. self-consistent dynamical equations,  turning the set of equations of motion to non-linear ODEs. The resulting equations of motion read
\beq
\ii\frac{\rm d}{{\rm d}t}\Gamma=
\left[\tilde{h}(\Gamma),\;\Gamma\right],
\label{eqn:RTE_int}
\eeq
where $\tilde{H}(\Gamma)=\sum_{\k\alpha,\,\q\beta}\tilde{h}_{\k\alpha,\,\q\beta}(\Gamma){c}_{\k\alpha}^\dagger{c}_{\q\beta}$ is the effective self-consistent Hamiltonian that depends explicitly on the correlation matrix itself. 
This same Hamiltonian is the one used for the effective ITE equations in the variational treatment of the interacting problem, which yields
\beq
\frac{\rm d}{{\rm d}\tau}\Gamma=\{\tilde{h}(\Gamma),\;\Gamma\}-2\Gamma \tilde{h}(\Gamma)\Gamma.
\label{eqn:ITE_int}
\eeq
An explicit example of this procedure for our particular model can be found in Appendix \ref{app:Gross-Neveu-Wilson}.

\section{\bf Large-$N_{\rm f}$ benchmarks of  imaginary-time fermionic Gaussian states}\label{app:Gross-Neveu-Wilson}

In this Appendix, we review the Wilson-type lattice discretization underlying Eq.~\eqref{eqn:int_ham_SP} of the main text, and  show how the use of FGSs allows to recover the static phase diagram of the model, benchmarking these results with those of a large-$N_{f}$ expansion presented in \cite{gross_neveu_wilson}. In this context,  $N_{\rm f}$ is the number of flavours or copies of the Dirac field, and one can obtain a non-perturbative prediction of the dynamically-generated mass and the fermion condensates by letting $N_{\rm f}\to\infty$.

For a flat and static spacetime, the scale factor is trivial $\mathsf{a}(\eta)=1$, such that $\psi(x)=\chi(x)$. The Hamiltonian  of the model thus coincides with Eq.~\eqref{eq:GN_ham_QFT}, considering a static mass and allowing for $N_{\rm f}$ flavors of  Dirac fermions, which get coupled via the  four-Fermi interaction  
\beq
  \label{eq:app_GN_ham_QFT}
  H=\int\!{\rm d}{\rm x}\bigg(\overline{\Psi}(x)\left(\gamma^1\partial_{\rm x}+m\right)\Psi(x)-\frac{g_0^2}{2N_{\rm f}}\big(\overline{\Psi}\,\Psi\big)^2\bigg),
  \eeq
where $\Psi(x)=(\psi_1(x),\psi_2(x),\cdots,\psi_{N_{\rm f}}(x))^t$ contains  $N_{\rm f}$ copies of the two-component Dirac spinor. We note that the Dirac matrices only act on the spinor components $\gamma^0\mapsto\mathbb{1}_{N_{\rm f}}\otimes\gamma^0, \gamma^1\mapsto\mathbb{1}_{N_{\rm f}}\otimes\gamma^1, \gamma^5\mapsto\mathbb{1}_{N_{\rm f}}\otimes\ii\gamma^0\gamma^1$.  
This Hamiltonian was proposed as a toy model that reproduces several features of quantum chromodynamics \cite{PhysRevLett.30.1343,PhysRevLett.30.1346}. For instance, in the large-$N_{\rm f}$ limit \cite{Coleman1982}, the model displays chiral symmetry breaking by dynamical mass generation, i.e., the theory dynamically acquires a mass when $g_0^2>0$ even though this would be discarded in perturbation theory in light of  the discrete chiral symmetry under $\Psi(x)\rightarrow\gamma^5\Psi(x)$.

We consider a Wilson-type lattice regularization of this Hamiltonian  field theory, in which only the spatial coordinates of the field are discretized $\Lambda_{\rm latt}=\{{\bf x}=ia, i\in\mathbb{Z}_{N_{\rm S}
}\}$ with $a$ being the lattice spacing. 
We introduce fermionic modes for each lattice site, each flavour, and each spinor component $\psi_{i,f}=\frac{1}{\sqrt{a}}\left({c}_{i,\uparrow,f},\;{c}_{i,\downarrow,f}\right)^{\rm t}$, where 
the scaling with the lattice constant allows to recover  the correct anti-commutation relations $\{\psi^\dagger_{i,f},\;\psi_{j,f'}\}=\frac{1}{a}\delta_{i,\,j}\delta_{f,f'}\mapsto\delta(x-x')\delta_{f,f'}$ in the continuum limit $a\rightarrow0$. The operators ${c}_{i,\alpha,f}$ are thus  dimensionless, making a direct connection to the notation of Appendix~\ref{app:FGS}. If the spatial derivative responsible for the kinetic energy in Eq.~\eqref{eq:app_GN_ham_QFT} is replaced by a symmetric finite difference $\partial_{\rm x}\Psi\mapsto(\Psi_{i+1}-\Psi_{i-1})/{2a}$, 
the Hamiltonian  
\beq
H_{\rm GNN}=a\sum_{i=1}^{N_{\rm S}}\left(\frac{\Psi^\dagger_i\ii\gamma^0\gamma^1\left(\Psi_{i+1}-\Psi_{i-1}\right)}{2a}+\frac{g_0^2}{2N_{\rm f}}\left(\Psi_i^\dagger\gamma^0\Psi_i\right)^{\!\!2}\!\right).
\eeq
would correspond to  a Gross-Neveu model with na{i}ve (GNN) fermions. Here, due to the modification of the dispersion relation with respect to that of the original QFT,  the continuum limit is afflicted by the so-called fermion doubling \cite{Gattringer:2010zz,Rothe:1992nt,PhysRevD.16.3031} resulting in the so-called doublers appearing at the boundaries of the Brillouin zone (BZ). 

Following Wilson's prescription \cite{PhysRevD.10.2445}, we introduce additional terms in the Hamiltonian 
that lead to a momentum-dependent mass. This so-called Wilson mass  vanishes around the center of the BZ, 
but gets very large at the boundaries, sending the doublers to the ultraviolet cutoff $\Lambda_{\rm c}\propto 1/a$. 
The full Gross-Neveu-Wilson (GNW) model is thus ${H}_{\rm GNW}=H_{\rm GNN}+H_{\rm W}$, where the Wilson term reads
\beq
{H}_{\rm W}=a\sum_{i=1}^{N_{\rm S}}\left( \frac{\ii m}{2}\Psi_i^\dagger\gamma^0\Psi_i+\frac{\ii r}{2a}\Psi_i^\dagger\gamma^0\left(\Psi_i-\Psi_{i+1}\right)+{\rm H.c.}\right).
\eeq
Here, $r\in\left[0,\;1\right]$ is the  Wilson parameter that shall be fixed to  $r=1$ in this work. These terms keep the mass of the theory   $m_0=m$ around the center of the Brillouin zone, ${\rm k}=0$, but introduce a large mass $m_{\pi}=m+2/a$ around ${\rm k}=\pm\pi/a$. 
The final step to connect to the physics of a FRW background spacetime is, according to the results of Appendix~\ref{app:dirac_in_CS}, to exchange the bare mass by $m\mapsto m\mathsf{a}(\eta)$, and rescale the fields such that $\Psi_i\mapsto \chi_i$. A standard Fourier transformation of the fields leads to Eq.~\eqref{eqn:ham_BZ_main} in the main text.

As a benchmark of the variational method based on FGSs discussed in Appendix~\ref{app:FGS}, we will start by comparing the static properties of the GNW model $\mathsf{a}(\eta)=1$
with predictions based on large-$N_{\rm f}$ techniques described in~\cite{gross_neveu_wilson}. In the $N_{\rm f}\to\infty$ limit, one can use path-integral techniques to derive a set of self-consistent gap equations for two different fermion condensates. The scalar or chiral condensate  is the vacuum expectation value of an auxiliary scalar field $\Sigma_0=\langle\sigma(x)\rangle$ that fulfills
\beq
\label{eq:scalar}
\Sigma_0=\frac{\ii g_0^2}{2N_{\rm S}}\sum_i\left\langle\Psi_i^\dagger\gamma^0\Psi_i\right\rangle.
\eeq
Couplings where $m+\Sigma_0=0$ or $m+2/a-\Sigma_0=0$ signal critical lines where one can recover a discrete chiral symmetry in the continuum limit. As a consequence of the Wilson discretization, these critical lines separate two different groundstates, which can be characterized by the presence or absence of a non-zero topological invariant and topological edge states when considering open boundary conditions.  This corresponds to a symmetry-protected topological (SPT) phase that belongs to the $\mathsf{BDI}$ class of topological insulators~\cite{RevModPhys.88.035005}, as noted in the main text.
In addition to the scalar condensate, the Wilson discretization brings in another possibility: a pseudo-scalar  condensate  can appear as the vacuum expectation value of an auxiliary pseudo-scalar field $\Pi_0=\langle \pi(x)\rangle$, namely
\beq
\label{eq:pseudo_scalar}
\Pi_0=\frac{ g_0^2}{2N_{\rm S}}\sum_i\left\langle\Psi_i^\dagger\gamma^1\Psi_i\right\rangle.
\eeq
This signals the spontaneous breakdown of parity $\Psi({ x})\mapsto\ee^{\ii\theta}\gamma^0\Psi(-x)$, where $\ee^{\ii\theta}$ is a phase factor,  and also appears in Wilson-type discretizations of lattice gauge theories~\cite{PhysRevB.87.165142, PhysRevB.87.205440}, leading to  a so-called Aoki phase and additional critical lines in parameter space.

 \begin{figure*}[t!]
    \centering
    \includegraphics[width=2\columnwidth]{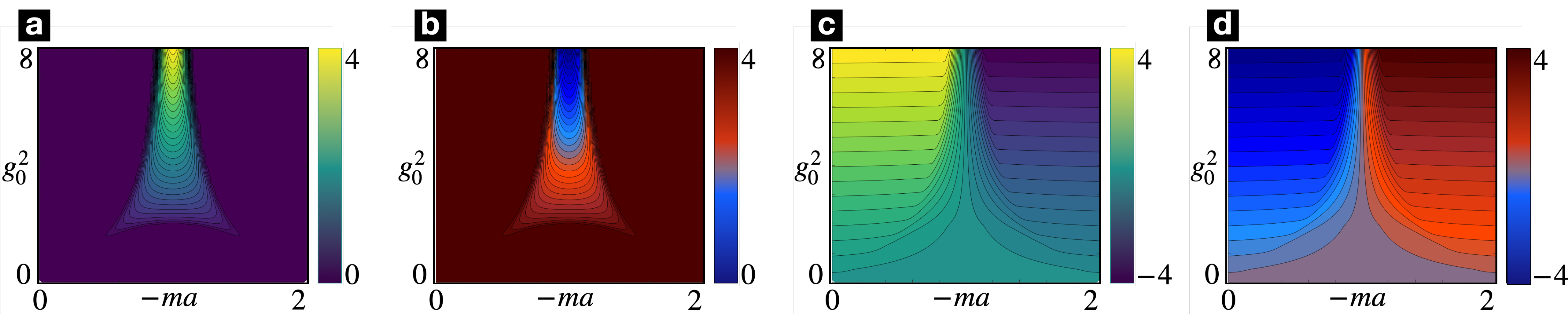}
    \caption{\textbf{Contour plots for the groundstate values of the condensates in parameter space}. \textbf{(a)} and \textbf{(c)} represent, respectively, the pseudo-scalar and scalar condensates obtained by solving the imaginary time evolution equations for the correlation matrix. \textbf{(b)} and \textbf{(d)} represent, respectively, the pseudo-scalar and scalar condensates obtained by minimizing the effective potential. We observe a perfect agreement between both approaches, thus validating the FGS formalism to describe the static properties of the Gross-Neveu-Wilson model.  The accuracy of the FGS approach depends on the maximum imaginary time $\tau_{\rm max}$ used when solving the ITE equations. In this work, we determine $\tau_{\rm max}$ dynamically when solving the ITE equations such that the convergence of the method is ensured, resulting in deviations from the large-$N$ predictions on the order of $10^{-6}$ to $10^{-7}$. We note that a longer $\tau_{\rm max}$ is required near the critical lines, where larger deviations are observed.}
    \label{fig:pi_FGS}
\end{figure*}

In particular, the large-$N_{\rm f}$ method leads to  a pair of coupled self-consistency gap equations, which can be obtained from the minimization of a so-called effective potential~\cite{Coleman1982}. Adapting the approach discussed in~\cite{10.21468/SciPostPhys.17.1.003} to the Gross-Neveu-Wilson model $H_{\rm GNW}$, the effective potential reads
\beq
\label{eq:eff_V}
V_{\rm eff}(\Sigma,\Pi)=\frac{1}{g_0^2}(\Sigma^2+\Pi^2)+\frac{1}{N_{\rm S}}\sum_{\k} \!\!\big(\epsilon_\k(\Sigma,\Pi)-\epsilon_\k(0,0)\big),
\eeq
where we have introduced the following single-particle energies renormalized by the presence of the fermion condensates
\beq
\label{eq:energies_v_eff}
\epsilon_\k(\Sigma,\Pi)=\sqrt{\frac{\sin^2(\k a)}{a^2}+\left(m+\Sigma+\frac{1-\cos(\k a)}{a}
\right)^2 +\Pi^2}.
\eeq
The fermion condensates in the limit $N_{\rm f}\to\infty$ are obtained by solving the non-linear minimization problem 
\beq
(\Sigma_0,\Pi_0)={\rm argmin}\big\{V_{\rm eff}(\Sigma,\Pi):\hspace{1ex}\Sigma\in\mathbb{R},\Pi\in\mathbb{R}\big\}.
\eeq
As described in detail in Fig.~\ref{fig:trajectories_explanation} of Sec.~\ref{sec:2}, this allows to distinguish three different regions in the phase diagram discussed in Eq.~\eqref{eq:phases} of the  main text.

We now use the large-$N_{\rm f}$ predictions as a benchmark of the ITE based on FGSs in Eq.~\eqref{eqn:ITE_int}, which require solving a set of non-linear ODEs for the correlation matrix~\eqref{eqn:def_correlation_matrix}. As it turns out, FGSs with a single flavour  is tantamount to a Hartree-Fock self-consistent approach, and yields the exact same results as the large-$N_{\rm f}$ analysis. As discussed in Appendix \ref{app:FGS}, in order to proceed with the ITE equation, we need  to find an effective  Hamiltonian that depends on the correlation matrix itself $\tilde{h}(\Gamma)$, which requires applying Wick's theorem to the quartic 
interactions. In reciprocal space, this  reads 
\beq
H_{\rm int}=\frac{g_0^2}{2aN_{\rm S}}\sum_{\k_1\k_2\k_3}\!\!\!\!\big({\psi}^{{\dagger}}_{\k_1}\gamma^0{\psi}^{\phantom{\dagger}}_{\k_2}\big)\!\big({\psi}^\dagger_{\k_3}\gamma^0{\psi}^{\phantom{\dagger}}_{\k_1-k_2+k_3}\big),
\label{eqn:ham_BZ_app}
\eeq
Guided by the previous large-$N_{\rm f}$ discussion, we use the scalar~\eqref{eq:scalar} and pseudo-scalar~\eqref{eq:pseudo_scalar} condensates as the main relevant components of the correlation matrix that enter in self-consistent FGS Hamiltonian
\beq
\tilde{h}(\Gamma)=\bigoplus_{\k}\left(h_{\k}-\ii\Sigma\gamma^0-\Pi\gamma^1\right)
,\label{eqn:int_MF}
\eeq
where we have made use of the single-particle Hamiltonian for Wilson fermions in Eq.~\eqref{eqn:hamSP_BZ_main} applied to the static problem, rewritten here for convenience
\beq
h_\k=\frac{\sin(\k a)}{a}\gamma^0\gamma^1+\ii\left(m+\frac{1-\cos(\k a)}{a}
\right)\gamma^0.
\eeq
We have also assumed homogeneous condensates, such that the resulting Hamiltonian is translationally-invariant and therefore can be casted into a block-diagonal form after performing a Fourier transform. 

We thus see that the scalar condensate $\Sigma_0$ amounts to a shift in the bare mass of the theory, while the pseudo-scalar condensate $\Pi_0$ introduces a new term in the Hamiltonian, proportional to $\gamma^1$, opening a different type of gap in the energy dispersion. This interpretation is fully consistent with the renormalized single-particle energies~\eqref{eq:energies_v_eff} entering in the large-$N_{\rm f}$ effective potential~\eqref{eq:eff_V}. 
To find the explicit self-consistent values of these condensates, and even the whole correlation matrix, we now  solve the equations of ITE~\eqref{eqn:ITE_int} for each combination of microscopic parameters $(ma,g_0^2)$, which forms the coupling space of the theory. 
To do so while benefiting from the translational invariance of the condensates, we construct the correlation matrix $\Gamma$ in momentum space which, in general,  is a $2N_{\rm S}\times2N_{\rm S}$ matrix with components $\Gamma_{\k\alpha,\q\beta}$. Components with different momenta are vanishing due to the translational invariance \cite{Surace22}, which allows us to reduce the complexity of the coupled ODEs. Just as the self-consistent Hamiltonian~\eqref{eqn:int_MF}, the correlation matrix is block-diagonal in momentum space with each sub-matrix being $2\times2$ due to the possible spinor components.  Each element of these correlation matrix blocks will be a variational parameter that is determined by solving  the imaginary-time evolution~\eqref{eqn:ITE_int} in the large-$\tau$ limit $\Gamma_0(m,g_0^2)=\lim_{\tau\to\infty}\Gamma(\tau)$ for a given set of values  of the microscopic  $(m,\;g_0^2)$ that spans the possible phases of the model in coupling space~\eqref{eq:phases}. As noted in Appendix~\ref{app:FGS}, 
this is equivalent to solving the variational FGS minimization $\Gamma_0={\rm argmin}\{\bra{\psi(\Gamma)}H_{\rm GNW}\ket{\psi(\Gamma)}/\langle{\psi(\Gamma)}\ket{\psi(\Gamma)}\}$, and allows us to infer the fermion condensates using
\beq
\begin{split}
\Sigma_0=\frac{g_0^2}{2aN_{\rm S}}\sum_\k\biggl( [\Gamma_0]_{\k,\uparrow,\k,\uparrow}-[\Gamma_0]_{\k,\downarrow,\k,\downarrow}\biggr),\\
\Pi_0=\frac{\ii g_0^2}{2aN_{\rm S}}\sum_\k\biggl( [\Gamma_0]_{\k,\downarrow,\k,\uparrow}-[\Gamma_0]_{\k,\uparrow,\k,\downarrow}\biggr).
\end{split}
\eeq

 The results from the large-$N_{\rm f}$ and the FGS with ITE are  shown in Figs. \ref{fig:pi_FGS}. We find that the region with a non-vanishing value of $\Pi_0$ corresponds exactly with the Aoki phase found using large-$N_{\rm f}$ methods, and that its boundaries also corresponds to the critical lines separating the trivial-Aoki phases, and the SPT-Aoki phases. We also find a non-vanishing value of the scalar condensate for all values of the parameters with $g_0^2>0$, corresponding to the aforementioned dynamical mass generation, except for the symmetry line $ma=-1$. With this benchmark, we can conclude that the presented FGS approach  is consistent with the static properties of the Gross-Neveu-Wilson model, which also supports its use  to study the dynamical properties in real time.

\section{\bf Mode-function benchmarks of  real-time fermionic Gaussian states}
\label{app:part_prod_FGS}

As a consistency check, we calculate in this appendix the spectra of produced particles for different  masses $m$ and setting $g_0^2=0$, such that  the formalism of FGSs should coincide with the exact results based on the evolution of mode functions \cite{FulgadoClaudio2023fermionproduction}. For the sake of completeness, we first review  the formalism of particle production with  mode functions. 

We begin by expanding the field operator $\chi$ in particle and anti-particle operators by using the spinors so-called mode functions, which will be denoted here by $|u_\k(\eta)\rangle$ and $|v_{-k}(\eta)\rangle$
\beq
\chi(\eta,\;\x)=\int\frac{d\k}{2\pi}\left(a_\k u_\k(\eta)+b_{-\k}^\dagger v_{-k}(\eta)\right)\ee^{\ii\k\x}.
\eeq

In this approach, the time-dependence of the field operator $\chi$ is fully encapsulated in the mode functions, such that these satisfy the Schrödinger-like equations
\beq
\begin{split}
    \ii\frac{du_\k(\eta)}{d\eta}=h_\k(\eta)u_\k(\eta),\\
    \ii\frac{dv_{\k}(\eta)}{d\eta}=h_{-\k}(\eta)v_\k(\eta),
    \label{eq:schrodinger_mode_functions}
\end{split}
\eeq
where $h_k(\eta)$ is the free single-particle Hamiltonian, which in this article corresponds to Eq.~\eqref{eqn:hamSP_BZ_main} in the main text. The fact that this Hamiltonian is time-dependent requires some care when defining the notion of particles and anti-particles. Since we consider spacetimes that are static both in the asymptotic in- and out- regions, we stick to the so-called diagonalization method \cite{Haro_2008}, which amounts to taking the mode functions as the instantaneous eigenstates of the single-particle Hamiltonian in order to define an instantaneous notion of particle and anti-particle. Therefore, at time $\eta=\eta_\star$, the operators associated to the correct notions of particle and anti-particle are those arising when expanding the field $\chi(x)$ using the instantaneous eigenstates of $h_\k(\eta_\star)$, denoted by ${\rm v}_{\k}^\pm(\eta_\star)$ and such that $h_\k(\eta_\star){\rm v}_{\k}^\pm(\eta_\star)=\pm\epsilon_\k(\eta_\star){\rm v}_{\k}^\pm(\eta_\star)$. 

In the in-region, $\eta=\eta_0$, the correct notion of particle is established by identifying 
\beq
\begin{split}
    &u_\k(\eta_0)={\rm v}_{\k}^{+}(\eta_0),\\
    &v_\k(\eta_0)={\rm v}_{\k}^{-}(\eta_0).
\end{split}
\eeq

If the adiabatic approximation is perfectly fulfilled, the evolution of the mode functions given by Eq.~\eqref{eq:schrodinger_mode_functions} will stick to the successive instantaneous groundstates. In this case, no particles are produced. However, in more general scenarios, the evolution of the mode functions will result in a mixing between the positive and the negative eigenstates. This mixing can be expressed by means of a Bogoliubov transformation
\beq
\begin{split}
    &u_\k(\eta_{\rm f})=\alpha_\k(\eta_{\rm f}){\rm v}_{\k}^{+}(\eta_0)+\beta_{\k}^{*}(\eta_{\rm f}){\rm v}_{\k}^{-}(\eta_0),\\
    &v_\k(\eta_{\rm f})=-\beta_\k(\eta_{\rm f}){\rm v}_{\k}^{+}(\eta_0)+\alpha_{\k}^{*}(\eta_{\rm f}){\rm v}_{\k}^{-}(\eta_0).
\end{split}
\label{eq:mode_functions_initial_conditions}
\eeq
This coefficients, so-called Bogoliubov, satisfy $|\alpha_\k(\eta)|^2+|\beta_\k(\eta)|^2=1$, and are precisely those appearing in Eq.~\ref{eq:evolutionop}. Therefore, in the free case, these coefficients can be obtained within the formalism of mode functions by solving Eqs.~\eqref{eq:schrodinger_mode_functions} with initial condition given by \eqref{eq:mode_functions_initial_conditions}. After solving them, the relevant quantity $|\beta_\k(\eta_{\rm f})|^2$ appearing in \eqref{eq:density_produced} is given by
\beq
|\beta_\k|^2=\left|\left({\rm v}_{\k}^{-}(\eta_f)\right)^\dagger\cdot u_\k(\eta_{\rm f})\right|^2.
\eeq

  \begin{figure}[t!]
    \centering
    \includegraphics[width=0.85\columnwidth]{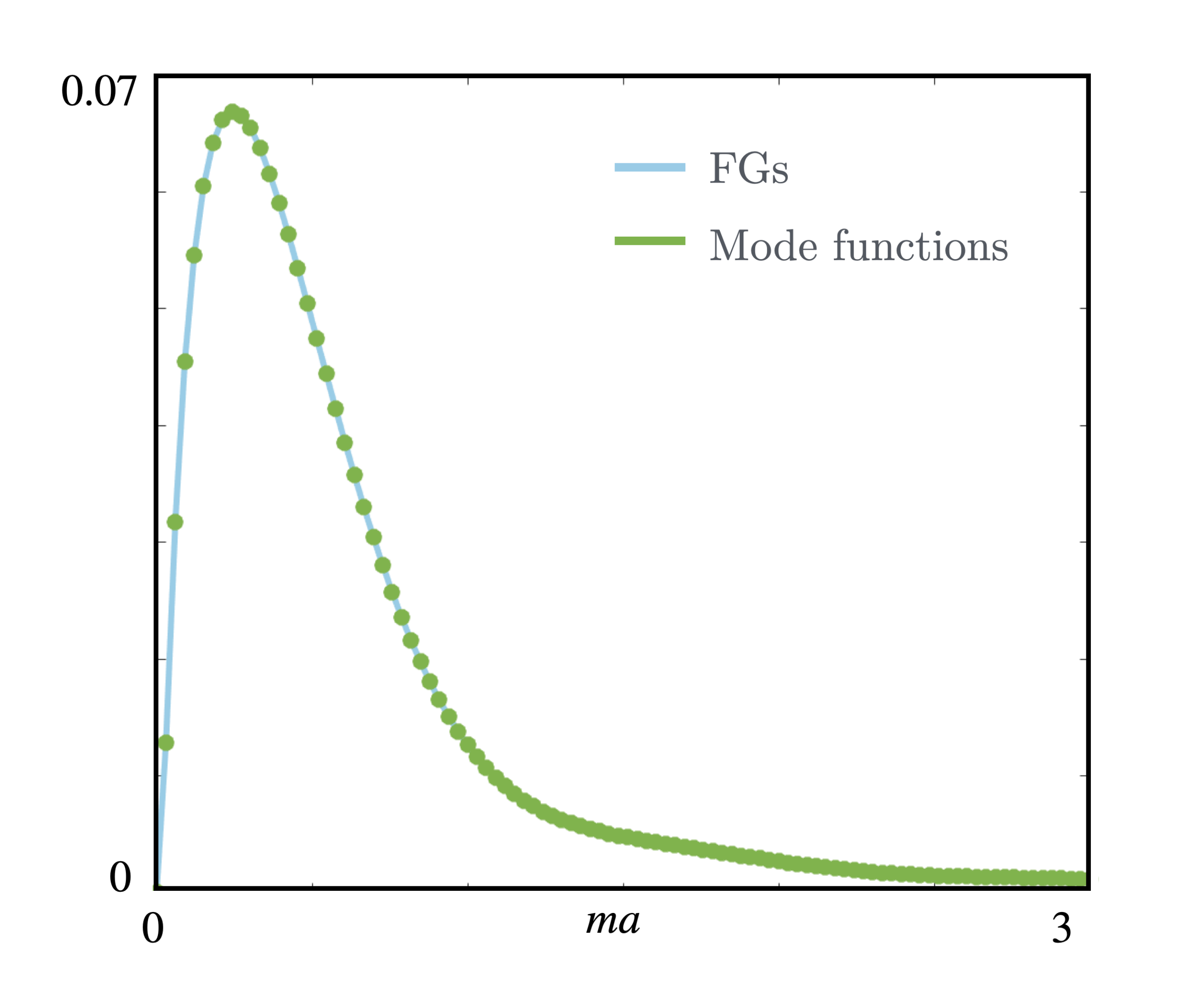}
    \caption{\textbf{Density of produced fermions as a function of the mass}. We recover the results that are obtained from the mode functions approach. $\mathsf{H}a=1$, $\mathsf{a}_0=0.1$, $\mathsf{a}_{\rm f}=1$, $N_{\rm S}=64$.}
    \label{fig:app_d_prod_vs_m}
\end{figure}
  Now, following this approach, we compute the density of produced particles as a function of the mass, and compare this result to that obtained by following the FGS approach described in the main text. The result for the density of particles is shown in Fig. \ref{fig:app_d_prod_vs_m}, where it is clear that there is a perfect agreement between the results using the protocol explained in Sec.~\ref{sec:FGS_part_prod}, and which are developed with the formalism of FGSs, and the one using the formalism of mode functions, as described in full detail in \cite{FulgadoClaudio2023fermionproduction}. 
  This equivalence between the results yielded from this method and our previous ones confirms that this approach using FGSs is suitable to treat the problem of particle production, and therefore allows us to generalize to the interacting regime.
  
\section{\bf Adiabaticity condition for the interacting case}
\label{app:adiab_int}

In this Appendix, we inspect how the adiabaticity condition~\eqref{eqn:adiab_cond_wilson_free} is modified in the interacting case. When interactions are switched on, the Hamiltonian is no longer quadratic and we cannot define a single-particle Hamiltonian independent of the fields.
Consequently, the analysis of the adiabaticity of the  free case cannot be  ported  to the interacting regime in full generality. However, since we are using  variational fermionic Gaussian states, which amounts to a self-consistent Hartree-Fock approximation, we can define a quadratic theory involving two auxiliary fields, the fermion condensates $\Sigma$ and $\Pi$, and gain some insight 
in the adiabatic parameter by introducing these condensates in a mean-field  Hamiltonian. 

First, let us consider the simpler case in which the condensates are constant throughout the whole evolution. This case corresponds to the limit of a quench. These effective time-independent condensates  enlarge the gap, and this fact implies that its static value does play a role in particle production, which becomes clear by inspecting how the adiabatic parameter is modified by including these static condensates. For simplicity, in the analytic expressions here we consider only the effect of the scalar condensate $\Sigma$, neglecting the presence of the pseudo-scalar condensate $\Pi$, so it would not be valid in the parity-breaking Aoki phase. However, the analysis can be easily extended numerically to situations with $\Pi\neq0$. For $\Pi=0$, $\Sigma=\Sigma_0\neq\Sigma(\eta)$, we obtain
\beq
\xi^{\rm int}_\k=Ha\left|\frac{\mathsf{a}(\eta)\sin(\k a) m_{\k}(\eta)a}{4\left({\sin^2(\k a)}+(m_{\k}(\eta)-\Sigma_0)^2a^2\right)^{3/2}}\right|.
\label{eqn:adiab_condition_wilson_int_quench}
\eeq

One can readily observe that setting $g_0^2=0$,  such that the scalar condensate vanishes $\Sigma_0=0$, the adiabatic parameter reduces to the free one \eqref{eqn:adiab_cond_wilson_free}. However, when 
$\Sigma_0$ is not zero, it can change the adiabaticity parameter depending on the sign of the condensate with respect to the bare mass. 
The form of this adiabaticity parameter offers some insight about the way in which the interactions affect particle production. In particular, we note that the static, non-vanishing value of the condensate $\Sigma_0$ only enters in the denominator of this parameter. Consequently, as the value of $\Sigma_0$ increases as the coupling strength $g_0^2$ increases, this results in the vanishing of the adiabatic parameter in the limit of large strength coupling, leading to a stricter fulfillment of the adiabatic approximation and therefore suppressing the production of particles.

When we depart from the quench limit, the time evolution of the condensates must be taken into account when studying the adiabaticity condition. In the most-adiabatic limit, the correlation matrix follows its groundstate values along the whole expansion, so the time-evolution of $\Sigma=\Sigma(\eta)$ can be understood as being effectively parametric, evolving through its groundstate values which can be directly read off from the groundstate phase diagram. 
For cases in which the evolution is neither purely adiabatic nor purely a quench, the exact time-evolution of the condensates must be numerically obtained. 
Considering a time-dependent condensate $\Sigma(\eta)$ (but not fixing its precise form, so that this analysis is applicable to different scenarios), the adiabatic parameter reads

\beq
\xi^{\rm int}_\k=\mathsf{H}a\left|\frac{\mathsf{a}(\eta){\sin(\k a)} \left(m_{\k}(\eta)-\frac{\dot\Sigma(\eta)}{\mathsf{a}(\eta)\mathsf{H}}\right)a}{4\left({\sin^2(\k a)}+\left(m_{\k}(\eta)-\Sigma(\eta)\right)^2a^2\right)^{3/2}}\right|,
\label{eqn:adiab_condition_wilson_int_non_quench}
\eeq
where $\dot\Sigma(\eta)=\frac{d\Sigma(\eta)}{d\eta}$. Note that, for a static condensate $\Sigma(\eta)=\Sigma_0$, the derivative $\dot\Sigma(\eta)$ is vanishing and we recover the quench result for the adiabatic parameter \eqref{eqn:adiab_condition_wilson_int_quench}. 
Now, the correlation matrix is not frozen on its initial value, and  neither is the value of the scalar condensate, whose dynamics acquire an importance from the point of view of the adiabaticity via the second term inside the parenthesis in the numerator of \eqref{eqn:adiab_condition_wilson_int_non_quench}.

Therefore, to understand the role played by the dynamical evolution of the condensates in the  production of particles, one has to depart from the quench limit, so that the evolution of the condensates becomes not-trivial. On the opposite limit, that of $\mathsf{H}a\rightarrow0$, the condensates acquire a self-consistent evolution given by its groundstate values, but the problem becomes trivial from the point of view of the production of particles, as the adiabatic approximation is perfectly fulfilled and no particles are generated. Therefore, the most interesting interplay between the dynamical evolution of the condensates and the production of particles will arise for values of $\mathsf{H}a$ that allow to explore regimes away from both the quench and the purely adiabatic limits. 

\bibliographystyle{apsrev4-1}
\bibliography{main.bib}

\begin{thebibliography}{143}%
\makeatletter
\providecommand \@ifxundefined [1]{%
 \@ifx{#1\undefined}
}%
\providecommand \@ifnum [1]{%
 \ifnum #1\expandafter \@firstoftwo
 \else \expandafter \@secondoftwo
 \fi
}%
\providecommand \@ifx [1]{%
 \ifx #1\expandafter \@firstoftwo
 \else \expandafter \@secondoftwo
 \fi
}%
\providecommand \natexlab [1]{#1}%
\providecommand \enquote  [1]{``#1''}%
\providecommand \bibnamefont  [1]{#1}%
\providecommand \bibfnamefont [1]{#1}%
\providecommand \citenamefont [1]{#1}%
\providecommand \href@noop [0]{\@secondoftwo}%
\providecommand \href [0]{\begingroup \@sanitize@url \@href}%
\providecommand \@href[1]{\@@startlink{#1}\@@href}%
\providecommand \@@href[1]{\endgroup#1\@@endlink}%
\providecommand \@sanitize@url [0]{\catcode `\\12\catcode `\$12\catcode
  `\&12\catcode `\#12\catcode `\^12\catcode `\_12\catcode `\%12\relax}%
\providecommand \@@startlink[1]{}%
\providecommand \@@endlink[0]{}%
\providecommand \url  [0]{\begingroup\@sanitize@url \@url }%
\providecommand \@url [1]{\endgroup\@href {#1}{\urlprefix }}%
\providecommand \urlprefix  [0]{URL }%
\providecommand \Eprint [0]{\href }%
\providecommand \doibase [0]{http://dx.doi.org/}%
\providecommand \selectlanguage [0]{\@gobble}%
\providecommand \bibinfo  [0]{\@secondoftwo}%
\providecommand \bibfield  [0]{\@secondoftwo}%
\providecommand \translation [1]{[#1]}%
\providecommand \BibitemOpen [0]{}%
\providecommand \bibitemStop [0]{}%
\providecommand \bibitemNoStop [0]{.\EOS\space}%
\providecommand \EOS [0]{\spacefactor3000\relax}%
\providecommand \BibitemShut  [1]{\csname bibitem#1\endcsname}%
\let\auto@bib@innerbib\@empty
\bibitem [{\citenamefont {Birrell}\ and\ \citenamefont
  {Davies}(1984)}]{Birrell:1982ix}%
  \BibitemOpen
  \bibfield  {author} {\bibinfo {author} {\bibfnamefont {N.~D.}\ \bibnamefont
  {Birrell}}\ and\ \bibinfo {author} {\bibfnamefont {P.~C.~W.}\ \bibnamefont
  {Davies}},\ }\href {\doibase 10.1017/CBO9780511622632} {\emph {\bibinfo
  {title} {{Quantum Fields in Curved Space}}}},\ Cambridge Monographs on
  Mathematical Physics\ (\bibinfo  {publisher} {Cambridge Univ. Press},\
  \bibinfo {address} {Cambridge, UK},\ \bibinfo {year} {1984})\BibitemShut
  {NoStop}%
\bibitem [{\citenamefont {Parker}\ and\ \citenamefont
  {Toms}(2009)}]{Parker:2009uva}%
  \BibitemOpen
  \bibfield  {author} {\bibinfo {author} {\bibfnamefont {L.~E.}\ \bibnamefont
  {Parker}}\ and\ \bibinfo {author} {\bibfnamefont {D.}~\bibnamefont {Toms}},\
  }\href {\doibase 10.1017/CBO9780511813924} {\emph {\bibinfo {title} {{Quantum
  Field Theory in Curved Spacetime}: {Quantized Field and Gravity}}}},\
  Cambridge Monographs on Mathematical Physics\ (\bibinfo  {publisher}
  {Cambridge University Press},\ \bibinfo {year} {2009})\BibitemShut {NoStop}%
\bibitem [{\citenamefont {Ford}(2021)}]{Ford_2021}%
  \BibitemOpen
  \bibfield  {author} {\bibinfo {author} {\bibfnamefont {L.~H.}\ \bibnamefont
  {Ford}},\ }\href {\doibase 10.1088/1361-6633/ac1b23} {\bibfield  {journal}
  {\bibinfo  {journal} {Reports on Progress in Physics}\ }\textbf {\bibinfo
  {volume} {84}},\ \bibinfo {pages} {116901} (\bibinfo {year}
  {2021})}\BibitemShut {NoStop}%
\bibitem [{\citenamefont {Hawking}(1974)}]{Hawking:1974rv}%
  \BibitemOpen
  \bibfield  {author} {\bibinfo {author} {\bibfnamefont {S.~W.}\ \bibnamefont
  {Hawking}},\ }\href {\doibase 10.1038/248030a0} {\bibfield  {journal}
  {\bibinfo  {journal} {Nature}\ }\textbf {\bibinfo {volume} {248}},\ \bibinfo
  {pages} {30} (\bibinfo {year} {1974})}\BibitemShut {NoStop}%
\bibitem [{\citenamefont {Unruh}(1976)}]{PhysRevD.14.870}%
  \BibitemOpen
  \bibfield  {author} {\bibinfo {author} {\bibfnamefont {W.~G.}\ \bibnamefont
  {Unruh}},\ }\href {\doibase 10.1103/PhysRevD.14.870} {\bibfield  {journal}
  {\bibinfo  {journal} {Phys. Rev. D}\ }\textbf {\bibinfo {volume} {14}},\
  \bibinfo {pages} {870} (\bibinfo {year} {1976})}\BibitemShut {NoStop}%
\bibitem [{\citenamefont {Parker}(1968)}]{PhysRevLett.21.562}%
  \BibitemOpen
  \bibfield  {author} {\bibinfo {author} {\bibfnamefont {L.}~\bibnamefont
  {Parker}},\ }\href {\doibase 10.1103/PhysRevLett.21.562} {\bibfield
  {journal} {\bibinfo  {journal} {Phys. Rev. Lett.}\ }\textbf {\bibinfo
  {volume} {21}},\ \bibinfo {pages} {562} (\bibinfo {year} {1968})}\BibitemShut
  {NoStop}%
\bibitem [{\citenamefont {Parker}(1971)}]{PhysRevD.3.346}%
  \BibitemOpen
  \bibfield  {author} {\bibinfo {author} {\bibfnamefont {L.}~\bibnamefont
  {Parker}},\ }\href {\doibase 10.1103/PhysRevD.3.346} {\bibfield  {journal}
  {\bibinfo  {journal} {Phys. Rev. D}\ }\textbf {\bibinfo {volume} {3}},\
  \bibinfo {pages} {346} (\bibinfo {year} {1971})}\BibitemShut {NoStop}%
\bibitem [{\citenamefont {Carroll}(2003)}]{carroll2003spacetime}%
  \BibitemOpen
  \bibfield  {author} {\bibinfo {author} {\bibfnamefont {S.}~\bibnamefont
  {Carroll}},\ }\href
  {http://www.amazon.com/Spacetime-Geometry-Introduction-General-Relativity/dp/0805387323}
  {\emph {\bibinfo {title} {Spacetime and Geometry: An Introduction to General
  Relativity}}}\ (\bibinfo  {publisher} {Benjamin Cummings},\ \bibinfo {year}
  {2003})\BibitemShut {NoStop}%
\bibitem [{\citenamefont {Barcelo}\ \emph {et~al.}(2005)\citenamefont
  {Barcelo}, \citenamefont {Liberati},\ and\ \citenamefont
  {Visser}}]{Barcelo:2005fc}%
  \BibitemOpen
  \bibfield  {author} {\bibinfo {author} {\bibfnamefont {C.}~\bibnamefont
  {Barcelo}}, \bibinfo {author} {\bibfnamefont {S.}~\bibnamefont {Liberati}}, \
  and\ \bibinfo {author} {\bibfnamefont {M.}~\bibnamefont {Visser}},\ }\href
  {\doibase 10.12942/lrr-2005-12} {\bibfield  {journal} {\bibinfo  {journal}
  {Living Rev. Rel.}\ }\textbf {\bibinfo {volume} {8}},\ \bibinfo {pages} {12}
  (\bibinfo {year} {2005})},\ \Eprint {http://arxiv.org/abs/gr-qc/0505065}
  {arXiv:gr-qc/0505065} \BibitemShut {NoStop}%
\bibitem [{\citenamefont {Jacquet}\ \emph {et~al.}(2020)\citenamefont
  {Jacquet}, \citenamefont {Weinfurtner},\ and\ \citenamefont
  {K\"onig}}]{Jacquet:2020bar}%
  \BibitemOpen
  \bibfield  {author} {\bibinfo {author} {\bibfnamefont {M.~J.}\ \bibnamefont
  {Jacquet}}, \bibinfo {author} {\bibfnamefont {S.}~\bibnamefont
  {Weinfurtner}}, \ and\ \bibinfo {author} {\bibfnamefont {F.}~\bibnamefont
  {K\"onig}},\ }\href {\doibase 10.1098/rsta.2019.0239} {\bibfield  {journal}
  {\bibinfo  {journal} {Phil. Trans. Roy. Soc. Lond. A}\ }\textbf {\bibinfo
  {volume} {378}},\ \bibinfo {pages} {20190239} (\bibinfo {year} {2020})},\
  \Eprint {http://arxiv.org/abs/2005.04027} {arXiv:2005.04027 [gr-qc]}
  \BibitemShut {NoStop}%
\bibitem [{\citenamefont {Philbin}\ \emph {et~al.}(2008)\citenamefont
  {Philbin}, \citenamefont {Kuklewicz}, \citenamefont {Robertson},
  \citenamefont {Hill}, \citenamefont {König},\ and\ \citenamefont
  {Leonhardt}}]{Philbin2008}%
  \BibitemOpen
  \bibfield  {author} {\bibinfo {author} {\bibfnamefont {T.~G.}\ \bibnamefont
  {Philbin}}, \bibinfo {author} {\bibfnamefont {C.}~\bibnamefont {Kuklewicz}},
  \bibinfo {author} {\bibfnamefont {S.}~\bibnamefont {Robertson}}, \bibinfo
  {author} {\bibfnamefont {S.}~\bibnamefont {Hill}}, \bibinfo {author}
  {\bibfnamefont {F.}~\bibnamefont {König}}, \ and\ \bibinfo {author}
  {\bibfnamefont {U.}~\bibnamefont {Leonhardt}},\ }\href {\doibase
  10.1126/science.1153625} {\bibfield  {journal} {\bibinfo  {journal}
  {Science}\ }\textbf {\bibinfo {volume} {319}},\ \bibinfo {pages} {1367}
  (\bibinfo {year} {2008})}\BibitemShut {NoStop}%
\bibitem [{\citenamefont {Belgiorno}\ \emph {et~al.}(2010)\citenamefont
  {Belgiorno}, \citenamefont {Cacciatori}, \citenamefont {Clerici},
  \citenamefont {Gorini}, \citenamefont {Ortenzi}, \citenamefont {Rizzi},
  \citenamefont {Rubino}, \citenamefont {Sala},\ and\ \citenamefont
  {Faccio}}]{Belgiorno2010}%
  \BibitemOpen
  \bibfield  {author} {\bibinfo {author} {\bibfnamefont {F.}~\bibnamefont
  {Belgiorno}}, \bibinfo {author} {\bibfnamefont {S.~L.}\ \bibnamefont
  {Cacciatori}}, \bibinfo {author} {\bibfnamefont {M.}~\bibnamefont {Clerici}},
  \bibinfo {author} {\bibfnamefont {V.}~\bibnamefont {Gorini}}, \bibinfo
  {author} {\bibfnamefont {G.}~\bibnamefont {Ortenzi}}, \bibinfo {author}
  {\bibfnamefont {L.}~\bibnamefont {Rizzi}}, \bibinfo {author} {\bibfnamefont
  {E.}~\bibnamefont {Rubino}}, \bibinfo {author} {\bibfnamefont {V.~G.}\
  \bibnamefont {Sala}}, \ and\ \bibinfo {author} {\bibfnamefont
  {D.}~\bibnamefont {Faccio}},\ }\href {\doibase
  10.1103/PhysRevLett.105.203901} {\bibfield  {journal} {\bibinfo  {journal}
  {Phys. Rev. Lett.}\ }\textbf {\bibinfo {volume} {105}},\ \bibinfo {pages}
  {203901} (\bibinfo {year} {2010})}\BibitemShut {NoStop}%
\bibitem [{\citenamefont {Drori}\ \emph {et~al.}(2019)\citenamefont {Drori},
  \citenamefont {Rosenberg}, \citenamefont {Bermudez}, \citenamefont
  {Silberberg},\ and\ \citenamefont {Leonhardt}}]{Drori2019}%
  \BibitemOpen
  \bibfield  {author} {\bibinfo {author} {\bibfnamefont {J.}~\bibnamefont
  {Drori}}, \bibinfo {author} {\bibfnamefont {Y.}~\bibnamefont {Rosenberg}},
  \bibinfo {author} {\bibfnamefont {D.}~\bibnamefont {Bermudez}}, \bibinfo
  {author} {\bibfnamefont {Y.}~\bibnamefont {Silberberg}}, \ and\ \bibinfo
  {author} {\bibfnamefont {U.}~\bibnamefont {Leonhardt}},\ }\href {\doibase
  10.1103/PhysRevLett.122.010404} {\bibfield  {journal} {\bibinfo  {journal}
  {Phys. Rev. Lett.}\ }\textbf {\bibinfo {volume} {122}},\ \bibinfo {pages}
  {010404} (\bibinfo {year} {2019})}\BibitemShut {NoStop}%
\bibitem [{\citenamefont {Garay}\ \emph {et~al.}(2000)\citenamefont {Garay},
  \citenamefont {Anglin}, \citenamefont {Cirac},\ and\ \citenamefont
  {Zoller}}]{Garay2000}%
  \BibitemOpen
  \bibfield  {author} {\bibinfo {author} {\bibfnamefont {L.~J.}\ \bibnamefont
  {Garay}}, \bibinfo {author} {\bibfnamefont {J.~R.}\ \bibnamefont {Anglin}},
  \bibinfo {author} {\bibfnamefont {J.~I.}\ \bibnamefont {Cirac}}, \ and\
  \bibinfo {author} {\bibfnamefont {P.}~\bibnamefont {Zoller}},\ }\href
  {\doibase 10.1103/PhysRevLett.85.4643} {\bibfield  {journal} {\bibinfo
  {journal} {Phys. Rev. Lett.}\ }\textbf {\bibinfo {volume} {85}},\ \bibinfo
  {pages} {4643} (\bibinfo {year} {2000})}\BibitemShut {NoStop}%
\bibitem [{\citenamefont {Lahav}\ \emph {et~al.}(2010)\citenamefont {Lahav},
  \citenamefont {Itah}, \citenamefont {Blumkin}, \citenamefont {Gordon},
  \citenamefont {Rinott}, \citenamefont {Zayats},\ and\ \citenamefont
  {Steinhauer}}]{Lahav2010}%
  \BibitemOpen
  \bibfield  {author} {\bibinfo {author} {\bibfnamefont {O.}~\bibnamefont
  {Lahav}}, \bibinfo {author} {\bibfnamefont {A.}~\bibnamefont {Itah}},
  \bibinfo {author} {\bibfnamefont {A.}~\bibnamefont {Blumkin}}, \bibinfo
  {author} {\bibfnamefont {C.}~\bibnamefont {Gordon}}, \bibinfo {author}
  {\bibfnamefont {S.}~\bibnamefont {Rinott}}, \bibinfo {author} {\bibfnamefont
  {A.}~\bibnamefont {Zayats}}, \ and\ \bibinfo {author} {\bibfnamefont
  {J.}~\bibnamefont {Steinhauer}},\ }\href {\doibase
  10.1103/PhysRevLett.105.240401} {\bibfield  {journal} {\bibinfo  {journal}
  {Phys. Rev. Lett.}\ }\textbf {\bibinfo {volume} {105}},\ \bibinfo {pages}
  {240401} (\bibinfo {year} {2010})}\BibitemShut {NoStop}%
\bibitem [{\citenamefont {Steinhauer}(2016)}]{Steinhauer2016}%
  \BibitemOpen
  \bibfield  {author} {\bibinfo {author} {\bibfnamefont {J.}~\bibnamefont
  {Steinhauer}},\ }\href {\doibase 10.1038/nphys3863} {\bibfield  {journal}
  {\bibinfo  {journal} {Nature Physics}\ }\textbf {\bibinfo {volume} {12}},\
  \bibinfo {pages} {959} (\bibinfo {year} {2016})}\BibitemShut {NoStop}%
\bibitem [{\citenamefont {Kolobov}\ \emph {et~al.}(2021)\citenamefont
  {Kolobov}, \citenamefont {Golubkov}, \citenamefont {Mu{\~{n}}oz~de Nova},\
  and\ \citenamefont {Steinhauer}}]{Kolobov2021}%
  \BibitemOpen
  \bibfield  {author} {\bibinfo {author} {\bibfnamefont {V.~I.}\ \bibnamefont
  {Kolobov}}, \bibinfo {author} {\bibfnamefont {K.}~\bibnamefont {Golubkov}},
  \bibinfo {author} {\bibfnamefont {J.~R.}\ \bibnamefont {Mu{\~{n}}oz~de
  Nova}}, \ and\ \bibinfo {author} {\bibfnamefont {J.}~\bibnamefont
  {Steinhauer}},\ }\href {\doibase 10.1038/s41567-020-01076-0} {\bibfield
  {journal} {\bibinfo  {journal} {Nature Physics}\ }\textbf {\bibinfo {volume}
  {17}},\ \bibinfo {pages} {362} (\bibinfo {year} {2021})}\BibitemShut
  {NoStop}%
\bibitem [{\citenamefont {Hung}\ \emph {et~al.}(2013)\citenamefont {Hung},
  \citenamefont {Gurarie},\ and\ \citenamefont {Chin}}]{Hung:2012nc}%
  \BibitemOpen
  \bibfield  {author} {\bibinfo {author} {\bibfnamefont {C.-L.}\ \bibnamefont
  {Hung}}, \bibinfo {author} {\bibfnamefont {V.}~\bibnamefont {Gurarie}}, \
  and\ \bibinfo {author} {\bibfnamefont {C.}~\bibnamefont {Chin}},\ }\href
  {\doibase 10.1126/science.1237557} {\bibfield  {journal} {\bibinfo  {journal}
  {Science}\ }\textbf {\bibinfo {volume} {341}},\ \bibinfo {pages} {1213}
  (\bibinfo {year} {2013})},\ \Eprint {http://arxiv.org/abs/1209.0011}
  {arXiv:1209.0011 [cond-mat.quant-gas]} \BibitemShut {NoStop}%
\bibitem [{\citenamefont {Eckel}\ \emph {et~al.}(2018)\citenamefont {Eckel},
  \citenamefont {Kumar}, \citenamefont {Jacobson}, \citenamefont {Spielman},\
  and\ \citenamefont {Campbell}}]{Eckel2018}%
  \BibitemOpen
  \bibfield  {author} {\bibinfo {author} {\bibfnamefont {S.}~\bibnamefont
  {Eckel}}, \bibinfo {author} {\bibfnamefont {A.}~\bibnamefont {Kumar}},
  \bibinfo {author} {\bibfnamefont {T.}~\bibnamefont {Jacobson}}, \bibinfo
  {author} {\bibfnamefont {I.~B.}\ \bibnamefont {Spielman}}, \ and\ \bibinfo
  {author} {\bibfnamefont {G.~K.}\ \bibnamefont {Campbell}},\ }\href {\doibase
  10.1103/PhysRevX.8.021021} {\bibfield  {journal} {\bibinfo  {journal} {Phys.
  Rev. X}\ }\textbf {\bibinfo {volume} {8}},\ \bibinfo {pages} {021021}
  (\bibinfo {year} {2018})}\BibitemShut {NoStop}%
\bibitem [{\citenamefont {Wang}\ \emph {et~al.}(2024)\citenamefont {Wang},
  \citenamefont {Fu},\ and\ \citenamefont {Levin}}]{PhysRevA.109.013316}%
  \BibitemOpen
  \bibfield  {author} {\bibinfo {author} {\bibfnamefont {K.}~\bibnamefont
  {Wang}}, \bibinfo {author} {\bibfnamefont {H.}~\bibnamefont {Fu}}, \ and\
  \bibinfo {author} {\bibfnamefont {K.}~\bibnamefont {Levin}},\ }\href
  {\doibase 10.1103/PhysRevA.109.013316} {\bibfield  {journal} {\bibinfo
  {journal} {Phys. Rev. A}\ }\textbf {\bibinfo {volume} {109}},\ \bibinfo
  {pages} {013316} (\bibinfo {year} {2024})}\BibitemShut {NoStop}%
\bibitem [{\citenamefont {Wittemer}\ \emph {et~al.}(2019)\citenamefont
  {Wittemer}, \citenamefont {Hakelberg}, \citenamefont {Kiefer}, \citenamefont
  {Schröder}, \citenamefont {Fey}, \citenamefont {Schützhold}, \citenamefont
  {Warring},\ and\ \citenamefont {Schaetz}}]{Wittemer2019}%
  \BibitemOpen
  \bibfield  {author} {\bibinfo {author} {\bibfnamefont {M.}~\bibnamefont
  {Wittemer}}, \bibinfo {author} {\bibfnamefont {F.}~\bibnamefont {Hakelberg}},
  \bibinfo {author} {\bibfnamefont {P.}~\bibnamefont {Kiefer}}, \bibinfo
  {author} {\bibfnamefont {J.-P.}\ \bibnamefont {Schröder}}, \bibinfo {author}
  {\bibfnamefont {C.}~\bibnamefont {Fey}}, \bibinfo {author} {\bibfnamefont
  {R.}~\bibnamefont {Schützhold}}, \bibinfo {author} {\bibfnamefont
  {U.}~\bibnamefont {Warring}}, \ and\ \bibinfo {author} {\bibfnamefont
  {T.}~\bibnamefont {Schaetz}},\ }\href {\doibase
  10.1103/PhysRevLett.123.180502} {\bibfield  {journal} {\bibinfo  {journal}
  {Phys. Rev. Lett.}\ }\textbf {\bibinfo {volume} {123}},\ \bibinfo {pages}
  {180502} (\bibinfo {year} {2019})}\BibitemShut {NoStop}%
\bibitem [{\citenamefont {Hu}\ \emph {et~al.}(2019)\citenamefont {Hu},
  \citenamefont {Feng}, \citenamefont {Zhang},\ and\ \citenamefont
  {Chin}}]{Hu2019}%
  \BibitemOpen
  \bibfield  {author} {\bibinfo {author} {\bibfnamefont {J.}~\bibnamefont
  {Hu}}, \bibinfo {author} {\bibfnamefont {L.}~\bibnamefont {Feng}}, \bibinfo
  {author} {\bibfnamefont {Z.}~\bibnamefont {Zhang}}, \ and\ \bibinfo {author}
  {\bibfnamefont {C.}~\bibnamefont {Chin}},\ }\href {\doibase
  10.1038/s41567-019-0537-1} {\bibfield  {journal} {\bibinfo  {journal} {Nature
  Physics}\ }\textbf {\bibinfo {volume} {15}},\ \bibinfo {pages} {785}
  (\bibinfo {year} {2019})}\BibitemShut {NoStop}%
\bibitem [{\citenamefont {Steinhauer}\ \emph {et~al.}(2022)\citenamefont
  {Steinhauer}, \citenamefont {Abuzarli}, \citenamefont {Aladjidi},
  \citenamefont {Bienaim\'e}, \citenamefont {Piekarski}, \citenamefont {Liu},
  \citenamefont {Giacobino}, \citenamefont {Bramati},\ and\ \citenamefont
  {Glorieux}}]{Steinhauer:2021fhb}%
  \BibitemOpen
  \bibfield  {author} {\bibinfo {author} {\bibfnamefont {J.}~\bibnamefont
  {Steinhauer}}, \bibinfo {author} {\bibfnamefont {M.}~\bibnamefont
  {Abuzarli}}, \bibinfo {author} {\bibfnamefont {T.}~\bibnamefont {Aladjidi}},
  \bibinfo {author} {\bibfnamefont {T.}~\bibnamefont {Bienaim\'e}}, \bibinfo
  {author} {\bibfnamefont {C.}~\bibnamefont {Piekarski}}, \bibinfo {author}
  {\bibfnamefont {W.}~\bibnamefont {Liu}}, \bibinfo {author} {\bibfnamefont
  {E.}~\bibnamefont {Giacobino}}, \bibinfo {author} {\bibfnamefont
  {A.}~\bibnamefont {Bramati}}, \ and\ \bibinfo {author} {\bibfnamefont
  {Q.}~\bibnamefont {Glorieux}},\ }\href {\doibase 10.1038/s41467-022-30603-1}
  {\bibfield  {journal} {\bibinfo  {journal} {Nature Commun.}\ }\textbf
  {\bibinfo {volume} {13}},\ \bibinfo {pages} {2890} (\bibinfo {year}
  {2022})},\ \Eprint {http://arxiv.org/abs/2102.08279} {arXiv:2102.08279
  [cond-mat.quant-gas]} \BibitemShut {NoStop}%
\bibitem [{\citenamefont {Viermann}\ \emph {et~al.}(2022)\citenamefont
  {Viermann}, \citenamefont {Sparn}, \citenamefont {Liebster}, \citenamefont
  {Hans}, \citenamefont {Kath}, \citenamefont {Parra-L{\'o}pez}, \citenamefont
  {Tolosa-Sime{\'o}n}, \citenamefont {S{\'a}nchez-Kuntz}, \citenamefont {Haas},
  \citenamefont {Strobel}, \citenamefont {Floerchinger},\ and\ \citenamefont
  {Oberthaler}}]{Viermann2022}%
  \BibitemOpen
  \bibfield  {author} {\bibinfo {author} {\bibfnamefont {C.}~\bibnamefont
  {Viermann}}, \bibinfo {author} {\bibfnamefont {M.}~\bibnamefont {Sparn}},
  \bibinfo {author} {\bibfnamefont {N.}~\bibnamefont {Liebster}}, \bibinfo
  {author} {\bibfnamefont {M.}~\bibnamefont {Hans}}, \bibinfo {author}
  {\bibfnamefont {E.}~\bibnamefont {Kath}}, \bibinfo {author} {\bibfnamefont
  {{\'A}.}~\bibnamefont {Parra-L{\'o}pez}}, \bibinfo {author} {\bibfnamefont
  {M.}~\bibnamefont {Tolosa-Sime{\'o}n}}, \bibinfo {author} {\bibfnamefont
  {N.}~\bibnamefont {S{\'a}nchez-Kuntz}}, \bibinfo {author} {\bibfnamefont
  {T.}~\bibnamefont {Haas}}, \bibinfo {author} {\bibfnamefont {H.}~\bibnamefont
  {Strobel}}, \bibinfo {author} {\bibfnamefont {S.}~\bibnamefont
  {Floerchinger}}, \ and\ \bibinfo {author} {\bibfnamefont {M.~K.}\
  \bibnamefont {Oberthaler}},\ }\href {\doibase 10.1038/s41586-022-05313-9}
  {\bibfield  {journal} {\bibinfo  {journal} {Nature}\ }\textbf {\bibinfo
  {volume} {611}},\ \bibinfo {pages} {260} (\bibinfo {year}
  {2022})}\BibitemShut {NoStop}%
\bibitem [{\citenamefont {Tolosa-Simeón}\ \emph {et~al.}(2022)\citenamefont
  {Tolosa-Simeón}, \citenamefont {Parra-López}, \citenamefont
  {Sánchez-Kuntz}, \citenamefont {Haas}, \citenamefont {Viermann},
  \citenamefont {Sparn}, \citenamefont {Liebster}, \citenamefont {Hans},
  \citenamefont {Kath}, \citenamefont {Strobel}, \citenamefont {Oberthaler},\
  and\ \citenamefont {Floerchinger}}]{Tolosa-Simeon2022}%
  \BibitemOpen
  \bibfield  {author} {\bibinfo {author} {\bibfnamefont {M.}~\bibnamefont
  {Tolosa-Simeón}}, \bibinfo {author} {\bibfnamefont {A.}~\bibnamefont
  {Parra-López}}, \bibinfo {author} {\bibfnamefont {N.}~\bibnamefont
  {Sánchez-Kuntz}}, \bibinfo {author} {\bibfnamefont {T.}~\bibnamefont
  {Haas}}, \bibinfo {author} {\bibfnamefont {C.}~\bibnamefont {Viermann}},
  \bibinfo {author} {\bibfnamefont {M.}~\bibnamefont {Sparn}}, \bibinfo
  {author} {\bibfnamefont {N.}~\bibnamefont {Liebster}}, \bibinfo {author}
  {\bibfnamefont {M.}~\bibnamefont {Hans}}, \bibinfo {author} {\bibfnamefont
  {E.}~\bibnamefont {Kath}}, \bibinfo {author} {\bibfnamefont {H.}~\bibnamefont
  {Strobel}}, \bibinfo {author} {\bibfnamefont {M.~K.}\ \bibnamefont
  {Oberthaler}}, \ and\ \bibinfo {author} {\bibfnamefont {S.}~\bibnamefont
  {Floerchinger}},\ }\href {\doibase 10.1103/PhysRevA.106.033313} {\bibfield
  {journal} {\bibinfo  {journal} {Phys. Rev. A}\ }\textbf {\bibinfo {volume}
  {106}},\ \bibinfo {pages} {033313} (\bibinfo {year} {2022})}\BibitemShut
  {NoStop}%
\bibitem [{\citenamefont {Jacquet}\ \emph {et~al.}(2023)\citenamefont
  {Jacquet}, \citenamefont {Giacomelli}, \citenamefont {Valnais}, \citenamefont
  {Joly}, \citenamefont {Claude}, \citenamefont {Giacobino}, \citenamefont
  {Glorieux}, \citenamefont {Carusotto},\ and\ \citenamefont
  {Bramati}}]{PhysRevLett.130.111501}%
  \BibitemOpen
  \bibfield  {author} {\bibinfo {author} {\bibfnamefont {M.~J.}\ \bibnamefont
  {Jacquet}}, \bibinfo {author} {\bibfnamefont {L.}~\bibnamefont {Giacomelli}},
  \bibinfo {author} {\bibfnamefont {Q.}~\bibnamefont {Valnais}}, \bibinfo
  {author} {\bibfnamefont {M.}~\bibnamefont {Joly}}, \bibinfo {author}
  {\bibfnamefont {F.}~\bibnamefont {Claude}}, \bibinfo {author} {\bibfnamefont
  {E.}~\bibnamefont {Giacobino}}, \bibinfo {author} {\bibfnamefont
  {Q.}~\bibnamefont {Glorieux}}, \bibinfo {author} {\bibfnamefont
  {I.}~\bibnamefont {Carusotto}}, \ and\ \bibinfo {author} {\bibfnamefont
  {A.}~\bibnamefont {Bramati}},\ }\href {\doibase
  10.1103/PhysRevLett.130.111501} {\bibfield  {journal} {\bibinfo  {journal}
  {Phys. Rev. Lett.}\ }\textbf {\bibinfo {volume} {130}},\ \bibinfo {pages}
  {111501} (\bibinfo {year} {2023})}\BibitemShut {NoStop}%
\bibitem [{\citenamefont {Tajik}\ \emph {et~al.}(2023)\citenamefont {Tajik},
  \citenamefont {Gluza}, \citenamefont {Sebe}, \citenamefont {Schüttelkopf},
  \citenamefont {Cataldini}, \citenamefont {Sabino}, \citenamefont {Møller},
  \citenamefont {Ji}, \citenamefont {Erne}, \citenamefont {Guarnieri},
  \citenamefont {Sotiriadis}, \citenamefont {Eisert},\ and\ \citenamefont
  {Schmiedmayer}}]{doi:10.1073/pnas.2301287120}%
  \BibitemOpen
  \bibfield  {author} {\bibinfo {author} {\bibfnamefont {M.}~\bibnamefont
  {Tajik}}, \bibinfo {author} {\bibfnamefont {M.}~\bibnamefont {Gluza}},
  \bibinfo {author} {\bibfnamefont {N.}~\bibnamefont {Sebe}}, \bibinfo {author}
  {\bibfnamefont {P.}~\bibnamefont {Schüttelkopf}}, \bibinfo {author}
  {\bibfnamefont {F.}~\bibnamefont {Cataldini}}, \bibinfo {author}
  {\bibfnamefont {J.}~\bibnamefont {Sabino}}, \bibinfo {author} {\bibfnamefont
  {F.}~\bibnamefont {Møller}}, \bibinfo {author} {\bibfnamefont {S.-C.}\
  \bibnamefont {Ji}}, \bibinfo {author} {\bibfnamefont {S.}~\bibnamefont
  {Erne}}, \bibinfo {author} {\bibfnamefont {G.}~\bibnamefont {Guarnieri}},
  \bibinfo {author} {\bibfnamefont {S.}~\bibnamefont {Sotiriadis}}, \bibinfo
  {author} {\bibfnamefont {J.}~\bibnamefont {Eisert}}, \ and\ \bibinfo {author}
  {\bibfnamefont {J.}~\bibnamefont {Schmiedmayer}},\ }\href {\doibase
  10.1073/pnas.2301287120} {\bibfield  {journal} {\bibinfo  {journal}
  {Proceedings of the National Academy of Sciences}\ }\textbf {\bibinfo
  {volume} {120}},\ \bibinfo {pages} {e2301287120} (\bibinfo {year} {2023})},\
  \Eprint
  {http://arxiv.org/abs/https://www.pnas.org/doi/pdf/10.1073/pnas.2301287120}
  {https://www.pnas.org/doi/pdf/10.1073/pnas.2301287120} \BibitemShut {NoStop}%
\bibitem [{\citenamefont {Jaksch}\ and\ \citenamefont
  {Zoller}(2005)}]{Jaksch_2005}%
  \BibitemOpen
  \bibfield  {author} {\bibinfo {author} {\bibfnamefont {D.}~\bibnamefont
  {Jaksch}}\ and\ \bibinfo {author} {\bibfnamefont {P.}~\bibnamefont
  {Zoller}},\ }\href {\doibase https://doi.org/10.1016/j.aop.2004.09.010}
  {\bibfield  {journal} {\bibinfo  {journal} {Annals of Physics}\ }\textbf
  {\bibinfo {volume} {315}},\ \bibinfo {pages} {52} (\bibinfo {year} {2005})},\
  \bibinfo {note} {special Issue}\BibitemShut {NoStop}%
\bibitem [{\citenamefont {Lewenstein}\ \emph {et~al.}(2007)\citenamefont
  {Lewenstein}, \citenamefont {Sanpera}, \citenamefont {Ahufinger},
  \citenamefont {Damski}, \citenamefont {Sen(De)},\ and\ \citenamefont
  {Sen}}]{Lewenstein_2007}%
  \BibitemOpen
  \bibfield  {author} {\bibinfo {author} {\bibfnamefont {M.}~\bibnamefont
  {Lewenstein}}, \bibinfo {author} {\bibfnamefont {A.}~\bibnamefont {Sanpera}},
  \bibinfo {author} {\bibfnamefont {V.}~\bibnamefont {Ahufinger}}, \bibinfo
  {author} {\bibfnamefont {B.}~\bibnamefont {Damski}}, \bibinfo {author}
  {\bibfnamefont {A.}~\bibnamefont {Sen(De)}}, \ and\ \bibinfo {author}
  {\bibfnamefont {U.}~\bibnamefont {Sen}},\ }\href {\doibase
  10.1080/00018730701223200} {\bibfield  {journal} {\bibinfo  {journal}
  {Advances in Physics}\ }\textbf {\bibinfo {volume} {56}},\ \bibinfo {pages}
  {243} (\bibinfo {year} {2007})}\BibitemShut {NoStop}%
\bibitem [{\citenamefont {Bloch}\ \emph {et~al.}(2008)\citenamefont {Bloch},
  \citenamefont {Dalibard},\ and\ \citenamefont {Zwerger}}]{RevModPhys.80.885}%
  \BibitemOpen
  \bibfield  {author} {\bibinfo {author} {\bibfnamefont {I.}~\bibnamefont
  {Bloch}}, \bibinfo {author} {\bibfnamefont {J.}~\bibnamefont {Dalibard}}, \
  and\ \bibinfo {author} {\bibfnamefont {W.}~\bibnamefont {Zwerger}},\ }\href
  {\doibase 10.1103/RevModPhys.80.885} {\bibfield  {journal} {\bibinfo
  {journal} {Rev. Mod. Phys.}\ }\textbf {\bibinfo {volume} {80}},\ \bibinfo
  {pages} {885} (\bibinfo {year} {2008})}\BibitemShut {NoStop}%
\bibitem [{\citenamefont {Gross}\ and\ \citenamefont
  {Bloch}(2017)}]{Gross_2017}%
  \BibitemOpen
  \bibfield  {author} {\bibinfo {author} {\bibfnamefont {C.}~\bibnamefont
  {Gross}}\ and\ \bibinfo {author} {\bibfnamefont {I.}~\bibnamefont {Bloch}},\
  }\href {\doibase 10.1126/science.aal3837} {\bibfield  {journal} {\bibinfo
  {journal} {Science}\ }\textbf {\bibinfo {volume} {357}},\ \bibinfo {pages}
  {995} (\bibinfo {year} {2017})}\BibitemShut {NoStop}%
\bibitem [{\citenamefont {Kogut}\ and\ \citenamefont
  {Susskind}(1975)}]{PhysRevD.11.395}%
  \BibitemOpen
  \bibfield  {author} {\bibinfo {author} {\bibfnamefont {J.}~\bibnamefont
  {Kogut}}\ and\ \bibinfo {author} {\bibfnamefont {L.}~\bibnamefont
  {Susskind}},\ }\href {\doibase 10.1103/PhysRevD.11.395} {\bibfield  {journal}
  {\bibinfo  {journal} {Phys. Rev. D}\ }\textbf {\bibinfo {volume} {11}},\
  \bibinfo {pages} {395} (\bibinfo {year} {1975})}\BibitemShut {NoStop}%
\bibitem [{\citenamefont {Wilson}(1974)}]{PhysRevD.10.2445}%
  \BibitemOpen
  \bibfield  {author} {\bibinfo {author} {\bibfnamefont {K.~G.}\ \bibnamefont
  {Wilson}},\ }\href {\doibase 10.1103/PhysRevD.10.2445} {\bibfield  {journal}
  {\bibinfo  {journal} {Phys. Rev. D}\ }\textbf {\bibinfo {volume} {10}},\
  \bibinfo {pages} {2445} (\bibinfo {year} {1974})}\BibitemShut {NoStop}%
\bibitem [{\citenamefont {Gattringer}\ and\ \citenamefont
  {Lang}(2010{\natexlab{a}})}]{gattringer_lang_2010}%
  \BibitemOpen
  \bibfield  {author} {\bibinfo {author} {\bibfnamefont {C.}~\bibnamefont
  {Gattringer}}\ and\ \bibinfo {author} {\bibfnamefont {C.~B.}\ \bibnamefont
  {Lang}},\ }\href@noop {} {\emph {\bibinfo {title} {Quantum chromodynamics on
  the lattice: an introductory presentation}}}\ (\bibinfo  {publisher}
  {Springer},\ \bibinfo {year} {2010})\BibitemShut {NoStop}%
\bibitem [{\citenamefont {Zhang}\ and\ \citenamefont {Liu}(2018)}]{Zhang_2018}%
  \BibitemOpen
  \bibfield  {author} {\bibinfo {author} {\bibfnamefont {L.}~\bibnamefont
  {Zhang}}\ and\ \bibinfo {author} {\bibfnamefont {X.-J.}\ \bibnamefont
  {Liu}},\ }in\ \href {\doibase 10.1142/9789813272538_0001} {\emph {\bibinfo
  {booktitle} {Synthetic Spin-Orbit Coupling in Cold Atoms}}}\ (\bibinfo
  {publisher} {{WORLD} {SCIENTIFIC}},\ \bibinfo {year} {2018})\ pp.\ \bibinfo
  {pages} {1--87}\BibitemShut {NoStop}%
\bibitem [{\citenamefont {Liu}\ \emph {et~al.}(2013)\citenamefont {Liu},
  \citenamefont {Liu},\ and\ \citenamefont {Cheng}}]{PhysRevLett.110.076401}%
  \BibitemOpen
  \bibfield  {author} {\bibinfo {author} {\bibfnamefont {X.-J.}\ \bibnamefont
  {Liu}}, \bibinfo {author} {\bibfnamefont {Z.-X.}\ \bibnamefont {Liu}}, \ and\
  \bibinfo {author} {\bibfnamefont {M.}~\bibnamefont {Cheng}},\ }\href
  {\doibase 10.1103/PhysRevLett.110.076401} {\bibfield  {journal} {\bibinfo
  {journal} {Phys. Rev. Lett.}\ }\textbf {\bibinfo {volume} {110}},\ \bibinfo
  {pages} {076401} (\bibinfo {year} {2013})}\BibitemShut {NoStop}%
\bibitem [{\citenamefont {Liu}\ \emph {et~al.}(2014{\natexlab{a}})\citenamefont
  {Liu}, \citenamefont {Law},\ and\ \citenamefont
  {Ng}}]{PhysRevLett.112.086401}%
  \BibitemOpen
  \bibfield  {author} {\bibinfo {author} {\bibfnamefont {X.-J.}\ \bibnamefont
  {Liu}}, \bibinfo {author} {\bibfnamefont {K.~T.}\ \bibnamefont {Law}}, \ and\
  \bibinfo {author} {\bibfnamefont {T.~K.}\ \bibnamefont {Ng}},\ }\href
  {\doibase 10.1103/PhysRevLett.112.086401} {\bibfield  {journal} {\bibinfo
  {journal} {Phys. Rev. Lett.}\ }\textbf {\bibinfo {volume} {112}},\ \bibinfo
  {pages} {086401} (\bibinfo {year} {2014}{\natexlab{a}})}\BibitemShut
  {NoStop}%
\bibitem [{\citenamefont {Liu}\ \emph {et~al.}(2014{\natexlab{b}})\citenamefont
  {Liu}, \citenamefont {Law},\ and\ \citenamefont
  {Ng}}]{PhysRevLett.113.059901}%
  \BibitemOpen
  \bibfield  {author} {\bibinfo {author} {\bibfnamefont {X.-J.}\ \bibnamefont
  {Liu}}, \bibinfo {author} {\bibfnamefont {K.~T.}\ \bibnamefont {Law}}, \ and\
  \bibinfo {author} {\bibfnamefont {T.~K.}\ \bibnamefont {Ng}},\ }\href
  {\doibase 10.1103/PhysRevLett.113.059901} {\bibfield  {journal} {\bibinfo
  {journal} {Phys. Rev. Lett.}\ }\textbf {\bibinfo {volume} {113}},\ \bibinfo
  {pages} {059901} (\bibinfo {year} {2014}{\natexlab{b}})}\BibitemShut
  {NoStop}%
\bibitem [{\citenamefont {Wu}\ \emph {et~al.}(2016)\citenamefont {Wu},
  \citenamefont {Zhang}, \citenamefont {Sun}, \citenamefont {Xu}, \citenamefont
  {Wang}, \citenamefont {Ji}, \citenamefont {Deng}, \citenamefont {Chen},
  \citenamefont {Liu},\ and\ \citenamefont
  {Pan}}]{doi:10.1126/science.aaf6689}%
  \BibitemOpen
  \bibfield  {author} {\bibinfo {author} {\bibfnamefont {Z.}~\bibnamefont
  {Wu}}, \bibinfo {author} {\bibfnamefont {L.}~\bibnamefont {Zhang}}, \bibinfo
  {author} {\bibfnamefont {W.}~\bibnamefont {Sun}}, \bibinfo {author}
  {\bibfnamefont {X.-T.}\ \bibnamefont {Xu}}, \bibinfo {author} {\bibfnamefont
  {B.-Z.}\ \bibnamefont {Wang}}, \bibinfo {author} {\bibfnamefont {S.-C.}\
  \bibnamefont {Ji}}, \bibinfo {author} {\bibfnamefont {Y.}~\bibnamefont
  {Deng}}, \bibinfo {author} {\bibfnamefont {S.}~\bibnamefont {Chen}}, \bibinfo
  {author} {\bibfnamefont {X.-J.}\ \bibnamefont {Liu}}, \ and\ \bibinfo
  {author} {\bibfnamefont {J.-W.}\ \bibnamefont {Pan}},\ }\href {\doibase
  10.1126/science.aaf6689} {\bibfield  {journal} {\bibinfo  {journal}
  {Science}\ }\textbf {\bibinfo {volume} {354}},\ \bibinfo {pages} {83}
  (\bibinfo {year} {2016})}\BibitemShut {NoStop}%
\bibitem [{\citenamefont {Sun}\ \emph {et~al.}(2018)\citenamefont {Sun},
  \citenamefont {Wang}, \citenamefont {Xu}, \citenamefont {Yi}, \citenamefont
  {Zhang}, \citenamefont {Wu}, \citenamefont {Deng}, \citenamefont {Liu},
  \citenamefont {Chen},\ and\ \citenamefont {Pan}}]{PhysRevLett.121.150401}%
  \BibitemOpen
  \bibfield  {author} {\bibinfo {author} {\bibfnamefont {W.}~\bibnamefont
  {Sun}}, \bibinfo {author} {\bibfnamefont {B.-Z.}\ \bibnamefont {Wang}},
  \bibinfo {author} {\bibfnamefont {X.-T.}\ \bibnamefont {Xu}}, \bibinfo
  {author} {\bibfnamefont {C.-R.}\ \bibnamefont {Yi}}, \bibinfo {author}
  {\bibfnamefont {L.}~\bibnamefont {Zhang}}, \bibinfo {author} {\bibfnamefont
  {Z.}~\bibnamefont {Wu}}, \bibinfo {author} {\bibfnamefont {Y.}~\bibnamefont
  {Deng}}, \bibinfo {author} {\bibfnamefont {X.-J.}\ \bibnamefont {Liu}},
  \bibinfo {author} {\bibfnamefont {S.}~\bibnamefont {Chen}}, \ and\ \bibinfo
  {author} {\bibfnamefont {J.-W.}\ \bibnamefont {Pan}},\ }\href {\doibase
  10.1103/PhysRevLett.121.150401} {\bibfield  {journal} {\bibinfo  {journal}
  {Phys. Rev. Lett.}\ }\textbf {\bibinfo {volume} {121}},\ \bibinfo {pages}
  {150401} (\bibinfo {year} {2018})}\BibitemShut {NoStop}%
\bibitem [{\citenamefont {Song}\ \emph {et~al.}(2018)\citenamefont {Song},
  \citenamefont {Zhang}, \citenamefont {He}, \citenamefont {Poon},
  \citenamefont {Hajiyev}, \citenamefont {Zhang}, \citenamefont {Liu},\ and\
  \citenamefont {Jo}}]{doi:10.1126/sciadv.aao4748}%
  \BibitemOpen
  \bibfield  {author} {\bibinfo {author} {\bibfnamefont {B.}~\bibnamefont
  {Song}}, \bibinfo {author} {\bibfnamefont {L.}~\bibnamefont {Zhang}},
  \bibinfo {author} {\bibfnamefont {C.}~\bibnamefont {He}}, \bibinfo {author}
  {\bibfnamefont {T.~F.~J.}\ \bibnamefont {Poon}}, \bibinfo {author}
  {\bibfnamefont {E.}~\bibnamefont {Hajiyev}}, \bibinfo {author} {\bibfnamefont
  {S.}~\bibnamefont {Zhang}}, \bibinfo {author} {\bibfnamefont {X.-J.}\
  \bibnamefont {Liu}}, \ and\ \bibinfo {author} {\bibfnamefont {G.-B.}\
  \bibnamefont {Jo}},\ }\href {\doibase 10.1126/sciadv.aao4748} {\bibfield
  {journal} {\bibinfo  {journal} {Science Advances}\ }\textbf {\bibinfo
  {volume} {4}},\ \bibinfo {pages} {eaao4748} (\bibinfo {year}
  {2018})}\BibitemShut {NoStop}%
\bibitem [{\citenamefont {Liang}\ \emph {et~al.}(2023)\citenamefont {Liang},
  \citenamefont {Wei}, \citenamefont {Zhang}, \citenamefont {Wang},
  \citenamefont {Zhang}, \citenamefont {Wang}, \citenamefont {Qi},
  \citenamefont {Liu},\ and\ \citenamefont
  {Zhang}}]{https://doi.org/10.48550/arxiv.2109.08885}%
  \BibitemOpen
  \bibfield  {author} {\bibinfo {author} {\bibfnamefont {M.-C.}\ \bibnamefont
  {Liang}}, \bibinfo {author} {\bibfnamefont {Y.-D.}\ \bibnamefont {Wei}},
  \bibinfo {author} {\bibfnamefont {L.}~\bibnamefont {Zhang}}, \bibinfo
  {author} {\bibfnamefont {X.-J.}\ \bibnamefont {Wang}}, \bibinfo {author}
  {\bibfnamefont {H.}~\bibnamefont {Zhang}}, \bibinfo {author} {\bibfnamefont
  {W.-W.}\ \bibnamefont {Wang}}, \bibinfo {author} {\bibfnamefont
  {W.}~\bibnamefont {Qi}}, \bibinfo {author} {\bibfnamefont {X.-J.}\
  \bibnamefont {Liu}}, \ and\ \bibinfo {author} {\bibfnamefont
  {X.}~\bibnamefont {Zhang}},\ }\href {\doibase
  10.1103/PhysRevResearch.5.L012006} {\bibfield  {journal} {\bibinfo  {journal}
  {Phys. Rev. Res.}\ }\textbf {\bibinfo {volume} {5}},\ \bibinfo {pages}
  {L012006} (\bibinfo {year} {2023})}\BibitemShut {NoStop}%
\bibitem [{\citenamefont {Fulgado-Claudio}\ \emph {et~al.}(2023)\citenamefont
  {Fulgado-Claudio}, \citenamefont {Vel{\'{a}}zquez},\ and\ \citenamefont
  {Bermudez}}]{FulgadoClaudio2023fermionproduction}%
  \BibitemOpen
  \bibfield  {author} {\bibinfo {author} {\bibfnamefont {C.}~\bibnamefont
  {Fulgado-Claudio}}, \bibinfo {author} {\bibfnamefont {J.~M.~S.}\ \bibnamefont
  {Vel{\'{a}}zquez}}, \ and\ \bibinfo {author} {\bibfnamefont {A.}~\bibnamefont
  {Bermudez}},\ }\href {\doibase 10.22331/q-2023-06-21-1042} {\bibfield
  {journal} {\bibinfo  {journal} {{Quantum}}\ }\textbf {\bibinfo {volume}
  {7}},\ \bibinfo {pages} {1042} (\bibinfo {year} {2023})}\BibitemShut
  {NoStop}%
\bibitem [{\citenamefont {Mukhanov}\ and\ \citenamefont
  {Winitzki}(2007)}]{Mukhanov}%
  \BibitemOpen
  \bibfield  {author} {\bibinfo {author} {\bibfnamefont {V.}~\bibnamefont
  {Mukhanov}}\ and\ \bibinfo {author} {\bibfnamefont {S.}~\bibnamefont
  {Winitzki}},\ }\href {\doibase https://doi.org/10.1017/CBO9780511809149}
  {\emph {\bibinfo {title} {{Introduction to Quantum Effects in Gravity}}}}\
  (\bibinfo  {publisher} {Cambridge University Press},\ \bibinfo {year}
  {2007})\BibitemShut {NoStop}%
\bibitem [{\citenamefont {Fermi}(1934)}]{Fermi1934}%
  \BibitemOpen
  \bibfield  {author} {\bibinfo {author} {\bibfnamefont {E.}~\bibnamefont
  {Fermi}},\ }\href {\doibase 10.1007/BF01351864} {\bibfield  {journal}
  {\bibinfo  {journal} {Zeitschrift f{\"u}r Physik}\ }\textbf {\bibinfo
  {volume} {88}},\ \bibinfo {pages} {161} (\bibinfo {year} {1934})}\BibitemShut
  {NoStop}%
\bibitem [{\citenamefont {Wilson}(1968)}]{doi:10.1119/1.1974382}%
  \BibitemOpen
  \bibfield  {author} {\bibinfo {author} {\bibfnamefont {F.~L.}\ \bibnamefont
  {Wilson}},\ }\href {\doibase 10.1119/1.1974382} {\bibfield  {journal}
  {\bibinfo  {journal} {American Journal of Physics}\ }\textbf {\bibinfo
  {volume} {36}},\ \bibinfo {pages} {1150} (\bibinfo {year}
  {1968})}\BibitemShut {NoStop}%
\bibitem [{\citenamefont {Nambu}\ and\ \citenamefont
  {Jona-Lasinio}(1961{\natexlab{a}})}]{PhysRev.122.345}%
  \BibitemOpen
  \bibfield  {author} {\bibinfo {author} {\bibfnamefont {Y.}~\bibnamefont
  {Nambu}}\ and\ \bibinfo {author} {\bibfnamefont {G.}~\bibnamefont
  {Jona-Lasinio}},\ }\href {\doibase 10.1103/PhysRev.122.345} {\bibfield
  {journal} {\bibinfo  {journal} {Phys. Rev.}\ }\textbf {\bibinfo {volume}
  {122}},\ \bibinfo {pages} {345} (\bibinfo {year}
  {1961}{\natexlab{a}})}\BibitemShut {NoStop}%
\bibitem [{\citenamefont {Nambu}\ and\ \citenamefont
  {Jona-Lasinio}(1961{\natexlab{b}})}]{PhysRev.124.246}%
  \BibitemOpen
  \bibfield  {author} {\bibinfo {author} {\bibfnamefont {Y.}~\bibnamefont
  {Nambu}}\ and\ \bibinfo {author} {\bibfnamefont {G.}~\bibnamefont
  {Jona-Lasinio}},\ }\href {\doibase 10.1103/PhysRev.124.246} {\bibfield
  {journal} {\bibinfo  {journal} {Phys. Rev.}\ }\textbf {\bibinfo {volume}
  {124}},\ \bibinfo {pages} {246} (\bibinfo {year}
  {1961}{\natexlab{b}})}\BibitemShut {NoStop}%
\bibitem [{\citenamefont {Gross}\ and\ \citenamefont
  {Neveu}(1974{\natexlab{a}})}]{PhysRevD.10.3235}%
  \BibitemOpen
  \bibfield  {author} {\bibinfo {author} {\bibfnamefont {D.~J.}\ \bibnamefont
  {Gross}}\ and\ \bibinfo {author} {\bibfnamefont {A.}~\bibnamefont {Neveu}},\
  }\href {\doibase 10.1103/PhysRevD.10.3235} {\bibfield  {journal} {\bibinfo
  {journal} {Phys. Rev. D}\ }\textbf {\bibinfo {volume} {10}},\ \bibinfo
  {pages} {3235} (\bibinfo {year} {1974}{\natexlab{a}})}\BibitemShut {NoStop}%
\bibitem [{\citenamefont {Hands}(1997)}]{hep-lat/9706018}%
  \BibitemOpen
  \bibfield  {author} {\bibinfo {author} {\bibfnamefont {S.}~\bibnamefont
  {Hands}},\ }\href@noop {} {\enquote {\bibinfo {title} {Fixed point four-fermi
  theories},}\ } (\bibinfo {year} {1997}),\ \Eprint
  {http://arxiv.org/abs/arXiv:hep-lat/9706018} {arXiv:hep-lat/9706018}
  \BibitemShut {NoStop}%
\bibitem [{\citenamefont {Tong}\ \emph {et~al.}(2024)\citenamefont {Tong},
  \citenamefont {Wang}, \citenamefont {Zhang},\ and\ \citenamefont
  {Zhu}}]{Tong_2024}%
  \BibitemOpen
  \bibfield  {author} {\bibinfo {author} {\bibfnamefont {X.}~\bibnamefont
  {Tong}}, \bibinfo {author} {\bibfnamefont {Y.}~\bibnamefont {Wang}}, \bibinfo
  {author} {\bibfnamefont {C.}~\bibnamefont {Zhang}}, \ and\ \bibinfo {author}
  {\bibfnamefont {Y.}~\bibnamefont {Zhu}},\ }\href {\doibase
  10.1088/1475-7516/2024/04/022} {\bibfield  {journal} {\bibinfo  {journal}
  {Journal of Cosmology and Astroparticle Physics}\ }\textbf {\bibinfo {volume}
  {2024}},\ \bibinfo {pages} {022} (\bibinfo {year} {2024})}\BibitemShut
  {NoStop}%
\bibitem [{\citenamefont {Cazalilla}\ and\ \citenamefont
  {Rey}(2014)}]{Cazalilla:2014wfa}%
  \BibitemOpen
  \bibfield  {author} {\bibinfo {author} {\bibfnamefont {M.}~\bibnamefont
  {Cazalilla}}\ and\ \bibinfo {author} {\bibfnamefont {A.}~\bibnamefont
  {Rey}},\ }\href {\doibase 10.1088/0034-4885/77/12/124401} {\bibfield
  {journal} {\bibinfo  {journal} {Rept. Prog. Phys.}\ }\textbf {\bibinfo
  {volume} {77}},\ \bibinfo {pages} {124401} (\bibinfo {year} {2014})},\
  \Eprint {http://arxiv.org/abs/1403.2792} {arXiv:1403.2792
  [cond-mat.quant-gas]} \BibitemShut {NoStop}%
\bibitem [{\citenamefont {Klevansky}(1992)}]{RevModPhys.64.649}%
  \BibitemOpen
  \bibfield  {author} {\bibinfo {author} {\bibfnamefont {S.~P.}\ \bibnamefont
  {Klevansky}},\ }\href {\doibase 10.1103/RevModPhys.64.649} {\bibfield
  {journal} {\bibinfo  {journal} {Rev. Mod. Phys.}\ }\textbf {\bibinfo {volume}
  {64}},\ \bibinfo {pages} {649} (\bibinfo {year} {1992})}\BibitemShut
  {NoStop}%
\bibitem [{\citenamefont {Wilson}(1977)}]{Wilson1977}%
  \BibitemOpen
  \bibfield  {author} {\bibinfo {author} {\bibfnamefont {K.~G.}\ \bibnamefont
  {Wilson}},\ }in\ \href {\doibase 10.1007/978-1-4613-4208-3_6} {\emph
  {\bibinfo {booktitle} {New Phenomena in Subnuclear Physics}}}\ (\bibinfo
  {publisher} {Springer {US}},\ \bibinfo {year} {1977})\ pp.\ \bibinfo {pages}
  {69--142}\BibitemShut {NoStop}%
\bibitem [{\citenamefont {Golterman}\ \emph {et~al.}(1993)\citenamefont
  {Golterman}, \citenamefont {Jansen},\ and\ \citenamefont
  {Kaplan}}]{GOLTERMAN1993219}%
  \BibitemOpen
  \bibfield  {author} {\bibinfo {author} {\bibfnamefont {M.~F.}\ \bibnamefont
  {Golterman}}, \bibinfo {author} {\bibfnamefont {K.}~\bibnamefont {Jansen}}, \
  and\ \bibinfo {author} {\bibfnamefont {D.~B.}\ \bibnamefont {Kaplan}},\
  }\href {\doibase https://doi.org/10.1016/0370-2693(93)90692-B} {\bibfield
  {journal} {\bibinfo  {journal} {Physics Letters B}\ }\textbf {\bibinfo
  {volume} {301}},\ \bibinfo {pages} {219 } (\bibinfo {year}
  {1993})}\BibitemShut {NoStop}%
\bibitem [{\citenamefont {Creutz}(1999)}]{PhysRevLett.83.2636}%
  \BibitemOpen
  \bibfield  {author} {\bibinfo {author} {\bibfnamefont {M.}~\bibnamefont
  {Creutz}},\ }\href {\doibase 10.1103/PhysRevLett.83.2636} {\bibfield
  {journal} {\bibinfo  {journal} {Phys. Rev. Lett.}\ }\textbf {\bibinfo
  {volume} {83}},\ \bibinfo {pages} {2636} (\bibinfo {year}
  {1999})}\BibitemShut {NoStop}%
\bibitem [{\citenamefont {Kaplan}(2009)}]{Kaplan:2009yg}%
  \BibitemOpen
  \bibfield  {author} {\bibinfo {author} {\bibfnamefont {D.~B.}\ \bibnamefont
  {Kaplan}},\ }in\ \href {\doibase 10.1093/acprof:oso/9780199691609.001.0001}
  {\emph {\bibinfo {booktitle} {{Les Houches Summer School: Session 93: Modern
  perspectives in lattice QCD: Quantum field theory and high performance
  computing}}}}\ (\bibinfo {year} {2009})\ pp.\ \bibinfo {pages} {223--272},\
  \Eprint {http://arxiv.org/abs/0912.2560} {arXiv:0912.2560 [hep-lat]}
  \BibitemShut {NoStop}%
\bibitem [{\citenamefont {Bermudez}\ \emph {et~al.}(2010)\citenamefont
  {Bermudez}, \citenamefont {Mazza}, \citenamefont {Rizzi}, \citenamefont
  {Goldman}, \citenamefont {Lewenstein},\ and\ \citenamefont
  {Martin-Delgado}}]{PhysRevLett.105.190404}%
  \BibitemOpen
  \bibfield  {author} {\bibinfo {author} {\bibfnamefont {A.}~\bibnamefont
  {Bermudez}}, \bibinfo {author} {\bibfnamefont {L.}~\bibnamefont {Mazza}},
  \bibinfo {author} {\bibfnamefont {M.}~\bibnamefont {Rizzi}}, \bibinfo
  {author} {\bibfnamefont {N.}~\bibnamefont {Goldman}}, \bibinfo {author}
  {\bibfnamefont {M.}~\bibnamefont {Lewenstein}}, \ and\ \bibinfo {author}
  {\bibfnamefont {M.~A.}\ \bibnamefont {Martin-Delgado}},\ }\href {\doibase
  10.1103/PhysRevLett.105.190404} {\bibfield  {journal} {\bibinfo  {journal}
  {Phys. Rev. Lett.}\ }\textbf {\bibinfo {volume} {105}},\ \bibinfo {pages}
  {190404} (\bibinfo {year} {2010})}\BibitemShut {NoStop}%
\bibitem [{\citenamefont {Kaplan}\ and\ \citenamefont
  {Sun}(2012)}]{PhysRevLett.108.181807}%
  \BibitemOpen
  \bibfield  {author} {\bibinfo {author} {\bibfnamefont {D.~B.}\ \bibnamefont
  {Kaplan}}\ and\ \bibinfo {author} {\bibfnamefont {S.}~\bibnamefont {Sun}},\
  }\href {\doibase 10.1103/PhysRevLett.108.181807} {\bibfield  {journal}
  {\bibinfo  {journal} {Phys. Rev. Lett.}\ }\textbf {\bibinfo {volume} {108}},\
  \bibinfo {pages} {181807} (\bibinfo {year} {2012})}\BibitemShut {NoStop}%
\bibitem [{\citenamefont {J\"unemann}\ \emph {et~al.}(2017)\citenamefont
  {J\"unemann}, \citenamefont {Piga}, \citenamefont {Ran}, \citenamefont
  {Lewenstein}, \citenamefont {Rizzi},\ and\ \citenamefont
  {Bermudez}}]{PhysRevX.7.031057}%
  \BibitemOpen
  \bibfield  {author} {\bibinfo {author} {\bibfnamefont {J.}~\bibnamefont
  {J\"unemann}}, \bibinfo {author} {\bibfnamefont {A.}~\bibnamefont {Piga}},
  \bibinfo {author} {\bibfnamefont {S.-J.}\ \bibnamefont {Ran}}, \bibinfo
  {author} {\bibfnamefont {M.}~\bibnamefont {Lewenstein}}, \bibinfo {author}
  {\bibfnamefont {M.}~\bibnamefont {Rizzi}}, \ and\ \bibinfo {author}
  {\bibfnamefont {A.}~\bibnamefont {Bermudez}},\ }\href {\doibase
  10.1103/PhysRevX.7.031057} {\bibfield  {journal} {\bibinfo  {journal} {Phys.
  Rev. X}\ }\textbf {\bibinfo {volume} {7}},\ \bibinfo {pages} {031057}
  (\bibinfo {year} {2017})}\BibitemShut {NoStop}%
\bibitem [{\citenamefont {Bermudez}\ \emph
  {et~al.}(2018{\natexlab{a}})\citenamefont {Bermudez}, \citenamefont
  {Tirrito}, \citenamefont {Rizzi}, \citenamefont {Lewenstein},\ and\
  \citenamefont {Hands}}]{alejandroGrossNeveu2018}%
  \BibitemOpen
  \bibfield  {author} {\bibinfo {author} {\bibfnamefont {A.}~\bibnamefont
  {Bermudez}}, \bibinfo {author} {\bibfnamefont {E.}~\bibnamefont {Tirrito}},
  \bibinfo {author} {\bibfnamefont {M.}~\bibnamefont {Rizzi}}, \bibinfo
  {author} {\bibfnamefont {M.}~\bibnamefont {Lewenstein}}, \ and\ \bibinfo
  {author} {\bibfnamefont {S.}~\bibnamefont {Hands}},\ }\href {\doibase
  10.1016/j.aop.2018.10.007} {\bibfield  {journal} {\bibinfo  {journal} {Annals
  of Physics}\ }\textbf {\bibinfo {volume} {399}},\ \bibinfo {pages} {149}
  (\bibinfo {year} {2018}{\natexlab{a}})}\BibitemShut {NoStop}%
\bibitem [{\citenamefont {Tirrito}\ \emph {et~al.}(2022)\citenamefont
  {Tirrito}, \citenamefont {Lewenstein},\ and\ \citenamefont
  {Bermudez}}]{PhysRevB.106.045147}%
  \BibitemOpen
  \bibfield  {author} {\bibinfo {author} {\bibfnamefont {E.}~\bibnamefont
  {Tirrito}}, \bibinfo {author} {\bibfnamefont {M.}~\bibnamefont {Lewenstein}},
  \ and\ \bibinfo {author} {\bibfnamefont {A.}~\bibnamefont {Bermudez}},\
  }\href {\doibase 10.1103/PhysRevB.106.045147} {\bibfield  {journal} {\bibinfo
   {journal} {Phys. Rev. B}\ }\textbf {\bibinfo {volume} {106}},\ \bibinfo
  {pages} {045147} (\bibinfo {year} {2022})}\BibitemShut {NoStop}%
\bibitem [{\citenamefont {Braunstein}\ and\ \citenamefont {van
  Loock}(2005)}]{RevModPhys.77.513}%
  \BibitemOpen
  \bibfield  {author} {\bibinfo {author} {\bibfnamefont {S.~L.}\ \bibnamefont
  {Braunstein}}\ and\ \bibinfo {author} {\bibfnamefont {P.}~\bibnamefont {van
  Loock}},\ }\href {\doibase 10.1103/RevModPhys.77.513} {\bibfield  {journal}
  {\bibinfo  {journal} {Rev. Mod. Phys.}\ }\textbf {\bibinfo {volume} {77}},\
  \bibinfo {pages} {513} (\bibinfo {year} {2005})}\BibitemShut {NoStop}%
\bibitem [{\citenamefont {Wang}\ \emph {et~al.}(2007)\citenamefont {Wang},
  \citenamefont {Hiroshima}, \citenamefont {Tomita},\ and\ \citenamefont
  {Hayashi}}]{WANG20071}%
  \BibitemOpen
  \bibfield  {author} {\bibinfo {author} {\bibfnamefont {X.-B.}\ \bibnamefont
  {Wang}}, \bibinfo {author} {\bibfnamefont {T.}~\bibnamefont {Hiroshima}},
  \bibinfo {author} {\bibfnamefont {A.}~\bibnamefont {Tomita}}, \ and\ \bibinfo
  {author} {\bibfnamefont {M.}~\bibnamefont {Hayashi}},\ }\href {\doibase
  https://doi.org/10.1016/j.physrep.2007.04.005} {\bibfield  {journal}
  {\bibinfo  {journal} {Physics Reports}\ }\textbf {\bibinfo {volume} {448}},\
  \bibinfo {pages} {1} (\bibinfo {year} {2007})}\BibitemShut {NoStop}%
\bibitem [{\citenamefont {Kraus}\ \emph {et~al.}(2009)\citenamefont {Kraus},
  \citenamefont {Wolf}, \citenamefont {Cirac},\ and\ \citenamefont
  {Giedke}}]{Kraus_2009}%
  \BibitemOpen
  \bibfield  {author} {\bibinfo {author} {\bibfnamefont {C.~V.}\ \bibnamefont
  {Kraus}}, \bibinfo {author} {\bibfnamefont {M.~M.}\ \bibnamefont {Wolf}},
  \bibinfo {author} {\bibfnamefont {J.~I.}\ \bibnamefont {Cirac}}, \ and\
  \bibinfo {author} {\bibfnamefont {G.}~\bibnamefont {Giedke}},\ }\href
  {\doibase 10.1103/physreva.79.012306} {\bibfield  {journal} {\bibinfo
  {journal} {Physical Review A}\ }\textbf {\bibinfo {volume} {79}} (\bibinfo
  {year} {2009}),\ 10.1103/physreva.79.012306}\BibitemShut {NoStop}%
\bibitem [{\citenamefont {Kraus}\ and\ \citenamefont
  {Cirac}(2010)}]{Kraus_2010}%
  \BibitemOpen
  \bibfield  {author} {\bibinfo {author} {\bibfnamefont {C.~V.}\ \bibnamefont
  {Kraus}}\ and\ \bibinfo {author} {\bibfnamefont {J.~I.}\ \bibnamefont
  {Cirac}},\ }\href {\doibase 10.1088/1367-2630/12/11/113004} {\bibfield
  {journal} {\bibinfo  {journal} {New Journal of Physics}\ }\textbf {\bibinfo
  {volume} {12}},\ \bibinfo {pages} {113004} (\bibinfo {year}
  {2010})}\BibitemShut {NoStop}%
\bibitem [{\citenamefont {Adesso}\ \emph {et~al.}(2014)\citenamefont {Adesso},
  \citenamefont {Ragy},\ and\ \citenamefont
  {Lee}}]{doi:10.1142/S1230161214400010}%
  \BibitemOpen
  \bibfield  {author} {\bibinfo {author} {\bibfnamefont {G.}~\bibnamefont
  {Adesso}}, \bibinfo {author} {\bibfnamefont {S.}~\bibnamefont {Ragy}}, \ and\
  \bibinfo {author} {\bibfnamefont {A.~R.}\ \bibnamefont {Lee}},\ }\href
  {\doibase 10.1142/S1230161214400010} {\bibfield  {journal} {\bibinfo
  {journal} {Open Systems \& Information Dynamics}\ }\textbf {\bibinfo {volume}
  {21}},\ \bibinfo {pages} {1440001} (\bibinfo {year} {2014})}\BibitemShut
  {NoStop}%
\bibitem [{\citenamefont {Shi}\ \emph {et~al.}(2018)\citenamefont {Shi},
  \citenamefont {Demler},\ and\ \citenamefont {{Ignacio Cirac}}}]{SHI2018245}%
  \BibitemOpen
  \bibfield  {author} {\bibinfo {author} {\bibfnamefont {T.}~\bibnamefont
  {Shi}}, \bibinfo {author} {\bibfnamefont {E.}~\bibnamefont {Demler}}, \ and\
  \bibinfo {author} {\bibfnamefont {J.}~\bibnamefont {{Ignacio Cirac}}},\
  }\href {\doibase https://doi.org/10.1016/j.aop.2017.11.014} {\bibfield
  {journal} {\bibinfo  {journal} {Annals of Physics}\ }\textbf {\bibinfo
  {volume} {390}},\ \bibinfo {pages} {245} (\bibinfo {year}
  {2018})}\BibitemShut {NoStop}%
\bibitem [{\citenamefont {Surace}\ and\ \citenamefont
  {Tagliacozzo}(2022)}]{Surace22}%
  \BibitemOpen
  \bibfield  {author} {\bibinfo {author} {\bibfnamefont {J.}~\bibnamefont
  {Surace}}\ and\ \bibinfo {author} {\bibfnamefont {L.}~\bibnamefont
  {Tagliacozzo}},\ }\href {\doibase 10.21468/SciPostPhysLectNotes.54}
  {\bibfield  {journal} {\bibinfo  {journal} {SciPost Phys. Lect. Notes}\ ,\
  \bibinfo {pages} {54}} (\bibinfo {year} {2022})}\BibitemShut {NoStop}%
\bibitem [{\citenamefont {Senthil}(2015)}]{originaltopophases}%
  \BibitemOpen
  \bibfield  {author} {\bibinfo {author} {\bibfnamefont {T.}~\bibnamefont
  {Senthil}},\ }\href {\doibase 10.1146/annurev-conmatphys-031214-014740}
  {\bibfield  {journal} {\bibinfo  {journal} {Annual Review of Condensed Matter
  Physics}\ }\textbf {\bibinfo {volume} {6}},\ \bibinfo {pages} {299} (\bibinfo
  {year} {2015})}\BibitemShut {NoStop}%
\bibitem [{\citenamefont {Kitaev}(2009)}]{doi:10.1063/1.3149495}%
  \BibitemOpen
  \bibfield  {author} {\bibinfo {author} {\bibfnamefont {A.}~\bibnamefont
  {Kitaev}},\ }\href {\doibase 10.1063/1.3149495} {\bibfield  {journal}
  {\bibinfo  {journal} {AIP Conference Proceedings}\ }\textbf {\bibinfo
  {volume} {1134}},\ \bibinfo {pages} {22} (\bibinfo {year}
  {2009})}\BibitemShut {NoStop}%
\bibitem [{\citenamefont {Chiu}\ \emph
  {et~al.}(2016{\natexlab{a}})\citenamefont {Chiu}, \citenamefont {Teo},
  \citenamefont {Schnyder},\ and\ \citenamefont {Ryu}}]{classification_spt}%
  \BibitemOpen
  \bibfield  {author} {\bibinfo {author} {\bibfnamefont {C.-K.}\ \bibnamefont
  {Chiu}}, \bibinfo {author} {\bibfnamefont {J.~C.~Y.}\ \bibnamefont {Teo}},
  \bibinfo {author} {\bibfnamefont {A.~P.}\ \bibnamefont {Schnyder}}, \ and\
  \bibinfo {author} {\bibfnamefont {S.}~\bibnamefont {Ryu}},\ }\href {\doibase
  10.1103/RevModPhys.88.035005} {\bibfield  {journal} {\bibinfo  {journal}
  {Rev. Mod. Phys.}\ }\textbf {\bibinfo {volume} {88}},\ \bibinfo {pages}
  {035005} (\bibinfo {year} {2016}{\natexlab{a}})}\BibitemShut {NoStop}%
\bibitem [{\citenamefont {Aoki}(1984)}]{PhysRevD.30.2653}%
  \BibitemOpen
  \bibfield  {author} {\bibinfo {author} {\bibfnamefont {S.}~\bibnamefont
  {Aoki}},\ }\href {\doibase 10.1103/PhysRevD.30.2653} {\bibfield  {journal}
  {\bibinfo  {journal} {Phys. Rev. D}\ }\textbf {\bibinfo {volume} {30}},\
  \bibinfo {pages} {2653} (\bibinfo {year} {1984})}\BibitemShut {NoStop}%
\bibitem [{\citenamefont {Schollw\"ock}(2005)}]{RevModPhys.77.259}%
  \BibitemOpen
  \bibfield  {author} {\bibinfo {author} {\bibfnamefont {U.}~\bibnamefont
  {Schollw\"ock}},\ }\href {\doibase 10.1103/RevModPhys.77.259} {\bibfield
  {journal} {\bibinfo  {journal} {Rev. Mod. Phys.}\ }\textbf {\bibinfo {volume}
  {77}},\ \bibinfo {pages} {259} (\bibinfo {year} {2005})}\BibitemShut
  {NoStop}%
\bibitem [{\citenamefont {Schollw{\"o}ck}(2011)}]{SCHOLLWOCK201196}%
  \BibitemOpen
  \bibfield  {author} {\bibinfo {author} {\bibfnamefont {U.}~\bibnamefont
  {Schollw{\"o}ck}},\ }\href {\doibase
  https://doi.org/10.1016/j.aop.2010.09.012} {\bibfield  {journal} {\bibinfo
  {journal} {Annals of Physics}\ }\textbf {\bibinfo {volume} {326}},\ \bibinfo
  {pages} {96 } (\bibinfo {year} {2011})},\ \bibinfo {note} {january 2011
  Special Issue}\BibitemShut {NoStop}%
\bibitem [{\citenamefont {Cirac}\ \emph {et~al.}(2021)\citenamefont {Cirac},
  \citenamefont {P\'erez-Garc\'{\i}a}, \citenamefont {Schuch},\ and\
  \citenamefont {Verstraete}}]{RevModPhys.93.045003}%
  \BibitemOpen
  \bibfield  {author} {\bibinfo {author} {\bibfnamefont {J.~I.}\ \bibnamefont
  {Cirac}}, \bibinfo {author} {\bibfnamefont {D.}~\bibnamefont
  {P\'erez-Garc\'{\i}a}}, \bibinfo {author} {\bibfnamefont {N.}~\bibnamefont
  {Schuch}}, \ and\ \bibinfo {author} {\bibfnamefont {F.}~\bibnamefont
  {Verstraete}},\ }\href {\doibase 10.1103/RevModPhys.93.045003} {\bibfield
  {journal} {\bibinfo  {journal} {Rev. Mod. Phys.}\ }\textbf {\bibinfo {volume}
  {93}},\ \bibinfo {pages} {045003} (\bibinfo {year} {2021})}\BibitemShut
  {NoStop}%
\bibitem [{\citenamefont {Shapiro}(2022)}]{Shapiro:2016pfm}%
  \BibitemOpen
  \bibfield  {author} {\bibinfo {author} {\bibfnamefont {I.~L.}\ \bibnamefont
  {Shapiro}},\ }\href {\doibase 10.3390/universe8110586} {\bibfield  {journal}
  {\bibinfo  {journal} {Universe}\ }\textbf {\bibinfo {volume} {8}},\ \bibinfo
  {pages} {586} (\bibinfo {year} {2022})},\ \Eprint
  {http://arxiv.org/abs/1611.02263} {arXiv:1611.02263 [gr-qc]} \BibitemShut
  {NoStop}%
\bibitem [{\citenamefont {Baumann}(2022)}]{Baumann:2022mni}%
  \BibitemOpen
  \bibfield  {author} {\bibinfo {author} {\bibfnamefont {D.}~\bibnamefont
  {Baumann}},\ }\href {\doibase 10.1017/9781108937092} {\emph {\bibinfo {title}
  {{Cosmology}}}}\ (\bibinfo  {publisher} {Cambridge University Press},\
  \bibinfo {year} {2022})\BibitemShut {NoStop}%
\bibitem [{\citenamefont {Peskin}\ and\ \citenamefont
  {Schroeder}(1995{\natexlab{a}})}]{Peskin}%
  \BibitemOpen
  \bibfield  {author} {\bibinfo {author} {\bibfnamefont {M.~E.}\ \bibnamefont
  {Peskin}}\ and\ \bibinfo {author} {\bibfnamefont {D.~V.}\ \bibnamefont
  {Schroeder}},\ }\href {\doibase 10.1201/9780429503559} {\emph {\bibinfo
  {title} {{An Introduction to quantum field theory}}}}\ (\bibinfo  {publisher}
  {Addison-Wesley},\ \bibinfo {address} {Reading, USA},\ \bibinfo {year}
  {1995})\BibitemShut {NoStop}%
\bibitem [{\citenamefont {Preskill}(1990)}]{preskill_notes_QFT_curved}%
  \BibitemOpen
  \bibfield  {author} {\bibinfo {author} {\bibfnamefont {J.}~\bibnamefont
  {Preskill}},\ }\href {http://theory.caltech.edu/~preskill/notes.html}
  {\enquote {\bibinfo {title} {Quantum field theory in curved spacetime},}\ }
  (\bibinfo {year} {1990})\BibitemShut {NoStop}%
\bibitem [{\citenamefont {Svozil}(1990)}]{PhysRevLett.65.3341}%
  \BibitemOpen
  \bibfield  {author} {\bibinfo {author} {\bibfnamefont {K.}~\bibnamefont
  {Svozil}},\ }\href {\doibase 10.1103/PhysRevLett.65.3341} {\bibfield
  {journal} {\bibinfo  {journal} {Phys. Rev. Lett.}\ }\textbf {\bibinfo
  {volume} {65}},\ \bibinfo {pages} {3341} (\bibinfo {year}
  {1990})}\BibitemShut {NoStop}%
\bibitem [{\citenamefont {Valatin}(1958)}]{valatin1958}%
  \BibitemOpen
  \bibfield  {author} {\bibinfo {author} {\bibfnamefont {J.~G.}\ \bibnamefont
  {Valatin}},\ }\href {\doibase 10.1007/BF02745589} {\bibfield  {journal}
  {\bibinfo  {journal} {Nuovo Cim.}\ }\textbf {\bibinfo {volume} {7}},\
  \bibinfo {pages} {843} (\bibinfo {year} {1958})}\BibitemShut {NoStop}%
\bibitem [{\citenamefont {Bogolyubov}(1958)}]{bogolyubov1958}%
  \BibitemOpen
  \bibfield  {author} {\bibinfo {author} {\bibfnamefont {N.~N.}\ \bibnamefont
  {Bogolyubov}},\ }\href {\doibase 10.1007/BF02745585} {\bibfield  {journal}
  {\bibinfo  {journal} {Nuovo Cim.}\ }\textbf {\bibinfo {volume} {7}},\
  \bibinfo {pages} {794} (\bibinfo {year} {1958})}\BibitemShut {NoStop}%
\bibitem [{Note1()}]{Note1}%
  \BibitemOpen
  \bibinfo {note} {As discussed in~\cite {FulgadoClaudio2023fermionproduction},
  the time-dependence of the mode functions is completely determined by a pair
  of exactly-solvable Bessel differential equations, which admit a solution in
  closed form.}\BibitemShut {Stop}%
\bibitem [{\citenamefont {Schnetz}\ \emph {et~al.}(2006)\citenamefont
  {Schnetz}, \citenamefont {Thies},\ and\ \citenamefont
  {Urlichs}}]{Schnetz:2005ih}%
  \BibitemOpen
  \bibfield  {author} {\bibinfo {author} {\bibfnamefont {O.}~\bibnamefont
  {Schnetz}}, \bibinfo {author} {\bibfnamefont {M.}~\bibnamefont {Thies}}, \
  and\ \bibinfo {author} {\bibfnamefont {K.}~\bibnamefont {Urlichs}},\ }\href
  {\doibase 10.1016/j.aop.2005.12.007} {\bibfield  {journal} {\bibinfo
  {journal} {Annals Phys.}\ }\textbf {\bibinfo {volume} {321}},\ \bibinfo
  {pages} {2604} (\bibinfo {year} {2006})},\ \Eprint
  {http://arxiv.org/abs/hep-th/0511206} {arXiv:hep-th/0511206} \BibitemShut
  {NoStop}%
\bibitem [{\citenamefont {Gross}\ and\ \citenamefont
  {Neveu}(1974{\natexlab{b}})}]{Gross-Neveu}%
  \BibitemOpen
  \bibfield  {author} {\bibinfo {author} {\bibfnamefont {D.~J.}\ \bibnamefont
  {Gross}}\ and\ \bibinfo {author} {\bibfnamefont {A.}~\bibnamefont {Neveu}},\
  }\href {\doibase 10.1103/PhysRevD.10.3235} {\bibfield  {journal} {\bibinfo
  {journal} {Phys. Rev. D}\ }\textbf {\bibinfo {volume} {10}},\ \bibinfo
  {pages} {3235} (\bibinfo {year} {1974}{\natexlab{b}})}\BibitemShut {NoStop}%
\bibitem [{\citenamefont {Peskin}\ and\ \citenamefont
  {Schroeder}(1995{\natexlab{b}})}]{Peskin:1995ev}%
  \BibitemOpen
  \bibfield  {author} {\bibinfo {author} {\bibfnamefont {M.~E.}\ \bibnamefont
  {Peskin}}\ and\ \bibinfo {author} {\bibfnamefont {D.~V.}\ \bibnamefont
  {Schroeder}},\ }\href {\doibase https://doi.org/10.1201/9780429503559} {\emph
  {\bibinfo {title} {{An Introduction to Quantum Field Theory}}}}\ (\bibinfo
  {publisher} {Addison-Wesley},\ \bibinfo {address} {Reading, USA},\ \bibinfo
  {year} {1995})\BibitemShut {NoStop}%
\bibitem [{\citenamefont {Coleman}(1982)}]{Coleman1982}%
  \BibitemOpen
  \bibfield  {author} {\bibinfo {author} {\bibfnamefont {S.}~\bibnamefont
  {Coleman}},\ }\enquote {\bibinfo {title} {1/n},}\ in\ \href {\doibase
  10.1007/978-1-4684-1065-5_2} {\emph {\bibinfo {booktitle} {Pointlike
  Structures Inside and Outside Hadrons}}},\ \bibinfo {editor} {edited by\
  \bibinfo {editor} {\bibfnamefont {A.}~\bibnamefont {Zichichi}}}\ (\bibinfo
  {publisher} {Springer US},\ \bibinfo {address} {Boston, MA},\ \bibinfo {year}
  {1982})\ pp.\ \bibinfo {pages} {11--97}\BibitemShut {NoStop}%
\bibitem [{\citenamefont {Gaudin}(1960)}]{GAUDIN196089}%
  \BibitemOpen
  \bibfield  {author} {\bibinfo {author} {\bibfnamefont {M.}~\bibnamefont
  {Gaudin}},\ }\href {\doibase https://doi.org/10.1016/0029-5582(60)90285-6}
  {\bibfield  {journal} {\bibinfo  {journal} {Nuclear Physics}\ }\textbf
  {\bibinfo {volume} {15}},\ \bibinfo {pages} {89} (\bibinfo {year}
  {1960})}\BibitemShut {NoStop}%
\bibitem [{\citenamefont {Peschel}(2003)}]{Peschel_2003}%
  \BibitemOpen
  \bibfield  {author} {\bibinfo {author} {\bibfnamefont {I.}~\bibnamefont
  {Peschel}},\ }\href {\doibase 10.1088/0305-4470/36/14/101} {\bibfield
  {journal} {\bibinfo  {journal} {Journal of Physics A: Mathematical and
  General}\ }\textbf {\bibinfo {volume} {36}},\ \bibinfo {pages} {L205–L208}
  (\bibinfo {year} {2003})}\BibitemShut {NoStop}%
\bibitem [{\citenamefont {Bermudez}\ \emph
  {et~al.}(2018{\natexlab{b}})\citenamefont {Bermudez}, \citenamefont
  {Tirrito}, \citenamefont {Rizzi}, \citenamefont {Lewenstein},\ and\
  \citenamefont {Hands}}]{gross_neveu_wilson}%
  \BibitemOpen
  \bibfield  {author} {\bibinfo {author} {\bibfnamefont {A.}~\bibnamefont
  {Bermudez}}, \bibinfo {author} {\bibfnamefont {E.}~\bibnamefont {Tirrito}},
  \bibinfo {author} {\bibfnamefont {M.}~\bibnamefont {Rizzi}}, \bibinfo
  {author} {\bibfnamefont {M.}~\bibnamefont {Lewenstein}}, \ and\ \bibinfo
  {author} {\bibfnamefont {S.}~\bibnamefont {Hands}},\ }\href {\doibase
  10.1016/j.aop.2018.10.007} {\bibfield  {journal} {\bibinfo  {journal} {Annals
  Phys.}\ }\textbf {\bibinfo {volume} {399}},\ \bibinfo {pages} {149} (\bibinfo
  {year} {2018}{\natexlab{b}})},\ \Eprint {http://arxiv.org/abs/1807.03202}
  {arXiv:1807.03202 [cond-mat.quant-gas]} \BibitemShut {NoStop}%
\bibitem [{\citenamefont {Calabrese}\ and\ \citenamefont
  {Cardy}(2005)}]{Calabrese_2005}%
  \BibitemOpen
  \bibfield  {author} {\bibinfo {author} {\bibfnamefont {P.}~\bibnamefont
  {Calabrese}}\ and\ \bibinfo {author} {\bibfnamefont {J.}~\bibnamefont
  {Cardy}},\ }\href {\doibase 10.1088/1742-5468/2005/04/P04010} {\bibfield
  {journal} {\bibinfo  {journal} {Journal of Statistical Mechanics: Theory and
  Experiment}\ }\textbf {\bibinfo {volume} {2005}},\ \bibinfo {pages} {P04010}
  (\bibinfo {year} {2005})}\BibitemShut {NoStop}%
\bibitem [{\citenamefont {Susskind}(1977)}]{PhysRevD.16.3031}%
  \BibitemOpen
  \bibfield  {author} {\bibinfo {author} {\bibfnamefont {L.}~\bibnamefont
  {Susskind}},\ }\href {\doibase 10.1103/PhysRevD.16.3031} {\bibfield
  {journal} {\bibinfo  {journal} {Phys. Rev. D}\ }\textbf {\bibinfo {volume}
  {16}},\ \bibinfo {pages} {3031} (\bibinfo {year} {1977})}\BibitemShut
  {NoStop}%
\bibitem [{\citenamefont {Ba\ifmmode~\mbox{\c{s}}\else \c{s}\fi{}ar}\ \emph
  {et~al.}(2009)\citenamefont {Ba\ifmmode~\mbox{\c{s}}\else \c{s}\fi{}ar},
  \citenamefont {Dunne},\ and\ \citenamefont {Thies}}]{PhysRevD.79.105012}%
  \BibitemOpen
  \bibfield  {author} {\bibinfo {author} {\bibfnamefont {G.~m.~c.}\
  \bibnamefont {Ba\ifmmode~\mbox{\c{s}}\else \c{s}\fi{}ar}}, \bibinfo {author}
  {\bibfnamefont {G.~V.}\ \bibnamefont {Dunne}}, \ and\ \bibinfo {author}
  {\bibfnamefont {M.}~\bibnamefont {Thies}},\ }\href {\doibase
  10.1103/PhysRevD.79.105012} {\bibfield  {journal} {\bibinfo  {journal} {Phys.
  Rev. D}\ }\textbf {\bibinfo {volume} {79}},\ \bibinfo {pages} {105012}
  (\bibinfo {year} {2009})}\BibitemShut {NoStop}%
\bibitem [{\citenamefont {Buballa}\ and\ \citenamefont
  {Carignano}(2015)}]{BUBALLA201539}%
  \BibitemOpen
  \bibfield  {author} {\bibinfo {author} {\bibfnamefont {M.}~\bibnamefont
  {Buballa}}\ and\ \bibinfo {author} {\bibfnamefont {S.}~\bibnamefont
  {Carignano}},\ }\href {\doibase https://doi.org/10.1016/j.ppnp.2014.11.001}
  {\bibfield  {journal} {\bibinfo  {journal} {Progress in Particle and Nuclear
  Physics}\ }\textbf {\bibinfo {volume} {81}},\ \bibinfo {pages} {39} (\bibinfo
  {year} {2015})}\BibitemShut {NoStop}%
\bibitem [{\citenamefont {Koenigstein}\ \emph {et~al.}(2022)\citenamefont
  {Koenigstein}, \citenamefont {Pannullo}, \citenamefont {Rechenberger},
  \citenamefont {Steil},\ and\ \citenamefont {Winstel}}]{Koenigstein_2022}%
  \BibitemOpen
  \bibfield  {author} {\bibinfo {author} {\bibfnamefont {A.}~\bibnamefont
  {Koenigstein}}, \bibinfo {author} {\bibfnamefont {L.}~\bibnamefont
  {Pannullo}}, \bibinfo {author} {\bibfnamefont {S.}~\bibnamefont
  {Rechenberger}}, \bibinfo {author} {\bibfnamefont {M.~J.}\ \bibnamefont
  {Steil}}, \ and\ \bibinfo {author} {\bibfnamefont {M.}~\bibnamefont
  {Winstel}},\ }\href {\doibase 10.1088/1751-8121/ac820a} {\bibfield  {journal}
  {\bibinfo  {journal} {Journal of Physics A: Mathematical and Theoretical}\
  }\textbf {\bibinfo {volume} {55}},\ \bibinfo {pages} {375402} (\bibinfo
  {year} {2022})}\BibitemShut {NoStop}%
\bibitem [{\citenamefont {Zak}(1989)}]{PhysRevLett.62.2747}%
  \BibitemOpen
  \bibfield  {author} {\bibinfo {author} {\bibfnamefont {J.}~\bibnamefont
  {Zak}},\ }\href {\doibase 10.1103/PhysRevLett.62.2747} {\bibfield  {journal}
  {\bibinfo  {journal} {Phys. Rev. Lett.}\ }\textbf {\bibinfo {volume} {62}},\
  \bibinfo {pages} {2747} (\bibinfo {year} {1989})}\BibitemShut {NoStop}%
\bibitem [{\citenamefont {Chiu}\ \emph
  {et~al.}(2016{\natexlab{b}})\citenamefont {Chiu}, \citenamefont {Teo},
  \citenamefont {Schnyder},\ and\ \citenamefont {Ryu}}]{RevModPhys.88.035005}%
  \BibitemOpen
  \bibfield  {author} {\bibinfo {author} {\bibfnamefont {C.-K.}\ \bibnamefont
  {Chiu}}, \bibinfo {author} {\bibfnamefont {J.~C.~Y.}\ \bibnamefont {Teo}},
  \bibinfo {author} {\bibfnamefont {A.~P.}\ \bibnamefont {Schnyder}}, \ and\
  \bibinfo {author} {\bibfnamefont {S.}~\bibnamefont {Ryu}},\ }\href {\doibase
  10.1103/RevModPhys.88.035005} {\bibfield  {journal} {\bibinfo  {journal}
  {Rev. Mod. Phys.}\ }\textbf {\bibinfo {volume} {88}},\ \bibinfo {pages}
  {035005} (\bibinfo {year} {2016}{\natexlab{b}})}\BibitemShut {NoStop}%
\bibitem [{\citenamefont {Schnyder}\ \emph {et~al.}(2009)\citenamefont
  {Schnyder}, \citenamefont {Ryu}, \citenamefont {Furusaki}, \citenamefont
  {Ludwig}, \citenamefont {Lebedev},\ and\ \citenamefont
  {Feigel’man}}]{Schnyder_2009}%
  \BibitemOpen
  \bibfield  {author} {\bibinfo {author} {\bibfnamefont {A.~P.}\ \bibnamefont
  {Schnyder}}, \bibinfo {author} {\bibfnamefont {S.}~\bibnamefont {Ryu}},
  \bibinfo {author} {\bibfnamefont {A.}~\bibnamefont {Furusaki}}, \bibinfo
  {author} {\bibfnamefont {A.~W.~W.}\ \bibnamefont {Ludwig}}, \bibinfo {author}
  {\bibfnamefont {V.}~\bibnamefont {Lebedev}}, \ and\ \bibinfo {author}
  {\bibfnamefont {M.}~\bibnamefont {Feigel’man}},\ }in\ \href {\doibase
  10.1063/1.3149481} {\emph {\bibinfo {booktitle} {AIP Conference
  Proceedings}}}\ (\bibinfo  {publisher} {AIP},\ \bibinfo {year}
  {2009})\BibitemShut {NoStop}%
\bibitem [{\citenamefont {Izubuchi}\ \emph {et~al.}(1998)\citenamefont
  {Izubuchi}, \citenamefont {Noaki},\ and\ \citenamefont
  {Ukawa}}]{PhysRevD.58.114507}%
  \BibitemOpen
  \bibfield  {author} {\bibinfo {author} {\bibfnamefont {T.}~\bibnamefont
  {Izubuchi}}, \bibinfo {author} {\bibfnamefont {J.}~\bibnamefont {Noaki}}, \
  and\ \bibinfo {author} {\bibfnamefont {A.}~\bibnamefont {Ukawa}},\ }\href
  {\doibase 10.1103/PhysRevD.58.114507} {\bibfield  {journal} {\bibinfo
  {journal} {Phys. Rev. D}\ }\textbf {\bibinfo {volume} {58}},\ \bibinfo
  {pages} {114507} (\bibinfo {year} {1998})}\BibitemShut {NoStop}%
\bibitem [{\citenamefont {Albash}\ and\ \citenamefont
  {Lidar}(2018)}]{RevModPhys.90.015002}%
  \BibitemOpen
  \bibfield  {author} {\bibinfo {author} {\bibfnamefont {T.}~\bibnamefont
  {Albash}}\ and\ \bibinfo {author} {\bibfnamefont {D.~A.}\ \bibnamefont
  {Lidar}},\ }\href {\doibase 10.1103/RevModPhys.90.015002} {\bibfield
  {journal} {\bibinfo  {journal} {Rev. Mod. Phys.}\ }\textbf {\bibinfo {volume}
  {90}},\ \bibinfo {pages} {015002} (\bibinfo {year} {2018})}\BibitemShut
  {NoStop}%
\bibitem [{\citenamefont {Calabrese}\ and\ \citenamefont
  {Cardy}(2006)}]{PhysRevLett.96.136801}%
  \BibitemOpen
  \bibfield  {author} {\bibinfo {author} {\bibfnamefont {P.}~\bibnamefont
  {Calabrese}}\ and\ \bibinfo {author} {\bibfnamefont {J.}~\bibnamefont
  {Cardy}},\ }\href {\doibase 10.1103/PhysRevLett.96.136801} {\bibfield
  {journal} {\bibinfo  {journal} {Phys. Rev. Lett.}\ }\textbf {\bibinfo
  {volume} {96}},\ \bibinfo {pages} {136801} (\bibinfo {year}
  {2006})}\BibitemShut {NoStop}%
\bibitem [{\citenamefont {Kibble}(1976)}]{Kibble:1976sj}%
  \BibitemOpen
  \bibfield  {author} {\bibinfo {author} {\bibfnamefont {T.~W.~B.}\
  \bibnamefont {Kibble}},\ }\href {\doibase 10.1088/0305-4470/9/8/029}
  {\bibfield  {journal} {\bibinfo  {journal} {J. Phys. A}\ }\textbf {\bibinfo
  {volume} {9}},\ \bibinfo {pages} {1387} (\bibinfo {year} {1976})}\BibitemShut
  {NoStop}%
\bibitem [{\citenamefont {Zurek}(1985)}]{Zurek:1985qw}%
  \BibitemOpen
  \bibfield  {author} {\bibinfo {author} {\bibfnamefont {W.~H.}\ \bibnamefont
  {Zurek}},\ }\href {\doibase 10.1038/317505a0} {\bibfield  {journal} {\bibinfo
   {journal} {Nature}\ }\textbf {\bibinfo {volume} {317}},\ \bibinfo {pages}
  {505} (\bibinfo {year} {1985})}\BibitemShut {NoStop}%
\bibitem [{\citenamefont {Vitanov}\ and\ \citenamefont
  {Garraway}(1996)}]{vitanov_landau-zener_1996}%
  \BibitemOpen
  \bibfield  {author} {\bibinfo {author} {\bibfnamefont {N.~V.}\ \bibnamefont
  {Vitanov}}\ and\ \bibinfo {author} {\bibfnamefont {B.~M.}\ \bibnamefont
  {Garraway}},\ }\href {\doibase 10.1103/PhysRevA.53.4288} {\bibfield
  {journal} {\bibinfo  {journal} {Physical Review A}\ }\textbf {\bibinfo
  {volume} {53}},\ \bibinfo {pages} {4288} (\bibinfo {year}
  {1996})}\BibitemShut {NoStop}%
\bibitem [{Note2()}]{Note2}%
  \BibitemOpen
  \bibinfo {note} {In the case of the Landau-Zener model, the Schrödinger
  equation can be analytically solved by means of parabolic cylinder functions
  \cite {abramowitz+stegun}. If the duration of the coupling lasts from
  $t_0\rightarrow -\infty $ to $t_{\protect \rm f}\rightarrow \infty $, one
  recovers the well known Landau-Zener formula \cite {Zener:1932ws}. However,
  when the duration of the coupling becomes finite, one finds the appearance of
  oscillations as a function of the detuning due to the behaviour of the
  parabolic cylinder functions.}\BibitemShut {Stop}%
\bibitem [{\citenamefont {Tirrito}\ \emph {et~al.}(2019)\citenamefont
  {Tirrito}, \citenamefont {Rizzi}, \citenamefont {Sierra}, \citenamefont
  {Lewenstein},\ and\ \citenamefont {Bermudez}}]{PhysRevB.99.125106}%
  \BibitemOpen
  \bibfield  {author} {\bibinfo {author} {\bibfnamefont {E.}~\bibnamefont
  {Tirrito}}, \bibinfo {author} {\bibfnamefont {M.}~\bibnamefont {Rizzi}},
  \bibinfo {author} {\bibfnamefont {G.}~\bibnamefont {Sierra}}, \bibinfo
  {author} {\bibfnamefont {M.}~\bibnamefont {Lewenstein}}, \ and\ \bibinfo
  {author} {\bibfnamefont {A.}~\bibnamefont {Bermudez}},\ }\href {\doibase
  10.1103/PhysRevB.99.125106} {\bibfield  {journal} {\bibinfo  {journal} {Phys.
  Rev. B}\ }\textbf {\bibinfo {volume} {99}},\ \bibinfo {pages} {125106}
  (\bibinfo {year} {2019})}\BibitemShut {NoStop}%
\bibitem [{\citenamefont {Ziegler}\ \emph
  {et~al.}(2022{\natexlab{a}})\citenamefont {Ziegler}, \citenamefont {Tirrito},
  \citenamefont {Lewenstein}, \citenamefont {Hands},\ and\ \citenamefont
  {Bermudez}}]{PhysRevResearch.4.L042012}%
  \BibitemOpen
  \bibfield  {author} {\bibinfo {author} {\bibfnamefont {L.}~\bibnamefont
  {Ziegler}}, \bibinfo {author} {\bibfnamefont {E.}~\bibnamefont {Tirrito}},
  \bibinfo {author} {\bibfnamefont {M.}~\bibnamefont {Lewenstein}}, \bibinfo
  {author} {\bibfnamefont {S.}~\bibnamefont {Hands}}, \ and\ \bibinfo {author}
  {\bibfnamefont {A.}~\bibnamefont {Bermudez}},\ }\href {\doibase
  10.1103/PhysRevResearch.4.L042012} {\bibfield  {journal} {\bibinfo  {journal}
  {Phys. Rev. Research}\ }\textbf {\bibinfo {volume} {4}},\ \bibinfo {pages}
  {L042012} (\bibinfo {year} {2022}{\natexlab{a}})}\BibitemShut {NoStop}%
\bibitem [{\citenamefont {Ziegler}\ \emph
  {et~al.}(2022{\natexlab{b}})\citenamefont {Ziegler}, \citenamefont {Tirrito},
  \citenamefont {Lewenstein}, \citenamefont {Hands},\ and\ \citenamefont
  {Bermudez}}]{ZIEGLER2022168763}%
  \BibitemOpen
  \bibfield  {author} {\bibinfo {author} {\bibfnamefont {L.}~\bibnamefont
  {Ziegler}}, \bibinfo {author} {\bibfnamefont {E.}~\bibnamefont {Tirrito}},
  \bibinfo {author} {\bibfnamefont {M.}~\bibnamefont {Lewenstein}}, \bibinfo
  {author} {\bibfnamefont {S.}~\bibnamefont {Hands}}, \ and\ \bibinfo {author}
  {\bibfnamefont {A.}~\bibnamefont {Bermudez}},\ }\href {\doibase
  https://doi.org/10.1016/j.aop.2022.168763} {\bibfield  {journal} {\bibinfo
  {journal} {Annals of Physics}\ }\textbf {\bibinfo {volume} {439}},\ \bibinfo
  {pages} {168763} (\bibinfo {year} {2022}{\natexlab{b}})}\BibitemShut
  {NoStop}%
\bibitem [{\citenamefont {Bermudez}\ \emph {et~al.}(2024)\citenamefont
  {Bermudez}, \citenamefont {González-Cuadra},\ and\ \citenamefont
  {Hands}}]{10.21468/SciPostPhys.17.1.003}%
  \BibitemOpen
  \bibfield  {author} {\bibinfo {author} {\bibfnamefont {A.}~\bibnamefont
  {Bermudez}}, \bibinfo {author} {\bibfnamefont {D.}~\bibnamefont
  {González-Cuadra}}, \ and\ \bibinfo {author} {\bibfnamefont
  {S.}~\bibnamefont {Hands}},\ }\href {\doibase 10.21468/SciPostPhys.17.1.003}
  {\bibfield  {journal} {\bibinfo  {journal} {SciPost Phys.}\ }\textbf
  {\bibinfo {volume} {17}},\ \bibinfo {pages} {003} (\bibinfo {year}
  {2024})}\BibitemShut {NoStop}%
\bibitem [{\citenamefont {Huang}\ and\ \citenamefont
  {Diehl}(2024)}]{huang2024interactioninducedtopologicalphasetransition}%
  \BibitemOpen
  \bibfield  {author} {\bibinfo {author} {\bibfnamefont {Z.-M.}\ \bibnamefont
  {Huang}}\ and\ \bibinfo {author} {\bibfnamefont {S.}~\bibnamefont {Diehl}},\
  }\href {https://arxiv.org/abs/2407.04779} {\enquote {\bibinfo {title}
  {Interaction-induced topological phase transition at finite temperature},}\ }
  (\bibinfo {year} {2024}),\ \Eprint {http://arxiv.org/abs/2407.04779}
  {arXiv:2407.04779 [cond-mat.quant-gas]} \BibitemShut {NoStop}%
\bibitem [{\citenamefont {Inagaki}\ \emph {et~al.}(1997)\citenamefont
  {Inagaki}, \citenamefont {Muta},\ and\ \citenamefont
  {Odintsov}}]{inagaki_dynamical_1997}%
  \BibitemOpen
  \bibfield  {author} {\bibinfo {author} {\bibfnamefont {T.}~\bibnamefont
  {Inagaki}}, \bibinfo {author} {\bibfnamefont {T.}~\bibnamefont {Muta}}, \
  and\ \bibinfo {author} {\bibfnamefont {S.~D.}\ \bibnamefont {Odintsov}},\
  }\href {\doibase 10.1143/PTPS.127.93} {\bibfield  {journal} {\bibinfo
  {journal} {Progress of Theoretical Physics Supplement}\ }\textbf {\bibinfo
  {volume} {127}},\ \bibinfo {pages} {93} (\bibinfo {year} {1997})}\BibitemShut
  {NoStop}%
\bibitem [{\citenamefont {Ebert}\ \emph {et~al.}(2008)\citenamefont {Ebert},
  \citenamefont {Klimenko}, \citenamefont {Tyukov},\ and\ \citenamefont
  {Zhukovsky}}]{ebert_pion_2008}%
  \BibitemOpen
  \bibfield  {author} {\bibinfo {author} {\bibfnamefont {D.}~\bibnamefont
  {Ebert}}, \bibinfo {author} {\bibfnamefont {K.~G.}\ \bibnamefont {Klimenko}},
  \bibinfo {author} {\bibfnamefont {A.~V.}\ \bibnamefont {Tyukov}}, \ and\
  \bibinfo {author} {\bibfnamefont {V.~C.}\ \bibnamefont {Zhukovsky}},\ }\href
  {\doibase 10.1140/epjc/s10052-008-0667-6} {\bibfield  {journal} {\bibinfo
  {journal} {The European Physical Journal C}\ }\textbf {\bibinfo {volume}
  {58}},\ \bibinfo {pages} {57} (\bibinfo {year} {2008})}\BibitemShut {NoStop}%
\bibitem [{\citenamefont {Flachi}\ and\ \citenamefont
  {Tanaka}(2011)}]{flachi_chiral_2011}%
  \BibitemOpen
  \bibfield  {author} {\bibinfo {author} {\bibfnamefont {A.}~\bibnamefont
  {Flachi}}\ and\ \bibinfo {author} {\bibfnamefont {T.}~\bibnamefont
  {Tanaka}},\ }\href {\doibase 10.1007/JHEP02(2011)026} {\bibfield  {journal}
  {\bibinfo  {journal} {Journal of High Energy Physics}\ }\textbf {\bibinfo
  {volume} {2011}},\ \bibinfo {pages} {26} (\bibinfo {year} {2011})},\ \bibinfo
  {note} {arXiv:1012.0463 [gr-qc, physics:hep-lat, physics:hep-ph,
  physics:hep-th, physics:math-ph]}\BibitemShut {NoStop}%
\bibitem [{\citenamefont {Forkel}(1992)}]{forkel_dynamical_1992}%
  \BibitemOpen
  \bibfield  {author} {\bibinfo {author} {\bibfnamefont {H.}~\bibnamefont
  {Forkel}},\ }\href {\doibase 10.1016/0370-2693(92)90763-T} {\bibfield
  {journal} {\bibinfo  {journal} {Physics Letters B}\ }\textbf {\bibinfo
  {volume} {280}},\ \bibinfo {pages} {5} (\bibinfo {year} {1992})}\BibitemShut
  {NoStop}%
\bibitem [{\citenamefont {BUCHBINDER}\ and\ \citenamefont
  {KIRILLOVA}(2012)}]{buchbinder_gross-neveu_2012}%
  \BibitemOpen
  \bibfield  {author} {\bibinfo {author} {\bibfnamefont {I.~L.}\ \bibnamefont
  {BUCHBINDER}}\ and\ \bibinfo {author} {\bibfnamefont {E.~N.}\ \bibnamefont
  {KIRILLOVA}},\ }\href {\doibase 10.1142/S0217751X89000054} {\bibfield
  {journal} {\bibinfo  {journal} {International Journal of Modern Physics A}\ }
  (\bibinfo {year} {2012}),\ 10.1142/S0217751X89000054},\ \bibinfo {note}
  {publisher: World Scientific Publishing Company}\BibitemShut {NoStop}%
\bibitem [{\citenamefont {Saharian}\ \emph {et~al.}(2021)\citenamefont
  {Saharian}, \citenamefont {de~Mello}, \citenamefont {Kotanjyan},\ and\
  \citenamefont {Petrosyan}}]{saharian_fermionic_2021}%
  \BibitemOpen
  \bibfield  {author} {\bibinfo {author} {\bibfnamefont {A.~A.}\ \bibnamefont
  {Saharian}}, \bibinfo {author} {\bibfnamefont {E.~R.~B.}\ \bibnamefont
  {de~Mello}}, \bibinfo {author} {\bibfnamefont {A.~S.}\ \bibnamefont
  {Kotanjyan}}, \ and\ \bibinfo {author} {\bibfnamefont {T.~A.}\ \bibnamefont
  {Petrosyan}},\ }\href {\doibase 10.1007/s10511-021-09713-z} {\bibfield
  {journal} {\bibinfo  {journal} {Astrophysics}\ }\textbf {\bibinfo {volume}
  {64}},\ \bibinfo {pages} {529} (\bibinfo {year} {2021})},\ \bibinfo {note}
  {arXiv:2110.12677 [gr-qc, physics:hep-th, physics:quant-ph]}\BibitemShut
  {NoStop}%
\bibitem [{\citenamefont {Flachi}(2013)}]{flachi_dual_2013}%
  \BibitemOpen
  \bibfield  {author} {\bibinfo {author} {\bibfnamefont {A.}~\bibnamefont
  {Flachi}},\ }\href {\doibase 10.1103/PhysRevD.88.085011} {\bibfield
  {journal} {\bibinfo  {journal} {Physical Review D}\ }\textbf {\bibinfo
  {volume} {88}},\ \bibinfo {pages} {085011} (\bibinfo {year} {2013})},\
  \bibinfo {note} {arXiv:1304.6880 [hep-lat, physics:hep-ph,
  physics:hep-th]}\BibitemShut {NoStop}%
\bibitem [{\citenamefont {Inagaki}\ \emph {et~al.}(1993)\citenamefont
  {Inagaki}, \citenamefont {Muta},\ and\ \citenamefont
  {Odinstov}}]{inagaki_nambu-jona-lasinio_1993}%
  \BibitemOpen
  \bibfield  {author} {\bibinfo {author} {\bibfnamefont {T.}~\bibnamefont
  {Inagaki}}, \bibinfo {author} {\bibfnamefont {T.}~\bibnamefont {Muta}}, \
  and\ \bibinfo {author} {\bibfnamefont {S.~D.}\ \bibnamefont {Odinstov}},\
  }\href {\doibase 10.1142/S0217732393001835} {\bibfield  {journal} {\bibinfo
  {journal} {Modern Physics Letters A}\ }\textbf {\bibinfo {volume} {08}},\
  \bibinfo {pages} {2117} (\bibinfo {year} {1993})},\ \bibinfo {note}
  {arXiv:hep-th/9306023}\BibitemShut {NoStop}%
\bibitem [{\citenamefont {Elizalde}\ \emph {et~al.}(1994)\citenamefont
  {Elizalde}, \citenamefont {Odintsov},\ and\ \citenamefont
  {Shil'nov}}]{elizalde_chiral_1994}%
  \BibitemOpen
  \bibfield  {author} {\bibinfo {author} {\bibfnamefont {E.}~\bibnamefont
  {Elizalde}}, \bibinfo {author} {\bibfnamefont {S.~D.}\ \bibnamefont
  {Odintsov}}, \ and\ \bibinfo {author} {\bibfnamefont {Y.~I.}\ \bibnamefont
  {Shil'nov}},\ }\href {\doibase 10.1142/S0217732394000733} {\bibfield
  {journal} {\bibinfo  {journal} {Modern Physics Letters A}\ }\textbf {\bibinfo
  {volume} {09}},\ \bibinfo {pages} {913} (\bibinfo {year} {1994})},\ \bibinfo
  {note} {arXiv:hep-th/9401056}\BibitemShut {NoStop}%
\bibitem [{\citenamefont {Friedmann}(1924)}]{Friedmann1924berDM}%
  \BibitemOpen
  \bibfield  {author} {\bibinfo {author} {\bibfnamefont {A.}~\bibnamefont
  {Friedmann}},\ }\href {https://api.semanticscholar.org/CorpusID:120551579}
  {\bibfield  {journal} {\bibinfo  {journal} {Zeitschrift f{\"u}r Physik}\
  }\textbf {\bibinfo {volume} {21}},\ \bibinfo {pages} {326} (\bibinfo {year}
  {1924})}\BibitemShut {NoStop}%
\bibitem [{\citenamefont {Friedman}(1922)}]{Friedman1922berDK}%
  \BibitemOpen
  \bibfield  {author} {\bibinfo {author} {\bibfnamefont {A.}~\bibnamefont
  {Friedman}},\ }\href {https://api.semanticscholar.org/CorpusID:125190902}
  {\bibfield  {journal} {\bibinfo  {journal} {Zeitschrift f{\"u}r Physik}\
  }\textbf {\bibinfo {volume} {10}},\ \bibinfo {pages} {377} (\bibinfo {year}
  {1922})}\BibitemShut {NoStop}%
\bibitem [{\citenamefont {Guth}(1981)}]{PhysRevD.23.347}%
  \BibitemOpen
  \bibfield  {author} {\bibinfo {author} {\bibfnamefont {A.~H.}\ \bibnamefont
  {Guth}},\ }\href {\doibase 10.1103/PhysRevD.23.347} {\bibfield  {journal}
  {\bibinfo  {journal} {Phys. Rev. D}\ }\textbf {\bibinfo {volume} {23}},\
  \bibinfo {pages} {347} (\bibinfo {year} {1981})}\BibitemShut {NoStop}%
\bibitem [{\citenamefont {Huang}\ and\ \citenamefont
  {Parker}(2009)}]{Huang:2008kh}%
  \BibitemOpen
  \bibfield  {author} {\bibinfo {author} {\bibfnamefont {X.}~\bibnamefont
  {Huang}}\ and\ \bibinfo {author} {\bibfnamefont {L.}~\bibnamefont {Parker}},\
  }\href {\doibase 10.1103/PhysRevD.79.024020} {\bibfield  {journal} {\bibinfo
  {journal} {Phys. Rev. D}\ }\textbf {\bibinfo {volume} {79}},\ \bibinfo
  {pages} {024020} (\bibinfo {year} {2009})},\ \Eprint
  {http://arxiv.org/abs/0811.2296} {arXiv:0811.2296 [hep-th]} \BibitemShut
  {NoStop}%
\bibitem [{\citenamefont {Min\'a\v{r}}\ and\ \citenamefont
  {Gr\'emaud}(2015)}]{Minar:2013hva}%
  \BibitemOpen
  \bibfield  {author} {\bibinfo {author} {\bibfnamefont {J.}~\bibnamefont
  {Min\'a\v{r}}}\ and\ \bibinfo {author} {\bibfnamefont {B.}~\bibnamefont
  {Gr\'emaud}},\ }\href {\doibase 10.1088/1751-8113/48/16/165001} {\bibfield
  {journal} {\bibinfo  {journal} {J. Phys. A}\ }\textbf {\bibinfo {volume}
  {48}},\ \bibinfo {pages} {165001} (\bibinfo {year} {2015})},\ \Eprint
  {http://arxiv.org/abs/1304.0889} {arXiv:1304.0889 [cond-mat.quant-gas]}
  \BibitemShut {NoStop}%
\bibitem [{\citenamefont {Gorbatenko}\ and\ \citenamefont
  {Neznamov}(2010)}]{Gorbatenko:2010bh}%
  \BibitemOpen
  \bibfield  {author} {\bibinfo {author} {\bibfnamefont {M.~V.}\ \bibnamefont
  {Gorbatenko}}\ and\ \bibinfo {author} {\bibfnamefont {V.~P.}\ \bibnamefont
  {Neznamov}},\ }\href {\doibase 10.1103/PhysRevD.82.104056} {\bibfield
  {journal} {\bibinfo  {journal} {Phys. Rev. D}\ }\textbf {\bibinfo {volume}
  {82}},\ \bibinfo {pages} {104056} (\bibinfo {year} {2010})},\ \Eprint
  {http://arxiv.org/abs/1007.4631} {arXiv:1007.4631 [gr-qc]} \BibitemShut
  {NoStop}%
\bibitem [{\citenamefont {Gorbatenko}\ and\ \citenamefont
  {Neznamov}(2011)}]{Gorbatenko:2011rd}%
  \BibitemOpen
  \bibfield  {author} {\bibinfo {author} {\bibfnamefont {M.~V.}\ \bibnamefont
  {Gorbatenko}}\ and\ \bibinfo {author} {\bibfnamefont {V.~P.}\ \bibnamefont
  {Neznamov}},\ }\href {\doibase 10.1103/PhysRevD.83.105002} {\bibfield
  {journal} {\bibinfo  {journal} {Phys. Rev. D}\ }\textbf {\bibinfo {volume}
  {83}},\ \bibinfo {pages} {105002} (\bibinfo {year} {2011})},\ \Eprint
  {http://arxiv.org/abs/1102.4067} {arXiv:1102.4067 [gr-qc]} \BibitemShut
  {NoStop}%
\bibitem [{\citenamefont {Parker}(1980)}]{Parker:1980kw}%
  \BibitemOpen
  \bibfield  {author} {\bibinfo {author} {\bibfnamefont {L.}~\bibnamefont
  {Parker}},\ }\href {\doibase 10.1103/PhysRevD.22.1922} {\bibfield  {journal}
  {\bibinfo  {journal} {Phys. Rev. D}\ }\textbf {\bibinfo {volume} {22}},\
  \bibinfo {pages} {1922} (\bibinfo {year} {1980})}\BibitemShut {NoStop}%
\bibitem [{\citenamefont {Shi}\ \emph {et~al.}(2023)\citenamefont {Shi},
  \citenamefont {He}, \citenamefont {Liu},\ and\ \citenamefont
  {Zhang}}]{Shi_2023}%
  \BibitemOpen
  \bibfield  {author} {\bibinfo {author} {\bibfnamefont {Y.-R.}\ \bibnamefont
  {Shi}}, \bibinfo {author} {\bibfnamefont {Y.-Y.}\ \bibnamefont {He}},
  \bibinfo {author} {\bibfnamefont {R.}~\bibnamefont {Liu}}, \ and\ \bibinfo
  {author} {\bibfnamefont {W.}~\bibnamefont {Zhang}},\ }\href {\doibase
  10.1088/1367-2630/acfbf2} {\bibfield  {journal} {\bibinfo  {journal} {New
  Journal of Physics}\ }\textbf {\bibinfo {volume} {25}},\ \bibinfo {pages}
  {093049} (\bibinfo {year} {2023})}\BibitemShut {NoStop}%
\bibitem [{\citenamefont {Bravyi}(2005)}]{Bravyi05}%
  \BibitemOpen
  \bibfield  {author} {\bibinfo {author} {\bibfnamefont {S.}~\bibnamefont
  {Bravyi}},\ }\href@noop {} {\bibfield  {journal} {\bibinfo  {journal}
  {Quantum Info. Comput.}\ }\textbf {\bibinfo {volume} {5}},\ \bibinfo {pages}
  {216–238} (\bibinfo {year} {2005})}\BibitemShut {NoStop}%
\bibitem [{\citenamefont {Motta}\ \emph {et~al.}(2019)\citenamefont {Motta},
  \citenamefont {Sun}, \citenamefont {Tan}, \citenamefont {Rourke},
  \citenamefont {Ye}, \citenamefont {Minnich}, \citenamefont {Brand\~ao},\ and\
  \citenamefont {Chan}}]{Motta:2019yya}%
  \BibitemOpen
  \bibfield  {author} {\bibinfo {author} {\bibfnamefont {M.}~\bibnamefont
  {Motta}}, \bibinfo {author} {\bibfnamefont {C.}~\bibnamefont {Sun}}, \bibinfo
  {author} {\bibfnamefont {A.~T.~K.}\ \bibnamefont {Tan}}, \bibinfo {author}
  {\bibfnamefont {M.~J.~O.}\ \bibnamefont {Rourke}}, \bibinfo {author}
  {\bibfnamefont {E.}~\bibnamefont {Ye}}, \bibinfo {author} {\bibfnamefont
  {A.~J.}\ \bibnamefont {Minnich}}, \bibinfo {author} {\bibfnamefont {F.~G.
  S.~L.}\ \bibnamefont {Brand\~ao}}, \ and\ \bibinfo {author} {\bibfnamefont
  {G.~K.-L.}\ \bibnamefont {Chan}},\ }\href {\doibase
  10.1038/s41567-019-0704-4} {\bibfield  {journal} {\bibinfo  {journal} {Nature
  Phys.}\ }\textbf {\bibinfo {volume} {16}},\ \bibinfo {pages} {205} (\bibinfo
  {year} {2019})},\ \Eprint {http://arxiv.org/abs/1901.07653} {arXiv:1901.07653
  [quant-ph]} \BibitemShut {NoStop}%
\bibitem [{\citenamefont {{Fock}}(1930)}]{1930ZPhy...61..126F}%
  \BibitemOpen
  \bibfield  {author} {\bibinfo {author} {\bibfnamefont {V.}~\bibnamefont
  {{Fock}}},\ }\href {\doibase 10.1007/BF01340294} {\bibfield  {journal}
  {\bibinfo  {journal} {Zeitschrift fur Physik}\ }\textbf {\bibinfo {volume}
  {61}},\ \bibinfo {pages} {126} (\bibinfo {year} {1930})}\BibitemShut
  {NoStop}%
\bibitem [{\citenamefont {{Hartree}}\ and\ \citenamefont
  {{Hartree}}(1935)}]{1935RSPSA.150....9H}%
  \BibitemOpen
  \bibfield  {author} {\bibinfo {author} {\bibfnamefont {D.~R.}\ \bibnamefont
  {{Hartree}}}\ and\ \bibinfo {author} {\bibfnamefont {W.}~\bibnamefont
  {{Hartree}}},\ }\href {\doibase 10.1098/rspa.1935.0085} {\bibfield  {journal}
  {\bibinfo  {journal} {Proceedings of the Royal Society of London Series A}\
  }\textbf {\bibinfo {volume} {150}},\ \bibinfo {pages} {9} (\bibinfo {year}
  {1935})}\BibitemShut {NoStop}%
\bibitem [{\citenamefont {Slater}(1951)}]{PhysRev.81.385}%
  \BibitemOpen
  \bibfield  {author} {\bibinfo {author} {\bibfnamefont {J.~C.}\ \bibnamefont
  {Slater}},\ }\href {\doibase 10.1103/PhysRev.81.385} {\bibfield  {journal}
  {\bibinfo  {journal} {Phys. Rev.}\ }\textbf {\bibinfo {volume} {81}},\
  \bibinfo {pages} {385} (\bibinfo {year} {1951})}\BibitemShut {NoStop}%
\bibitem [{\citenamefont {Gross}\ and\ \citenamefont
  {Wilczek}(1973)}]{PhysRevLett.30.1343}%
  \BibitemOpen
  \bibfield  {author} {\bibinfo {author} {\bibfnamefont {D.~J.}\ \bibnamefont
  {Gross}}\ and\ \bibinfo {author} {\bibfnamefont {F.}~\bibnamefont
  {Wilczek}},\ }\href {\doibase 10.1103/PhysRevLett.30.1343} {\bibfield
  {journal} {\bibinfo  {journal} {Phys. Rev. Lett.}\ }\textbf {\bibinfo
  {volume} {30}},\ \bibinfo {pages} {1343} (\bibinfo {year}
  {1973})}\BibitemShut {NoStop}%
\bibitem [{\citenamefont {Politzer}(1973)}]{PhysRevLett.30.1346}%
  \BibitemOpen
  \bibfield  {author} {\bibinfo {author} {\bibfnamefont {H.~D.}\ \bibnamefont
  {Politzer}},\ }\href {\doibase 10.1103/PhysRevLett.30.1346} {\bibfield
  {journal} {\bibinfo  {journal} {Phys. Rev. Lett.}\ }\textbf {\bibinfo
  {volume} {30}},\ \bibinfo {pages} {1346} (\bibinfo {year}
  {1973})}\BibitemShut {NoStop}%
\bibitem [{\citenamefont {Gattringer}\ and\ \citenamefont
  {Lang}(2010{\natexlab{b}})}]{Gattringer:2010zz}%
  \BibitemOpen
  \bibfield  {author} {\bibinfo {author} {\bibfnamefont {C.}~\bibnamefont
  {Gattringer}}\ and\ \bibinfo {author} {\bibfnamefont {C.~B.}\ \bibnamefont
  {Lang}},\ }\href {\doibase 10.1007/978-3-642-01850-3} {\emph {\bibinfo
  {title} {{Quantum chromodynamics on the lattice}}}},\ Vol.\ \bibinfo {volume}
  {788}\ (\bibinfo  {publisher} {Springer},\ \bibinfo {address} {Berlin},\
  \bibinfo {year} {2010})\BibitemShut {NoStop}%
\bibitem [{\citenamefont {Rothe}(2012)}]{Rothe:1992nt}%
  \BibitemOpen
  \bibfield  {author} {\bibinfo {author} {\bibfnamefont {H.~J.}\ \bibnamefont
  {Rothe}},\ }\href {\doibase 10.1142/8229} {\emph {\bibinfo {title} {{Lattice
  Gauge Theories : An Introduction (Fourth Edition)}}}},\ Vol.~\bibinfo
  {volume} {43}\ (\bibinfo  {publisher} {World Scientific Publishing Company},\
  \bibinfo {year} {2012})\BibitemShut {NoStop}%
\bibitem [{\citenamefont {Sekine}\ \emph {et~al.}(2013)\citenamefont {Sekine},
  \citenamefont {Nakano}, \citenamefont {Araki},\ and\ \citenamefont
  {Nomura}}]{PhysRevB.87.165142}%
  \BibitemOpen
  \bibfield  {author} {\bibinfo {author} {\bibfnamefont {A.}~\bibnamefont
  {Sekine}}, \bibinfo {author} {\bibfnamefont {T.~Z.}\ \bibnamefont {Nakano}},
  \bibinfo {author} {\bibfnamefont {Y.}~\bibnamefont {Araki}}, \ and\ \bibinfo
  {author} {\bibfnamefont {K.}~\bibnamefont {Nomura}},\ }\href {\doibase
  10.1103/PhysRevB.87.165142} {\bibfield  {journal} {\bibinfo  {journal} {Phys.
  Rev. B}\ }\textbf {\bibinfo {volume} {87}},\ \bibinfo {pages} {165142}
  (\bibinfo {year} {2013})}\BibitemShut {NoStop}%
\bibitem [{\citenamefont {Araki}\ and\ \citenamefont
  {Kimura}(2013)}]{PhysRevB.87.205440}%
  \BibitemOpen
  \bibfield  {author} {\bibinfo {author} {\bibfnamefont {Y.}~\bibnamefont
  {Araki}}\ and\ \bibinfo {author} {\bibfnamefont {T.}~\bibnamefont {Kimura}},\
  }\href {\doibase 10.1103/PhysRevB.87.205440} {\bibfield  {journal} {\bibinfo
  {journal} {Phys. Rev. B}\ }\textbf {\bibinfo {volume} {87}},\ \bibinfo
  {pages} {205440} (\bibinfo {year} {2013})}\BibitemShut {NoStop}%
\bibitem [{\citenamefont {Haro}\ and\ \citenamefont
  {Elizalde}(2008)}]{Haro_2008}%
  \BibitemOpen
  \bibfield  {author} {\bibinfo {author} {\bibfnamefont {J.}~\bibnamefont
  {Haro}}\ and\ \bibinfo {author} {\bibfnamefont {E.}~\bibnamefont
  {Elizalde}},\ }\href {\doibase 10.1088/1751-8113/41/37/372003} {\bibfield
  {journal} {\bibinfo  {journal} {Journal of Physics A: Mathematical and
  Theoretical}\ }\textbf {\bibinfo {volume} {41}},\ \bibinfo {pages} {372003}
  (\bibinfo {year} {2008})}\BibitemShut {NoStop}%
\bibitem [{\citenamefont {Abramowitz}\ and\ \citenamefont
  {Stegun}(1964)}]{abramowitz+stegun}%
  \BibitemOpen
  \bibfield  {author} {\bibinfo {author} {\bibfnamefont {M.}~\bibnamefont
  {Abramowitz}}\ and\ \bibinfo {author} {\bibfnamefont {I.~A.}\ \bibnamefont
  {Stegun}},\ }\href@noop {} {\emph {\bibinfo {title} {Handbook of Mathematical
  Functions with Formulas, Graphs, and Mathematical Tables}}},\ \bibinfo
  {edition} {ninth dover printing, tenth gpo printing}\ ed.\ (\bibinfo
  {publisher} {Dover},\ \bibinfo {address} {New York},\ \bibinfo {year}
  {1964})\BibitemShut {NoStop}%
\bibitem [{\citenamefont {Zener}(1932)}]{Zener:1932ws}%
  \BibitemOpen
  \bibfield  {author} {\bibinfo {author} {\bibfnamefont {C.}~\bibnamefont
  {Zener}},\ }\href {\doibase 10.1098/rspa.1932.0165} {\bibfield  {journal}
  {\bibinfo  {journal} {Proc. Roy. Soc. Lond. A}\ }\textbf {\bibinfo {volume}
  {137}},\ \bibinfo {pages} {696} (\bibinfo {year} {1932})}\BibitemShut
  {NoStop}%
\end{thebibliography}%

\end{document}